\documentclass[12pt,a4paper]{article}
\usepackage{amsmath}
\usepackage{amssymb}

\newenvironment{inmargins}[1]{
    \begin{list}{}{
        \leftmargin=#1
        \rightmargin=#1
        \parsep=0pt
        \partopsep=0pt
    }
    \item[]
}{\end{list}}

\newcommand{\be}{\begin{equation}}
\newcommand{\ee}{\end{equation}}
\newcommand{\bea}{\begin{eqnarray}}
\newcommand{\eea}{\end{eqnarray}}
\newcommand{\nn}{\nonumber}

\newcommand{\ba}{\begin{eqnarray}}
\newcommand{\ea}{\end{eqnarray}}

\makeatletter
\def\section{\@startsection{section}{1}{\z@}{-3.5ex plus -1ex minus -.2ex}{2.3ex plus .2ex}{\normalsize \bf}}
\def\subsection{\@startsection{subsection}{2}{\z@}{-3.25ex plus -1ex minus -.2ex}{1.5ex plus .2ex}{\normalsize \sl}}
\def\subsubsection{\@startsection{subsubsection}{3}{\z@}{-3.25ex plus -1ex minus -.2ex}{1.5ex plus .2ex}{\normalsize}}
\makeatother
\newcommand{\eqn}[1]{(\ref{#1})}
\newcommand{\eq}[1]{(\ref{#1})}

\makeatletter
\def\@evenhead{
    \vbox{\hbox to\hsize{\bf \thepage \hfill \sl \evenmark}
    }
}
\def\@oddhead{
    \vbox{
        \hbox to\hsize{\oddmarkA \oddmarkB \hfill \oddmarkC}
        \hbox to\hsize{\hfill \oddmarkD}
        \vspace{.03in}
        \hbox to\hsize{\hfill \oddmarkE}
        \vspace{.03in}
        \hbox to\hsize{\hfill \oddmarkF}
    }
}
\def\@evenfoot{\vbox{\hbox to\hsize{\hrulefill}
    \vspace{-40pt} \hbox to \hsize{\today \sl \footmsg \hfill}
    }
}
\def\@oddfoot{
    \vspace{-40pt}
    \hbox to \hsize{ \footmsgA \hfill }
}
\makeatother

\def\evenmark{\bf entropy}
\def\oddmarkA{{} {} {}}
\def\oddmarkB{\thepagerange}
\def\oddmarkC{}
\def\oddmarkD{{\Huge \bf entropy}}
\def\oddmarkE{\normalsize {\bf ISSN 1099-4300}}
\def\oddmarkF{{www.mdpi.org/entropy/}}

\def\thedoctitle{\bf { {Extremal Black Holes in Supergravity
and the Bekenstein--Hawking Entropy}}}

\def\theauthorname{L. Andrianopoli$^1$ and R. D'Auria$^2$}

\def\authoraddress{
1. Dipartimento di Fisica, Politecnico di Torino,
 Corso Duca degli Abruzzi 24, I-10129 Torino
and Istituto Nazionale di Fisica Nucleare (INFN) - Sezione di
Torino, Italy
\\ E-mail: andrianopoli@athena.polito.it
\\
2. Dipartimento di Fisica, Politecnico di Torino,
 Corso Duca degli Abruzzi 24, I-10129 Torino
and Istituto Nazionale di Fisica Nucleare (INFN) - Sezione di
Torino, Italy
\\ E-mail: dauria@polito.it}

\begin{document}

{
    \noindent {\\ \\}
}

{
    \LARGE \noindent \thedoctitle
} \\

{
    \noindent {\bf {\theauthorname}}
} \\

{
    \normalsize \noindent {\authoraddress}
} \\


\vspace{2pt} \hbox to \hsize{\hrulefill}
\vspace{.1in}

\begin{inmargins}{1cm}
\noindent {\bf Abstract:} We review some results on the
connection among supergravity central charges, BPS states and
Bekenstein--Hawking entropy. In particular,  $N=2$ supergravity
in four dimensions is studied in detail. For higher $N$
supergravities we just give an account of the general theory
specializing the discussion to the $N=8$ case when one half of
supersymmetry is preserved. We stress the fact that for extremal
supergravity black holes the entropy formula is topological, that
is the entropy turns out to be a moduli independent quantity and
can be written in terms of invariants of the duality group of the
supergravity theory.
\\

\end{inmargins}

\vspace{2pt} \hbox to \hsize{\hrulefill}

\vfill


\def\oddmarkC{\thepage}
\def\oddmarkB{}
\def\oddmarkD{}
\def\oddmarkE{}
\def\oddmarkF{}
\def\footmsgA{}
\def\sk{\vskip .4cm}
\def\ib{{\bar \imath}}
\def\jb{{\bar \jmath}}
\def\Im{{\rm Im ~}}
\def\Re{{\rm Re ~}}
\def\IP{\relax{\rm I\kern-.18em P}}
\def\arccosh{{\rm arccosh ~}}
\def\Es{\bf E_{7(7)}}
\def\muun{\underline \mu}
\def\mun{\underline m}
\def\nuun{\underline \nu}
\def\nun{\underline n}
\def\buun{\underline \bullet}
\def\Eb{{\bf E}}
\def\bu{\bullet}
\def\we{\wedge}
\font\cmss=cmss10 \font\cmsss=cmss10 at 7pt
\def\twomat#1#2#3#4{\left(\matrix{#1 & #2 \cr #3 & #4}\right)}
\def\inbar{\vrule height1.5ex width.4pt depth0pt}
\def\IC{\relax\,\hbox{$\inbar\kern-.3em{\rm C}$}}
\def\IG{\relax\,\hbox{$\inbar\kern-.3em{\rm G}$}}
\def\IB{\relax{\rm I\kern-.18em B}}
\def\ID{\relax{\rm I\kern-.18em D}}
\def\IL{\relax{\rm I\kern-.18em L}}
\def\IF{\relax{\rm I\kern-.18em F}}
\def\IH{\relax{\rm I\kern-.18em H}}
\def\II{\relax{\rm I\kern-.17em I}}
\def\IN{\relax{\rm I\kern-.18em N}}
\def\IP{\relax{\rm I\kern-.18em P}}
\def\IQ{\relax\,\hbox{$\inbar\kern-.3em{\rm Q}$}}
\def\bfzero{\relax\,\hbox{$\inbar\kern-.3em{\rm 0}$}}
\def\IK{\relax{\rm I\kern-.18em K}}
\def\IG{\relax\,\hbox{$\inbar\kern-.3em{\rm G}$}}
 \font\cmss=cmss10 \font\cmsss=cmss10 at 7pt
\def\IR{\relax{\rm I\kern-.18em R}}
\def\ZZ{\relax\ifmmode\mathchoice
{\hbox{\cmss Z\kern-.4em Z}}{\hbox{\cmss Z\kern-.4em Z}}
{\lower.9pt\hbox{\cmsss Z\kern-.4em Z}} {\lower1.2pt\hbox{\cmsss
Z\kern-.4em Z}}\else{\cmss Z\kern-.4em Z}\fi}
\def\bfone{\relax{\rm 1\kern-.35em 1}}
\def\dop{{\rm d}\hskip -1pt}
\def\real{{\rm Re}\hskip 1pt}
\def\trace{{\rm Tr}\hskip 1pt}
\def\ii{{\rm i}}
\def\diag{{\rm diag}}
\def\sch#1#2{\{#1;#2\}}
\def\IU{\relax\,\hbox{$\inbar\kern-.3em{\rm U}$}}
\def\sk{\vskip .4cm}
\def\noi{\noindent}
\def\om{\omega}
\def\Om{\Omega}
\def\al{\alpha}
\def\la{\lambda}
\def\be{\beta}
\def\ga{\gamma}
\def\Ga{\Gamma}
\def\de{\delta}
\def\epsi{\varepsilon}
\def\we{\wedge}
\def\part{\partial}
\def\bu{\bullet}
\def\ci{\circ}
\def\square{{\,\lower0.9pt\vbox{\hrule \hbox{\vrule height 0.2 cm
\hskip 0.2 cm \vrule height 0.2 cm}\hrule}\,}}
\def\muun{\underline \mu}
\def\mun{\underline m}
\def\nuun{\underline \nu}
\def\nun{\underline n}
\def\buun{\underline \bullet}
\def\Rb{{\bf R}}
\def\Eb{{\bf E}}
\def\gb{{\bf g}}
\def\dt{{\tilde d}}
\def\Dt{{\tilde D}}
\def\Dcal{{\cal D}}
\def\R#1#2{ R^{#1}_{~~~#2} }
\def\ome#1#2{\om^{#1}_{~#2}}
\def\Rf#1#2{ R^{\underline #1}_{~~~{\underline #2}} }
\def\Rfu#1#2{ R^{{\underline #1}{\underline #2}} }
\def\Rfd#1#2{ R_{{\underline #1}{\underline #2}} }
\def\Rfb#1#2{ {\bf R}^{\underline #1}_{~~~{\underline #2}} }
\def\omef#1#2{\om^{\underline #1}_{~{\underline #2}}}
\def\omefb#1#2{{ \omb}^{\underline #1}_{~{\underline #2}}}
\def\omefu#1#2{\om^{{\underline #1} {\underline #2}}}
\def\omefub#1#2{{\omb}^{{\underline #1} {\underline #2}}}
\def\Ef#1{E^{\underline #1}}
\def\Efb#1{{\bf E}^{\underline #1}}
\def\omb{\bf \mbox{\boldmath $\om$}}
\def\bfone{\relax{\rm 1\kern-.35em 1}}
\font\cmss=cmss10 \font\cmsss=cmss10 at 7pt
\def\a{\alpha} \def\b{\beta} \def\d{\delta}
\def\e{\epsilon} \def\c{\gamma}
\def\G{\Gamma} \def\l{\lambda}
\def\L{\Lambda} \def\s{\sigma}
\def\cA{{\cal A}} \def\cB{{\cal B}}
\def\cC{{\cal C}} \def\cD{{\cal D}}
\def\cF{{\cal F}} \def\cG{{\cal G}}
\def\cH{{\cal H}} \def\cI{{\cal I}}
\def\cJ{{\cal J}} \def\cK{{\cal K}}
\def\cL{{\cal L}} \def\cM{{\cal M}}
\def\cN{{\cal N}} \def\cO{{\cal O}}
\def\cP{{\cal P}} \def\cQ{{\cal Q}}
\def\cR{{\cal R}} \def\cV{{\cal V}}\def\cW{{\cal W}}
%
%
%
\def\crr{\crcr\noalign{\vskip {8.3333pt}}}
\def\tilde{\widetilde}
\def\bar{\overline}
\def\us#1{\underline{#1}}
\let\shat=\hat
\def\hat{\widehat}
\def\hyp{\vrule height 2.3pt width 2.5pt depth -1.5pt}
\def\Coeff#1#2{\frac{#1}{ #2}}
\def\Coe#1.#2.{\frac{#1}{ #2}}
\def\coeff#1#2{\relax{\textstyle {#1 \over #2}}\displaystyle}
\def\coe#1.#2.{\relax{\textstyle {#1 \over #2}}\displaystyle}
\def\half{{1 \over 2}}
\def\shalf{\relax{\textstyle \frac{1}{ 2}}\displaystyle}
\def\dag#1{#1\!\!\!/\,\,\,}
\def\to{\rightarrow}
\def\notin{\hbox{{$\in$}\kern-.51em\hbox{/}}}
\def\shdot{\!\cdot\!}
\def\ket#1{\,\big|\,#1\,\big>\,}
\def\bra#1{\,\big<\,#1\,\big|\,}
\def\equaltop#1{\mathrel{\mathop=^{#1}}}
\def\Trbel#1{\mathop{{\rm Tr}}_{#1}}
\def\inserteq#1{\noalign{\vskip-.2truecm\hbox{#1\hfil}
\vskip-.2cm}}
\def\attac#1{\Bigl\vert
{\phantom{X}\atop{{\rm\scriptstyle #1}}\phantom{X}}}
\def\exx#1{e^{{\displaystyle #1}}}
\def\del{\partial}
\def\delbar{\bar\partial}
\def\nex#1{$N\!=\!#1$}
\def\dex#1{$d\!=\!#1$}
\def\cex#1{$c\!=\!#1$}
\def\eg{{\it e.g.}} \def\ie{{\it i.e.}}
\def\IE{\relax{{\rm I\kern-.18em E}}}
\def\cE{{\cal E}}
\def\rt{{\cR^{(3)}}}
\def\IGam{\relax{{\rm I}\kern-.18em \Gamma}}
\def\IGa{\IA}
\def\ii{{\rm i}}
\def\inbar{\vrule height1.5ex width.4pt depth0pt}
\def\bfzero{\relax{\rm I\kern-.18em 0}}
\def\bfone{\relax{\rm 1\kern-.35em 1}}
\def\twomat#1#2#3#4{\left(\begin{array}{cc}
 {#1}&{#2}\\ {#3}&{#4}\\
\end{array}
\right)}
\def\twovec#1#2{\left(\begin{array}{c}
{#1}\\ {#2}\\
\end{array}
\right)}
\def\o#1#2{{{#1}\over{#2}}}
\newcommand{\La}{{\Lambda}}
\newcommand{\Si}{{\Sigma}}
\newcommand{\im}{{\rm Im\ }}
\def\cA{{\cal A}} \def\cB{{\cal B}}
\def\cC{{\cal C}} \def\cD{{\cal D}}
\def\cF{{\cal F}} \def\cG{{\cal G}}
\def\cH{{\cal H}} \def\cI{{\cal I}}
\def\cJ{{\cal J}} \def\cK{{\cal K}}
\def\cL{{\cal L}} \def\cM{{\cal M}}
\def\cN{{\cal N}} \def\cO{{\cal O}}
\def\cP{{\cal P}} \def\cQ{{\cal Q}}
\def\cR{{\cal R}} \def\cV{{\cal V}}\def\cW{{\cal W}}
\def\a{\alpha} \def\b{\beta} \def\d{\delta}
\def\e{\epsilon} \def\c{\gamma}
\def\G{\Gamma} \def\l{\lambda}
\def\L{\Lambda} \def\s{\sigma}
\def\T{T}


\section{Introduction:
Extremal Black Holes from Classical General Relativity to String
Theory}
\label{intro1} Black hole physics has many aspects of
great interest to physicists with very different cultural
backgrounds. These range from astrophysics to classical general
relativity, to quantum field theory in curved space--times,
particle physics  and finally string theory and supergravity.
This is not surprising since black holes are one of the basic
consequences of a fundamental theory, namely Einstein general
relativity. Furthermore black holes have fascinating
thermodynamical properties that seem to encode the deepest
properties of the so far unestablished  fundamental theory of
quantum gravity. Central in this context is the
Bekenstein--Hawking entropy:
\begin{equation}
S_{BH} = \frac{k_B}{G \hbar} \, \frac{1}{4} \, \mbox{Area}_H
\label{bekhaw}
\end{equation}
where $k_B$ is the Boltzman constant, $G$ is Newton's constant,
$\hbar $ is Planck's constant and $\mbox{Area}_H$ denotes the
area of the horizon surface.
\par
This very precise relation between a thermodynamical quantity and
a geometrical quantity such as the horizon area is a puzzle that
stimulated the interest of theoretical phisicists for more than
twenty years. Indeed a microscopic statistical explanation of the
area law for the black hole entropy has been correctly regarded
as possible only within a solid formulation of quantum gravity.
Superstring theory is the most serious candidate for a theory of
quantum gravity and as such should eventually provide such a
microscopic explanation of the area law. Although superstrings
have been around for more than twenty years, a significant
progress in this direction came only recently \cite{strovaf},
after the so called second string revolution (1995). Indeed black
holes are a typical non--perturbative phenomenon and perturbative
string theory could say very little about their entropy: only non
perturbative string theory can have a handle on it. Progresses in
this direction came after 1995 through the recognition of the
role of string dualities. These dualities allow to relate the
strong coupling regime of one superstring model to the weak
coupling regime of another one and are all encoded in the
symmetry group (the $U$--duality group) of the low energy {\it
supergravity effective action}.
\par
What we want to emphasize is that the first instance of a
microscopic explanation of the area law within string theory has
been limited to what in the language of general relativity would
be an {\it extremal black hole}. Indeed the extremality
condition, namely the coincidence of two horizons, obtains, in
the context of a supersymmetric theory, a profound
reinterpretation that makes extremal black holes the most
interesting objects to study. To introduce the concept consider
the usual Reissner Nordstrom metric describing a black--hole of
mass $m$ and electric (or magnetic) charge $q$:
\begin{equation}
ds^2 \, = \, -dt^2 \, \left( 1- \frac{2
m}{\rho}+\frac{q^2}{\rho^2} \right) + d\rho^2 \, \left( 1-
\frac{2 m}{\rho}+\frac{q^2}{\rho^2} \right)^{-1} + \rho^2 \,
d\Omega^2 \label{reinor}
\end{equation}
where $d\Omega^2= (d\theta^2 + \sin^2\theta \, d\phi^2)$ is the
metric on a $2$--sphere. As it is well known the metric
\eqn{reinor} admits two Killing horizons where the norm of the
Killing vector $\frac{\partial}{\partial t}$  changes sign. The
horizons are at the two roots of the quadratic form $\Delta
\equiv  -2  m  \rho + q^2 + \rho^2$ namely at:
\begin{equation}
 \rho_\pm = m \pm \sqrt{m^2-q^2}
 \label{2hor}
\end{equation}
If $m < |q|$ the two horizons disappear and we have a naked
singularity. For this reason in the context of classical general
relativity the {\it cosmic censorship} conjecture was advanced
that singularities should always be hidden inside horizon and
this conjecture was formulated as the bound:
\begin{equation}
m \, \ge \, |q| \label{censur}
\end{equation}
 Of
particular interest are the states that saturate the bound
\eqn{censur}. If $m=|q|$ the two horizons coincide and, setting:
\begin{equation}
m=|q|    \quad ; \quad \rho=r+m \quad ; \quad r^2 = {\vec x}\,
\cdot \, {\vec x} \label{rhor}
\end{equation}
 the metric \eqn{reinor} can be rewritten as:
\begin{eqnarray}
ds^2 & =& -dt^2 \, \left( 1+ \frac{q}{r}\right)^{-2} + \left( 1+
\frac{q}{r}\right)^{ 2} \, \left( dr^2 +r^2 \,
d\Omega^2\right)\nonumber\\
& =& - \, H^{-2}({\vec x})\, dt^2  + H^{2}({\vec x})\, d{\vec x}
\, \cdot \, d{\vec x} \label{guarda!}
\end{eqnarray}
where by:
\begin{equation}
\label{given}
 H ({\vec x})= \left(1 + \frac{q}{\sqrt{{\vec x}\, \cdot \, {\vec
 x}}}\right)
\end{equation}
we have denoted a harmonic function in a three--dimensional space
spanned by the three cartesian coordinates ${\vec x}$ with the
boundary condition that $H ({\vec x})$  goes to $1$ at infinity.
\par
Moreover, the extremal Reissner--Nordstr\"om configuration is a
soliton of classical general relativity, interpolating between the
flat Minkowski space--time, asymptotically reached at spatial
infinity $\rho\to \infty$, and the Bertotti--Robinson metric
describing the conformally flat geometry near the horizon $r \to
0$ \cite{black}:
 \begin{equation}ds^2_{BR} = - \frac{r^2}{M_{BR}^2}  dt^2 +
\frac{M_{BR}^2}{r^2}\left( dr^2+r^2 d\Omega \right). \label{br}
\end{equation} Last, let us note the the condition $m= |q|$ can be
written as a {\it no force condition} between the gravitational
interaction $F_g=\frac{m}{r^2}$ and the electric repulsion
$F_q=-\frac{q}{r^2}$.
\par
Extremal black-hole configurations are  embedded  in a natural way
in supergravity theories. Indeed supergravity, being invariant
under local super-Poincar\'e transformations, includes general
relativity, that is it describes gravitation coupled to other
fields in a supersymmetric framework. Therefore it admits, among
its classical solutions, black holes.
\par
Thinking of a black-hole configuration as a particular bosonic
background of an $N$-extended locally supersymmetric theory gives
a simple and natural understanding to the cosmic censorship
conjecture. Indeed, in  theories with extended supersymmetry the
bound \eqn{rhor} is just a consequence of the supersymmetry
algebra, and this ensures that in these theories the cosmic
censorship conjecture is always verified, that is there are
no naked singularities.\\
For extremal configurations, supersymmetry imposes that if the
bound $m=|q|$ is saturated in the classical theory, the same must
be true  also when quantum corrections are taken into account.
This is particularly relevant, since the quantum physics of black
holes is described by Hawking theory, which states that quantum
black holes are not stable, they radiate a termic radiation as a
black--body, and correspondingly they lose their energy (mass).
The only stable black-hole configurations are then the extremal
ones, because they have the minimal possible energy compatible
with the relation \eqn{censur} and so they cannot radiate.
\par
When the black hole is embedded in an $N$-extended supergravity
background the solutions depend in general also on scalar fields.
In this case,
 the electric charge $q$ has to be replaced by the maximum eigenvalue
of the central charge appearing in the supersymmetry algebra
(depending on the expectation value of scalar fields and on the
electric and magnetic charges). The Reissner--Nordstr\"om metric
takes in general a more complicated form.
\\
However, extremal black holes preserving some supersymmetries have
a peculiar feature: the event horizon loses all information about
scalar fields. Indeed, also in the scalar-dependent case, the near
horizon geometry is still described by a conformally flat,
Bertotti--Robinson-type geometry, with a mass parameter $M_{BR}$
depending on the electric and magnetic charges but not on the
scalars. \footnote{This in fact can be thought as one aspect of
the more general fact (true also for non-supersymmetric black
holes) that near the horizon black holes loose their `hair' (no
hair theorems). That is if one tries to perturb the black hole
with whatever additional hair (some slight mass anisotropy, or a
long-range field, like a scalar) all these features disappear near
the horizon, except for those associated with the conserved
quantities of general relativity, namely, for a
non-rotating black hole, its mass and charge.\\
Again, as it happened for the BPS bound, supersymmetry gives a
deeper and more natural understanding of this general feature.}
\\
This peculiarity has a counterpart in the fact that extremal
solutions of supergravity have to satisfy a set of first order
differential equations imposed  by the existence of at least one
supercovariant spinor, leaving invariant a fraction of the initial
supersymmetry. First order differential equations
$\frac{d\Phi}{dr}= f(\Phi)$ have in general fixed points,
corresponding to the values of $r$ for which
$f(\Phi)=0$.\\
It is possible to show \cite{fekast} that the first order
differential equations expressing the Killing spinors equations
for extremal black holes have as fixed point exactly the event
horizon. The horizon is an attractor point \cite{moore}. Scalar
fields, independently of their boundary conditions at spatial
infinity, approaching the horizon flow to a fixed point given by
a certain ratio of electric and magnetic charges.
\\
Remembering now that the black-hole entropy is given by the
area--entropy Bekenstein--Hawking relation \eq{bekhaw}, we see
that the entropy of extremal black holes is a topological
quantity, in the sense that it is fixed in terms of the quantized
electric and magnetic charges while it does not depend on
continuous parameters as scalars. The horizon mass parameter
$M_{BR}$ turns out to be given in this case (extremal
configurations) by the maximum eigenvalue $Z_M$ of the central
charge appearing in the supersymmetry algebra, evaluated at the
fixed point:
\begin{equation}M_{BR}=M_{BR}(e,g)=Z_M(\phi_{fix},e,g)
\end{equation} that gives, for the Bekenstein--Hawking entropy:
\begin{equation}S_{B-H}=\frac{A_{BR}(e,g)}{4}=\pi \vert
Z_M(\phi_{fix},e,g)\vert ^2 \label{bert} \end{equation} \vskip 5mm
Many efforts were spent in the course of the years to give an
explanation for the large, topological entropy of extremal black
holes in the context of a quantum theory of gravity, like string
theory. In particular, one would like to give a microscopical,
statistical mechanics interpretation of this thermodynamical
quantity. Although we will not treat at all the microscopical
point of view throughout this paper, it is important to mention
that such an interpretation became possible after the introduction
of D-branes in the context of string theory \cite{blackmicro}.
Following this approach, extremal black holes are interpreted as
bound states of D-branes in a space--time compactified to four or
five dimensions, and the different microstates giving rise to the
Bekenstein--Hawking entropy come from the different ways of
wrapping branes in the
internal directions.\\
It is important to note that all calculations made in particular
cases using this approach furnished values, for the
Bekenstein--Hawking entropy, compatible with those performed with
the supergravity, macroscopical techniques. The entropy formula
turns out to be in all cases a U-duality invariant expression
(homogeneous of degree two) built out of electric and magnetic
charges and as such can be in fact also computed through certain
(moduli independent) topological quantities which only depend on
the nature of the U-duality groups and the appropriate
representations of electric and magnetic charges. For example in
the $N=8$ theory the entropy can be shown to correspond to the
unique quartic $E_7$ invariant built in terms of the $56$
dimensional representation. Actually, one can derive for all $N
\geq 2$ theories topological U-invariants constructed in terms of
the (moduli dependent) central charges and matter charges and
show that, as expected, they coincide with the squared ADM mass at
fixed scalars.
\par
In the next section we shall interpret black holes of this form as
BPS saturated states namely as quantum states   filling special
irreducible representations of the supersymmetry algebra, the so
called {\it short supermultiplets}, the shortening condition being
precisely the saturation of the cosmic censorship bound
\eqn{censur}. Indeed such a bound can be restated as the equality
of the mass with the central charge which occurs when a certain
fraction of the supersymmetry charges identically annihilate  the
state. The remaining supercharges applied to the BPS state build
up a unitary irreducible representation of supersymmetry that is
shorter than the typical one since it contains  less states. As we
stress in the next section it is precisely this interpretation
what makes extremal black holes relevant to the string theory.
Indeed these classical solutions of supergravity belong to the non
perturbative particle  spectrum of superstring theory, and are not
accessible to perturbative string theory.
\section{Extremal Black--Holes as quantum  BPS states}
\label{introgen} In the previous section we have reviewed the idea
of extremal black holes as it arises in classical general
relativity. Extremal black--holes have become objects of utmost
relevance in the context of superstrings after {\it the second
string revolution} has taken place in 1995. Indeed supersymmetric
extremal black--holes have been studied in depth in a vast recent
literature \cite{malda,black,ortino}. This interest is just part
of a more general interest in the $p$--brane classical solutions
of supergravity theories in all dimensions $4 \le D \le 11$
\cite{mbrastelle,duffrep}. This interest streams from the
interpretation of the classical solutions of supergravity that
preserve a fraction of the original supersymmetries as the BPS non
perturbative states necessary to complete the perturbative string
spectrum and make it invariant under the many conjectured duality
symmetries \cite{schw,sesch,huto2,huto,vasch}. Extremal
black--holes and their parent $p$--branes in higher dimensions are
therefore viewed as additional {\it particle--like} states that
compose the  spectrum of a fundamental quantum theory. The reader
should be advised that the holes we are discussing here are
neither stellar--mass, nor mini--black holes: their mass  is
typically of the order of the Planck--mass:
\begin{equation}
M_{Black~Hole} \, \sim \, M_{Planck}
\end{equation}
The Schwarzschild radius is therefore microscopic.
\par
Yet, as the monopoles in gauge theories, these  non--perturbative
quantum states originate from  regular solutions of the classical
field equations, the same Einstein equations one deals with in
classical general relativity and astrophysics. The essential new
ingredient, in this respect, is supersymmetry that requires the
presence of {\it vector fields} and {\it scalar fields} in
appropriate proportions. Hence the black--holes we are going to
discuss are solutions of generalized Einstein--Maxwell--dilaton
equations.
\par
From an abstract viewpoint BPS saturated states are characterized
by the fact that they preserve a fraction, $1/2$ or $1/4$ or
$1/8$ of the original supersymmetries. What this actually means
is that there is a suitable projection operator $\IP^2_{BPS}
=\IP_{BPS}$ acting on the supersymmetry charge $Q_{SUSY}$, such
that:
\begin{equation}
 \left(\IP_{BPS} \,Q_{SUSY} \right) \, \vert \, \mbox{BPS state} \, >
 \,=  \, 0
 \label{bstato}
\end{equation}
Since the supersymmetry transformation rules of any supersymmetric
field theory are linear in the first derivatives of the fields
eq.\eqn{bstato} is actually a {\it system of first order
differential equations}. This system has to be combined with the
second order field equations of supergravity and the common
solutions to both system of equations is a classical BPS
saturated state. That it is actually an exact state of
non--perturbative string theory follows from supersymmetry
representation theory. The classical BPS state is by definition
an element of a {\it short supermultiplet} and, if supersymmetry
is unbroken, it cannot be renormalized to a {\it long
supermultiplet}.
\par
Translating eq. \eqn{bstato} into an explicit first order
differential system requires knowledge of the supersymmetry
transformation rules of supergravity. These latter have a rich
geometrical structure that is the purpose of the present paper to
illustrate. Indeed the geometrical structure of supergravity which
originates in its scalar sector is transferred into the physics of
extremal black holes by the BPS saturation condition.
\par
To fulfill the above program, it is necessary first to review the
formalism of $D=4$ $N$-extended supergravity theories. We begin
by recalling the algebraic definition of $D=4$ BPS states in a
theory with an even number of supercharges $N=2\nu$ \footnote{The
case with $N=\mbox{odd}$ can be similarly treated but needs some
minor modifications due to the fact the eigenvalues of an
antisymmetric matrix in odd dimensions are $\{ \pm \mbox{\rm
i}\lambda_i , 0 \}$.}.
\subsection{General definition of BPS states in a 4D theory with
$N=2 \, \times \, p$ supersymmetries} The $D=4$ supersymmetry
algebra with $N=2\, \times \, p$ supersymmetry charges  is given
by
\begin{eqnarray}
&\left\{ {\bar Q}_{A  \alpha }\, , \,{\bar Q}_{B  \beta}
\right\}\, = \,  {\rm i} \left( {\bf C} \, \gamma^\mu
\right)_{\alpha \beta} \, P_\mu \, \delta_{AB} \, \, - \, {\bf
C}_{\alpha \beta} \,
{\bf Z}_{AB}& \nonumber\\
&\left( A,B = 1,\dots, 2p \right)& \label{susyeven2}
\end{eqnarray}
where the SUSY charges ${\bar Q}_{A}\equiv Q_{A}^\dagger \gamma_0=
Q^T_{A} \, {\bf C}$ are Majorana spinors, $\bf C$ is the charge
conjugation matrix, $P_\mu$ is the 4--momentum operator and
 the antisymmetric tensor
${\bf Z}_{AB}=-{\bf Z}_{BA}$ is the central charge operator. It
can always be reduced to normal form
\begin{equation}
{\bf Z}_{AB} ~~~~= \left(\begin{matrix}\epsilon Z_1&0& \dots&
0\cr 0&\epsilon Z_2& \dots &0\cr \dots &\dots & \dots & \dots \cr
0&0& \dots &\epsilon Z_p\cr\end{matrix}\right) \label{skewZ}
\end{equation}
where $\epsilon$ is the $2\times 2$ antisymmetric matrix, (every
zero is a $2\times2$ zero matrix) and the $p$ skew eigenvalues
$Z_I$ of ${\bf Z}_{AB}$ are the central charges.
\par
If we identify each index $A,B,\dots$ with a pair of indices
\begin{equation}
 A=(a,I) \quad ; \quad a,b,\dots =1,2 \quad ; \quad I,J,\dots =1,
 \dots, p
 \label{indrang}
\end{equation}
then the superalgebra \eqn{susyeven2} can be rewritten as:
\begin{eqnarray}
&\left\{ {\bar Q}_{aI \vert \alpha }\, , \,{\bar Q}_{bJ \vert
\beta} \right\}\, = \,  {\rm i} \left( C \, \gamma^\mu
\right)_{\alpha \beta} \, P_\mu \, \delta_{ab} \, \delta_{IJ} \,
- \, C_{\alpha \beta} \, \epsilon_{ab} \, \times \, \ZZ_{IJ}&
\label{susyeven}
\end{eqnarray}
where the SUSY charges ${\bar Q}_{aI}\equiv Q_{aI}^\dagger
\gamma_0= Q^T_{aI} \, C$ are Majorana spinors, $C$ is the charge
conjugation matrix, $P_\mu$ is the 4--momentum operator,
$\epsilon_{ab}$ is the two--dimensional Levi Civita symbol and
the central charge operator is now represented by the {\it
symmetric tensor} $\ZZ_{IJ}=\ZZ_{JI}$ which can always be
diagonalized $\ZZ_{IJ}=\delta_{IJ} \, Z_J$. The $p$ eigenvalues
$Z_J$ are the skew eigenvalues introduced in equation \eqn{skewZ}.
\par
The Bogomolny bound on the mass of a generalized monopole state:
\begin{equation}
M \, \ge \, \vert \, Z_I \vert \qquad \forall Z_I \, , \,
I=1,\dots,p \label{bogobound}
\end{equation}
is an elementary consequence of the supersymmetry algebra and of
the identification between {\it central charges} and {\it
topological charges}. To see this it is convenient to introduce
the following reduced supercharges:
\begin{equation}
{\bar S}^{\pm}_{aI \vert \alpha }=\frac{1}{2} \, \left( {\bar
Q}_{aI} \gamma _0 \pm \mbox{i} \, \epsilon_{ab} \,  {\bar
Q}_{bI}\, \right)_\alpha \label{redchar}
\end{equation}
They can be regarded as the result of applying a projection
operator to the supersymmetry charges:
\begin{eqnarray}
{\bar S}^{\pm}_{aI} &=& {\bar Q}_{bI} \, \IP^\pm_{ba} \nonumber\\
 \IP^\pm_{ba}&=&\frac{1}{2}\, \left({\bf 1}\delta_{ba} \pm \mbox{i} \epsilon_{ba}
 \gamma_0 \right)
 \label{projop}
\end{eqnarray}
Combining eq.\eqn{susyeven} with the definition \eqn{redchar} and
choosing the rest frame where the four momentum is $P_\mu$
=$(M,0,0,0)$, we obtain the algebra:
\begin{equation}
\left\{ {\bar S}^{\pm}_{aI}  \, , \, {\bar S}^{\pm}_{bJ} \right\}
= \pm \epsilon_{ac}\, C \, \IP^\pm_{cb} \, \left( M \mp Z_I
\right)\, \delta_{IJ} \label{salgeb}
\end{equation}
By positivity of the operator $\left\{ {\bar S}^{\pm}_{aI}  \, ,
\, {\bar S}^{\pm}_{bJ} \right\} $ it follows that on a generic
state the Bogomolny bound \eqn{bogobound} is fulfilled.
Furthermore it also follows that the states which saturate the
bounds:
\begin{equation}
\left( M\pm Z_I \right) \, \vert \mbox{BPS state,} i\rangle = 0
\label{bpstate1}
\end{equation}
are those which are annihilated by the corresponding reduced
supercharges:
\begin{equation}
{\bar S}^{\pm}_{aI}   \, \vert \mbox{BPS state,} i\rangle = 0
\label{susinvbps}
\end{equation}
On one hand eq.\eqn{susinvbps} defines {\sl short multiplet
representations} of the original algebra \eqn{susyeven} in the
following sense: one constructs a linear representation of
\eqn{susyeven} where all states are identically annihilated by
the operators ${\bar S}^{\pm}_{aI}$ for $I=1,\dots,n_{max}$. If
$n_{max}=1$ we have the minimum shortening, if $n_{max}=p$ we
have the maximum shortening. On the other hand eq.\eqn{susinvbps}
can be translated into a first order differential equation on the
bosonic fields of supergravity.
 \par
Indeed, let us consider a configuration where all the fermionic
fields are zero. In order for a configuration to be
supersymmetric we have to impose that the supersymmetry variations
of all the fields are zero in the background. Since the bosonic
fields transform into spinors, they are automatically zero in the
background; for the fermionic fields, instead, this condition
gives a differential equation for the bosonic fields which is
called ``Killing spinor'' equation. Indeed,  the gravitino
transformation law contains the covariant derivative of the
supersymmetry parameter $\epsilon_A$ and the differential
equation one obtains in this case determines the functional
dependence of the parameter on the space--time coordinates. The
corresponding solution for $\epsilon_A$ is called a Killing
spinor.
\par
Setting the fermionic SUSY rules appropriate to such a background
equal to zero we find the following Killing spinor equation:
\begin{equation}
0=\delta \mbox{fermions} = \mbox{SUSY rule} \left(
\mbox{bosons},\epsilon_{AI} \right) \label{fermboserule}
\end{equation}
where the SUSY parameter satisfies the following conditions
($\xi^\mu$ denotes a time--like Killing vector):
\begin{equation}
\begin{array}{rclcl}
\xi^\mu \, \gamma_\mu \,\epsilon_{aI} &=& \mbox{\rm i}\,
\varepsilon_{ab}
\,  \epsilon^{bI}   & ; &   I=1,\dots,n_{max}\\
\epsilon_{aI} &=& 0  &;&   I > n_{max} \\
\end{array}
\label{rollato}
\end{equation}
Hence eq.s \eqn{fermboserule} with a parameter satisfying the
condition \eqn{rollato} will be our operative definition of BPS
states.
\section{Four-dimensional BPS black--holes
and the general form of the supergravity action}
 \label{gen4d} In this section we begin
the study of BPS black--hole solutions in four space-time
dimensions. To this aim we first have to introduce the main
features of four dimensional supergravities. Four dimensional
supergravity theories contain in the bosonic sector, besides the
metric, a number of vectors and scalars. The relevant bosonic
action is known to have the following general form:
\begin{eqnarray}
\label{genact} {\cal S}&=&\int\sqrt{-g}\, d^4x\left(2R+\Im{\cal
N}_{\Lambda \Gamma}F_{\mu\nu}^{~~\Lambda }F^{\Gamma |\mu\nu}+
\frac 16 g_{IJ}(\phi) \partial_{\mu}
\phi^{I}\partial^{\mu}\phi^{J}+\right.\nonumber\\
&+&\left.\frac 12\Re{\cal N}_{\Lambda \Gamma  }
\frac{\epsilon^{\mu\nu\rho\sigma}}{\sqrt{-g}}
F_{\mu\nu}^{~~\Lambda }F^{\Gamma }_{~~\rho\sigma}\right)
\end{eqnarray}
where $g_{IJ}(\phi)$ ($I,J,\cdots =1,\cdots ,m$ is the scalar
metric on the $\sigma$-model described by the $m$--dimensional
scalar manifold ${\cal M}_{scalar}$ and the vectors kinetic matrix
${\cal N}_{\Lambda\Sigma}(\phi) $ is a complex, symmetric, ${n_V}
\, \times \, {n_V}$ matrix depending on the scalar fields. The
 number  of vectors and scalars, that is ${
n_V}$ and $m$, and the geometrical properties of the scalar
manifold ${\cal M}_{scalar}$ depend on the number $N$ of
supersymmetries and are resumed in Table \ref{topotable}.
 The relation between this scalar geometry and the kinetic
matrix ${\cal N}$ has a very general and universal form. Indeed it
is related to the solution of a general problem, namely  how to
lift the action of the scalar manifold isometries from the scalar
to the vector fields. Such a lift is necessary because of
supersymmetry since scalars and vectors generically belong to the
same supermultiplet and must rotate coherently under symmetry
operations. This problem has been solved in a general (non
supersymmetric) framework in reference \cite{gaizum} by
considering   the possible extension of the Dirac
electric--magnetic duality to more general theories involving
scalars. In the next subsection we review this approach and in
particular we show how enforcing covariance with respect to such
duality rotations leads to a determination of the kinetic matrix
${\cal N}$. The structure of ${\cal N}$ enters the black--hole
equations in a crucial way so that the topological invariant
associated with the hole, that is its {\it entropy}, is an
invariant of the group of electro-magnetic duality rotations, the
U--duality group.
{\footnotesize
\begin{table}
\begin{center}
\caption{\sl Scalar Manifolds of Extended Supergravities}
\label{topotable}
\begin{tabular}{|c||c|c|c| }
\hline N & U-Duality group & ${\cal M}_{scalar}$ & $n_V , m$   \\
\hline \hline
$1$  &   ${\cal U} \subset Sp(2n,\IR)$ & K\"ahler & $n_V , m$ \\
\hline $2$  &  ${\cal U} \subset Sp(2n+2,\IR)$ & $\cM^Q(n_H)
\otimes
 \cM^{SK}(n)$ &$ n+1, 2n +4n_H$\\
\hline $3$  &  $SU(3,n)\subset Sp(2n+6,\IR)$ & $\frac{SU(3,n)}
{S(U(3)\times U(n))}$ & $3+n$, $6n$ \\
\hline $4$  &   $SU(1,1)\otimes SO(6,n)\subset Sp(2n+12,\IR)$ &
$\frac{SU(1,1)}{U(1)} \otimes \frac{SO(6,n)}{SO(6)\times SO(n)}$ & $6+ n$, $6n +2$ \\
\hline $5$  &  $SU(1,5)\subset Sp(20,\IR)$ & $\frac{SU(1,5)}
{S(U(1)\times U(5))}$ & 10, 10 \\
\hline $6$  &  $SO^\star(12)\subset Sp(32,\IR)$ &
$\frac{SO^\star(12)}{U(1)\times SU(6)}$ & 16, 30 \\
\hline $7,8$&  $E_{7(-7)}\subset Sp(56,\IR)$ &
$\frac{ E_{7(-7)} }{SU(8)}$ & 28, 70 \\
\hline
\end{tabular}
\end{center}
{\it $\cM^Q(n_H)$ denotes a quaternionic manifold of
quaternionic  dimension $n_H$ and $\cM^{SK}(n )$ a Special
K\"ahler manifold of complex dimension $n$.}
\end{table}
}
\subsection{Duality Rotations and Symplectic Covariance}
\label{LL1}  Let us review the general structure of an abelian
theory of vectors and scalars displaying covariance under a group
of duality rotations. The basic reference is the 1981 paper by
Gaillard and Zumino \cite{gaizum}. A general presentation in
$D=2p$ dimensions can be found in \cite{castdauriafre}. Here we
fix
 $D=4$.
\par
We consider a theory of  $n_V$ gauge fields $A^\Lambda_{\mu}$, in
a $D=4$  space--time with Lorentz signature. They  correspond to a
set of  $n_V$ differential $1$--forms
\begin{equation}
A^\Lambda ~ \equiv ~ A^\Lambda_{\mu} \, dx^{\mu} \quad \quad
 \left ( \Lambda = 1,
\dots , {n_V} \right )
\end{equation}
The corresponding field strengths and their Hodge duals are
defined by
\begin{eqnarray}
{ F}^\Lambda & \equiv & d \, A^\Lambda \,  \equiv   \,
   {\cal F}^\Lambda_{\mu  \nu} \,
dx^{\mu } \, \wedge \, dx^{\nu} \nonumber\\
{\cal F}^\Lambda_{\mu \nu} & \equiv & {\frac{1}{2 }} \,\left (
\partial_{\mu } A^\Lambda_{\nu} \, - \, \partial_{\nu }
A^\Lambda_{\mu} \right ) \nonumber\\ ^{\star}{ F}^{\Lambda} &
\equiv &\, {\tilde {\cal F}}^\Lambda_{\mu \nu} \, dx^{\mu } \,
\wedge \,
 dx^{\nu} \nonumber\\
{\tilde {\cal F}}^\Lambda_{\mu \nu} & \equiv &{\frac{1}{2}}
\varepsilon_{\mu  \nu \rho\sigma}\, {\cal F}^{\Lambda \vert \rho
\sigma} \label{campfort}
\end{eqnarray}
In addition to the gauge fields let us also introduce a set of
real scalar fields $\phi^I$ ( $I=1,\dots , m$) spanning an ${
m}$--dimensional manifold ${\cal M}_{scalar}$ endowed with a
metric $g_{IJ}(\phi)$. Utilizing the above field content we can
write the following action functional:
\begin{equation}
{\cal S}\, = \,  \int \, \left \{ -\, \gamma_{\Lambda\Sigma}(\phi)
\, F^\Lambda_{\mu\nu} \, F^{\Sigma \mu\nu} \, +  \,
\theta_{\Lambda\Sigma}(\phi) \, F^\Lambda_{\mu\nu}  \,\star
F^{\Sigma\mu\nu}
     \,  + \,{\frac 12}\,
g_{IJ}(\phi) \, \partial_\mu \phi^I \, \partial^\mu \phi^J \right
\}  \, \mbox{d}^4 x \label{gaiazuma}
\end{equation}
where the scalar fields dependent ${n_V} \times {n_V}$ matrix
$\gamma_{\Lambda\Sigma}(\phi)$  generalizes the inverse of the
squared coupling constant $\frac{1}{g^2}$ appearing in ordinary
 gauge theories. The field dependent matrix
$\theta_{\Lambda\Sigma}(\phi)$ is instead a generalization of the
$theta$--angle of quantum chromodynamics. Both $\gamma$ and
$\theta$ are symmetric matrices. Finally, we have introduced the
operator $\star$ that maps a field strength into its Hodge dual
\begin{equation} \left ( \star \, {\cal
F}^\Lambda \right )_{\mu \nu} \, \equiv \, {\frac{1}{ 2 }} \,
\epsilon_{\mu\nu \rho \sigma} \, {\cal F}^{\Lambda
\vert\rho\sigma}.
\end{equation}
Introducing self--dual and antiself--dual combinations
\begin{eqnarray}
  {\cal F}^{\pm} &=& {\frac 12}\left({\cal F}\, \pm {\rm i} \,
  \star
{\cal F}\right) \nonumber \\
\star \, {\cal F}^{\pm}& =& \mp \mbox{i} {\cal F}^{\pm}
\label{selfduals}
\end{eqnarray}
and the field--dependent symmetric matrices
\begin{eqnarray}
  {\cal N} & = & \theta \, - \,
\mbox{i} \gamma \nonumber\\
{\bar {\cal N}} & = & \theta + \mbox{i} \gamma\ , \label{scripten}
\end{eqnarray}
the vector part of the Lagrangian ~\eqn{gaiazuma} can be rewritten
as
\begin{equation}
{\cal L}_{vec} \,  = \, {\mbox{i}} \, \left [{\cal F}^{-T} {\bar
{\cal N}} {\cal F}^{-}- {\cal F}^{+T} {\cal N} {\cal F}^{+}
\right] \label{lagrapm}
\end{equation}
Introducing further the new tensors
\begin{equation}
  {\tilde{\cal
G}}^\Lambda_{\mu\nu} \, \equiv \,  {\frac 12 } { \frac{\partial
{\cal L}}{\partial {\cal F}^\Lambda_{\mu\nu}}}=
\left(-\gamma_{\Lambda\Sigma} +\theta_{\Lambda\Sigma}\right){\cal
F}^\Sigma_{\mu\nu} \leftrightarrow {\cal G}^{\mp \Lambda}_{\mu\nu}
\, \equiv \, \mp{\frac {\rm i}{2} } { \frac{\partial {\cal
L}}{\partial {\cal F}^{\mp \Lambda}_{\mu\nu}}} \label{gtensor}
\end{equation}
the Bianchi identities and field equations associated with the
Lagrangian ~\eqn{gaiazuma} can be written as \bea
\partial^{\mu }{\tilde {\cal F}}^{\Lambda}_{\mu\nu} &=& 0 \\
\partial^{\mu }{\tilde {\cal G}}^{\Lambda}_{\mu\nu} &=& 0
\label{biafieq} \eea or equivalently \bea
 \partial^{\mu }
{\rm Im}{\cal F}^{\pm \Lambda}_{\mu\nu} &=& 0 \\
\partial^{\mu } {\rm Im}{\cal G}^{\pm \Lambda}_{\mu\nu} &=&
0 \ . \label{biafieqpm} \eea This suggests that we introduce the
$2{n_V}$ column vector
\begin{equation}
{\bf V} \, \equiv \, \left ( \begin{matrix} \star \, {\cal F}\cr
\star\, {\cal G}\cr\end{matrix}\right ) \label{sympvec}
\end{equation}
and that we consider general linear transformations on such a
vector
\begin{equation}
\left ( \begin{matrix} \star \, {\cal F}\cr\star\, {\cal
G}\cr\end{matrix}\right )^\prime \, =\, \left (\begin{matrix} A &
B \cr C & D \cr\end{matrix} \right ) \left ( \begin{matrix}  \star
\, {\cal F}\cr \star \, {\cal G}\cr\end{matrix}\right )
\label{dualrot}
\end{equation}
For any matrix ${\cal S} =\left (\begin{matrix} A & B \cr C & D
\cr\end{matrix} \right ) \, \in \, GL(2{ n_V},\IR )$ the new
vector ${\bf V}^\prime$ of magnetic and electric field--strengths
satisfies the same equations ~\eqn{biafieq} as the old one. In a
condensed notation we can write
\begin{equation}
\partial \, {\bf V}\, = \, 0 \quad \Longleftrightarrow \quad
\partial \, {\bf V}^\prime \, = \, 0
\label{dualdue}
\end{equation}
Separating the self--dual and anti--self--dual parts
\begin{equation}
{\cal F}=\left ({\cal F}^+ +{\cal F}^- \right ) \qquad ; \qquad
{\cal G}=\left ({\cal G}^+ +{\cal G}^- \right ) \label{divorzio}
\end{equation}
and taking into account that   we have
\begin{equation}
{\cal G}^+ \, = \, {\cal N}{\cal F}^+  \quad {\cal G}^- \, = \,
{\bar {\cal N}}{\cal F}^- \label{gigiuno}
\end{equation}
the duality rotation of eq.~\eqn{dualrot} can be rewritten as
\begin{equation}
\left ( \begin{matrix}   {\cal F}^+ \cr
 {\cal G}^+\cr\end{matrix}\right )^\prime  \, = \,
\left (\begin{matrix} A & B \cr C & D \cr\end{matrix} \right )
\left ( \begin{matrix} {\cal F}^+\cr {\cal N} {\cal
F}^+\cr\end{matrix}\right ) \qquad ; \qquad \left ( \begin{matrix}
 {\cal F}^- \cr
 {\cal G}^-\cr\end{matrix}\right )^\prime \, = \,
\left (\begin{matrix} A & B \cr C & D \cr\end{matrix} \right )
\left ( \begin{matrix}  {\cal F}^-\cr {\bar {\cal N}} {\cal
F}^-\cr\end{matrix}\right ) . \label{trasform}
\end{equation}
Now, let  us note that, since $\Phi^I$ are the scalar partners of
$A^\Lambda$, when a duality rotation is performed on the vector
field strengths and their duals, also the scalars get transformed
correspondingly, through the action of some diffeomorphism on the
scalar manifold $\cM_{scal}$. In particular, also the kinetic
matrix $\cN(\Phi)$, that in supersymmetric theories is a function
of scalars, transforms under a duality rotation. That is, a
duality transformation $\xi$ acts in the following way on the
supersymmetric system:
\begin{equation}
\xi : \left\{\begin{matrix}\Phi &\rightarrow &\Phi^\prime
=\xi(\Phi) \cr \cN(\Phi) & \rightarrow & \cN^\prime
\left(\xi(\Phi)\right) \cr V & \rightarrow & V^{\prime\mp}=S_\xi
V^\mp \cr\end{matrix}\right. \label{xi}
\end{equation}
Thus, the transformation laws of the equations of motion and of
$\cN$, and so also the matrix ${\cal S}_\xi$, will be induced by a
diffeomorphism of the scalar fields.

Focusing in particular on the third relation in \eq{xi}, that
explicitly reads:
\begin{equation}\left(\begin{matrix}\cF^{\pm\prime} \cr
\cG^{\pm\prime}\cr\end{matrix}\right)=
\left(\begin{matrix}A_\xi\cF^\pm +B_\xi \cG^\pm \cr C_\xi\cF^\pm
+D_\xi \cG^\pm\cr\end{matrix}\right)
\end{equation}let us note that it contains the magnetic field
strength $\cG^\mp_\Lambda$, which is defined as a variation  of
the kinetic lagrangian. Under the transformations (\ref{xi}) the
lagrangian transforms in the following way:
\begin{eqnarray}
{\cal L}^\prime &=&{\rm i} \Bigl[
 \left(A_\xi + B_\xi\cN\right)^{\ \Lambda}_{\Gamma}
\left(A_\xi + B_\xi \cN\right)^\Sigma_{\ \Delta}
 \cN^\prime_{\Lambda\Sigma}(\Phi )\cF^{+\Gamma} \cF^{+\Delta} \nn\\
&-& \left(A_\xi + B_\xi\bar\cN\right)^{\ \Lambda}_{\Gamma}
\left(A_\xi + B_\xi\bar\cN\right)^\Sigma_{\ \Delta}
\bar\cN^\prime_{\Lambda\Sigma}(\Phi )\cF^{-\Gamma} \cF^{-\Delta}
\Bigr] ; \label{lageven}
 \end{eqnarray}
Equations \eq{xi} must be consistent with the definition of
$\cG^\mp$ as a variation of the lagrangian  \eq{lageven}:
\begin{equation}
\cG^{\prime +}_{\Lambda} = \left(C_\xi +
D_\xi\cN\right)_{\Lambda\Sigma}F^{+\Sigma} \equiv - \frac{\rm i }{
2} \frac{\partial {\cal L}^\prime}{  \partial \cF^{\prime +
\Lambda}}= \left(A_\xi + B_\xi\cN\right)^\Delta_{\ \Sigma}
 \cN^\prime_{\Lambda\Delta}\cF^{+\Sigma}
\end{equation}
that implies:
\begin{equation}
\cN^\prime_{\Lambda\Sigma}(\Phi^\prime) = \left[\left(C_\xi +
D_\xi\cN\right) \cdot \left(A_\xi +
B_\xi\cN\right)^{-1}\right]_{\Lambda\Sigma} ; \label{ntrasfeven}
\end{equation}
Recalling now that the matrix $\cN$ is symmetric, and that this
property must be true also in the duality transformed system, it
follows that the matrix ${\cal S}_\xi$ must satisfy a constraint
that allow to fix
  the duality  groups.
  \\
  Indeed, imposing that $\cN$ and $\cN^\prime$ be both
  symmetric matrices, gives the constraint:
  \begin{equation}  {\cal S}\in Sp(2n_V, \IR)\subset GL(2n_V, \IR).
  \end{equation}  \par
This observation has important implications on the scalar manifold
${\cal M}_{scal}$. Indeed, all the above discussion implies that
on the scalar manifold the following homomorphism is defined:
\begin{equation}Diff(\cM_{scal}) \to  Sp(2n,\IR)
\end{equation}
In particular, the presence on the manifold of a function of
scalars transforming with a fractional linear transformation under
a duality rotation (that is a diffeomorphism) on scalars, induces
the existence on $\cM_{scal}$ of a linear structure (inherited
from vectors). In particular, as we will see in section
\ref{sugras}, for the $N=2$ four dimensional theory this implies
that the scalar manifold be a {\it special manifold}, that is a
K\"ahler--Hodge manifold endowed with a flat symplectic bundle.
\par
As it is necessary for a duality rotation, the transformation
\eq{xi}, that is a duality symmetry on the system
field-equations/Bianchi-identities, cannot be extended to a
symmetry of the lagrangian. The scalar lagrangian ${\cal
L}_{scal}$ is left invariant under the action of the isometry
group of the metric $g_{IJ}$, but the vector part is in general
not invariant. Indeed, the transformed  lagrangian under the
action of ${\cal S}\in Sp(2n_V, \IR)$  can be rewritten:
\begin{eqnarray}
{\mbox Im}\left(\cF^{-\Lambda}\cG^-_\Lambda\right)& \rightarrow &
{\mbox Im}\left(\cF^{\prime-\Lambda}\cG^{\prime-}_\Lambda\right)\nonumber\\
&=& {\mbox Im}\bigl[ \cF^{-\Lambda}\cG^-_\Lambda
+2(C^TB)_\Lambda^{\ \Sigma}
\cF^{-\Lambda}\cG^-_\Sigma + \nonumber\\
&+&(C^TA)_{\Lambda\Sigma}\cF^{-\Lambda}\cF^{-\Sigma}+
(D^TB)^{\Lambda\Sigma} \cG^-_\Lambda \cG^-_\Sigma
\bigr]\label{trasfvett}
\end{eqnarray}
It is evident from \eq{trasfvett} that only the transformations
with $B=C=0$
are symmetries.\\
If $C\not= 0$, $B=0$ the lagrangian varies for a topological term:
\begin{equation}
(C^TA)_{\Lambda\Sigma}\cF^\Lambda_{\mu\nu}{}^\star
\cF^{\Sigma\vert{\mu\nu} }
\end{equation}
 corresponding to a redefinition of the function
$\theta_{\Lambda\Sigma}$; such a transformation leaves classical
physics invariant,  being a total derivative, but it is relevant
in the quantum theory. It is a symmetry of the partition function
only if
 $\Delta\theta=\frac{1 }{ 2}(C^TA)$
is an integer multiple of $2\pi$, and this implies that  $S\in
Sp(2n_V,\ZZ) \subset Sp(2n_V,\IR)$.\\
For $B\not= 0$ neither the action  nor the perturbative partition
function are invariant. Let us observe that, for $B\not= 0$, the
transformation law of the kinetic matrix $\cN=\theta -{\rm i}
\gamma$ contains the transformation $\cN\rightarrow -\frac{1}{
\cN}$ that is it exchanges the weak and strong coupling regimes of
the theory. One can then think a supersymmetric  quantum field
theory being described by a collection of local lagrangians, each
defined in a local patch. They are all equivalent once one defines
for each of them what is {\it electric} and what is {\it
magnetic}. Duality transformations map this set of lagrangians one
into the other. At this point we observe that the supergravity
bosonic lagrangian (\ref{genact}) is exactly of the form
considered in this section as far as the matter content is
concerned, so that we may apply the above considerations about
duality rotations to the supergravity case. In particular, the
U-duality acts in all theories with $N\geq 2$ supercharges, where
the vector supermultiplets contain both vectors and scalars. For
$N=1$ supergravity, instead, vectors and scalars are still present
but they are not related by supersymmetry, and as a consequence
they are not related by U-duality rotations, so that the previous
formalism does not apply. In the next section we will discuss in a
geometrical framework the structure of the supergravity theories
for $N\geq 2$. In particular, we will give the expression for the
kinetic vector matrix $\cN_{\Lambda\Sigma}$ in terms of the
$Sp(2n_V)$ coset representatives embedding the U-duality group for
theories whose $\sigma$-model is a coset space, namely $N>2$.
Furthermore we will show that the $N=2$ case can be treated in a
completely analogous way even if the $\sigma$-model of the
scalars is not in general a coset space.
\subsection{Duality symmetries and central charges in four
dimensions} \label{sugras}  Let us restrict our attention to
$N$-extended supersymmetric theories coupled to the gravitational
field, that is to supergravity theories, whose bosonic action has
been given in \eqn{genact}. For any theory we analyze the group
theoretical structure and find the expression of the central
charges, and the properties they obey. We will see that in each
theory all fields are in some representation of the isometry
group ${\rm U}$  of scalar fields or of its maximal compact
subgroup $H$. This is just a consequence of the Gaillard-Zumino
duality acting on the 2-forms and their duals, discussed in the
preceding section, so that a restriction to the integers of ${\rm
U}$ is the duality group.\par As we have already mentioned, all
$D=4$ supergravity theories contain scalar fields whose kinetic
Lagrangian is described by $\sigma$--models of the form ${\rm
U}/H$, with the exception of $D=4$, $N=1,2$. We begin to examine
the theories with $N > 2$, and then we will generalize the
results to the $N=2$ case
(The $N=1$ case, as explained before, is of no concern to us.).\\
Here ${\rm U}$ is a non compact group acting as an isometry group
on the scalar manifold while $H$, the isotropy subgroup, is of the
form:
\begin{equation}
H=H_{Aut} \otimes H_{matter}
\end{equation}
 $H_{Aut}$ being the automorphism group of the supersymmetry algebra
 while $H_{matter}$ is related to the matter multiplets.
 (Of course $H_{matter}=\bfone$ in all cases where supersymmetric matter
 doesn't exist, namely $N>4$).
The coset manifolds ${\rm U}/H$ and the automorphism groups for
various
 supergravity theories for any $D$ and $N$  can be found in the literature
 (see for instance \cite{cj},
\cite{sase}, \cite{castdauriafre}, \cite{noi}).
 As it was discussed in the previous section,
 the group ${\rm U}$ acts linearly on the
 field strengths ${\cal F}^\Lambda _{\mu\nu}$ appearing
 in the gravitational and matter
 multiplets. Here and in the following the index $\Lambda$ runs over
 the dimensions of some representation of the duality group ${\rm U}$.
The true duality symmetry (U--duality), acting on integral
quantized electric and magnetics charges,
 is  the restriction of  the continuous group ${\rm U}$ to the integers
 \cite{huto}. The moduli space of these theories is $ {\rm U}(\ZZ)
 \backslash {\rm U}/H$.
 \par
 All the properties of the given supergravity theories
 are completely fixed in terms of the geometry of ${\rm U}/H$,
 namely in terms of the coset representatives $L$ satisfying the
 relation:
 \begin{equation}
 L(\Phi^\prime) =g L(\Phi) h (g,\Phi)
\end{equation}
where $g\in {\rm U}$, $h\in H$  and $\Phi ^\prime =   \Phi ^\prime
 (\Phi)$,
 $\Phi$ being the coordinates of ${\rm U}/H$.
Note that the  scalar fields in ${\rm U}/H$ can be assigned, in
the linearized theory,  to linear representations
 $R_H$ of the local isotropy group  $H$ so that dim $R_H$ = dim ${\rm U}$
 $-$ dim $H$ (in the full theory, $R_H $ is the representation
 which the vielbein of ${\rm U}/H$ belongs to).
\par
 As explained in the following, the kinetic matrix $\cN_{\Lambda\Sigma}$ for
 the 2--forms ${\cal F}^\Lambda$ is fixed in terms of $L$ and the
 physical field strengths of the interacting theories are "dressed"
 with scalar fields in terms of the coset representatives.
 This allows us to write down the central charges associated to the
 vectors in the gravitational multiplet in a neat way in terms
 of the geometrical structure of the moduli space.
In an analogous way also the vectors of the matter multiplets give
rise to charges which, as we will see, are closely related to the
central charges.
\par To any field--strength $\cF^\Lambda$ we may associate a magnetic
charge $g^\Lambda$ and  an electric charge $e_\Lambda$ given
respectively by:
\begin{equation}
g^\Lambda = \int _{S^2} \cF^\Lambda \qquad \qquad e_\Lambda = \int
_ {S^2}  \cG _\Lambda
\end{equation}
These charges however are not the physical charges of the
interacting theory; the latter ones can be computed by looking at
the transformation laws of the fermion fields, where the physical
field--strengths appear dressed with the scalar fields
\cite{noi},\cite{noi1}. Let us first introduce the central
charges: they are associated to the dressed 2--form $T_{AB}$
appearing in the supersymmetry transformation law of the
gravitino 1-form. We have indeed:
\begin{equation}
\delta \psi_A = \nabla\epsilon_A + \alpha T_ {AB\vert
\mu\nu}\gamma^a\gamma^{ \mu\nu} \epsilon^B V_a+ \cdots
\label{tragra}
\end{equation}
Here $\nabla$ is the covariant derivative in terms of the
space--time spin connection and the composite connection of the
automorphism group $H_{Aut}$, $\alpha$ is a coefficient fixed by
supersymmetry, $V^a$ is the space--time vielbein, $A=1,\cdots ,N$
is the index acted on by the automorphism group. Here and in the
following
 the dots denote trilinear fermion terms which are characteristic
 of any supersymmetric theory but do not play any role in the following discussion.
The field-strength $T_{AB}$ will be constructed by dressing the
bare field-strengths  $\cF^\Lambda$ with the coset representative
$L(\Phi)$ of ${\rm U}/H$, $\Phi$ denoting a set of coordinates of
${\rm U}/H$.
\par
 Note that the same field strength $T_{AB}$ which appears in the gravitino
 transformation laws is also present in the dilatino transformation laws
  in the following way:
  \begin{equation}
\delta \chi_{ABC} = P_{ABCD,\ell}\partial_\mu \phi^\ell \gamma^\mu
\epsilon^D +\beta T_{[AB\vert \mu\nu}\gamma^{\mu\nu} \epsilon_{C]}
+\cdots   \label{tradil}
\end{equation}
 In an analogous way, when vector multiplets are present,
 the matter vector field
 strengths appearing in the transformation laws of the gaugino fields
 are dressed with the scalars:
  \begin{equation}
\delta \lambda_{IA}= {\rm i} P_{IAB,i}\partial_\mu \Phi^i
\gamma^\mu \epsilon^B +\gamma T_{I\vert \mu\nu}\gamma^{\mu\nu}
\epsilon_A +\cdots \label{tragau}
\end{equation}
  where $P_{ABCD}= P_{ABCD,\ell}d\phi^\ell$ and  $ P^I_{AB }=P^I_{AB,i }d\Phi^i$
  are the vielbein of the  scalar manifolds
  spanned by the scalar fields of the gravitational and vector multiplets respectively
  (more precise definitions are given below),
  and $\beta$ and $\gamma$ are constants fixed by supersymmetry.
  \par
In order to give the explicit dependence on scalars of $T_{AB}$,
$T^I$, it is necessary to recall from the previous subsection
that, according to the Gaillard--Zumino construction, the isometry
group ${\rm U}$ of the scalar manifold acts on the
  vector $(F^{- \Lambda},\cG^{-}_\Lambda)$
  (or its complex conjugate) as a subgroup of
  $Sp(2 n_V,\IR)$ ($n_V$ is the number of vector fields)
with duality transformations interchanging electric and magnetic
 field--strengths:
 \begin{equation}
{\cal S} \left(\begin{matrix}\cF^{-\Lambda} \cr
\cG^-_\Lambda\cr\end{matrix}\right)=
\left(\begin{matrix}\cF^{-\Lambda} \cr
\cG^-_\Lambda\cr\end{matrix}\right)^\prime
\end{equation}
according to the discussion in the previous subsection.
\\
If $L(\Phi)$ is the coset representative of ${\rm U}$ in some
representation, ${\cal S}$ represents the embedded coset
representative belonging to $Sp(2n_V,\IR)$ and in each theory,
$A,B,C,D$ can be constructed in terms of $L(\Phi)$. Using a
complex basis in the vector space of $Sp(2 n_V)$, we may rewrite
the symplectic matrix as a pseudo-unitary symplectic matrix of the
following form:
\begin{equation}
U = \frac{1 }{\sqrt{2}}\left(\begin{matrix}f+{\rm i}h & \bar
f+{\rm i}\bar h \cr f-{\rm i}h &\bar f-{\rm i}\bar h \cr
\end{matrix}\right) =
  \cA^{-1} {\cal S} \cA
  \label{defu}
\end{equation}
where:
\begin{eqnarray}
  f&=&\frac{1}{\sqrt{2}} (A-{\rm i} B) \nonumber\\
h &=&\frac{1}{\sqrt{2}} (C-{\rm i} D) \nonumber\\
 \cA &=& \left(\begin{matrix}1 & 1 \cr -{\rm i} & {\rm i}\cr\end{matrix}\right)
\end{eqnarray}
We will denote the $Sp(2n_V)$ group in the pseudo-unitary basis
as $Usp(n_V,n_V)$.
 The requirement that $ {U}\in Usp(n_V,n_V)$ must satisfy is:
\begin{eqnarray}
U^t \IC U &=& \IC \, ,\quad  \IC ^t = -\IC  \quad \mbox{($U$
symplectic)}\\
U^\dagger \eta U &=& \eta \, ,\quad  \eta =
\left(\begin{matrix}\bfone_{n_V \times n_V} & 0 \cr 0 & -
\bfone_{n_V \times n_V}\cr\end{matrix}\right) \quad \mbox{($U$
pseudo-unitary)}
\end{eqnarray}
 which implies, on the sub-blocks $f$ and $h$:
 \begin{equation}
\left\lbrace\begin{matrix}{\rm i}(f^\dagger h - h^\dagger f) &=&
\bfone \cr (f^t  h - h^t f) &=& 0\cr\end{matrix} \right.
\label{specdef}
\end{equation}
The $n_V\times n_V$ subblocks of $U$ are submatrices $f,h$ which
can be decomposed with respect to the isotropy group $H_{Aut}
\times H_{matter}$ as:
\begin{eqnarray}
  f&=& (f^\Lambda_{AB} , f^\Lambda _I) \nonumber\\
h&=& (h_{\Lambda AB} , h_{\Lambda I}) \label{deffh}
\end{eqnarray}
where $AB$ are indices in the antisymmetric representation of
$H_{Aut}= SU(N) \times U(1)$ and $I$ is an index of the
fundamental representation of $H_{matter}$. Upper $SU(N)$ indices
label objects in the complex conjugate representation of $SU(N)$:
$(f^\Lambda_{AB})^* = f^{\Lambda AB}$ etc.
\par
Note that we can consider $(f^\Lambda_{AB}, h_{\Lambda AB})$ and
$(f^\Lambda_{I}, h_{\Lambda I})$ as symplectic sections of a
$Sp(2n_V, \IR)$ bundle over ${\rm U}/H$. We will see in the
following that this bundle is actually flat. The real embedding
given by $S$ is appropriate for duality transformations  of
$\cF^\pm$
 and their duals $\cG^\pm$, according to equations (\ref{trasform}),
  while
the complex embedding in the matrix $U$ is appropriate in writing
down the fermion transformation laws and supercovariant
field--strengths. The kinetic matrix $\cN$, according to
Gaillard--Zumino \cite{gaizum}, can be written in terms of the
sub-blocks $f$, $h$, and turns out to be:
\begin{equation}
\cN= hf^{-1}, \quad\quad \cN = \cN^t \label{nfh-1}
\end{equation}
transforming projectively under $Sp(2n_V ,\IR)$ duality rotations
as already shown in the previous section. By using
(\ref{specdef})and (\ref{nfh-1}) we find that
   \begin{equation}
   (f^t)^{-1} = {\rm i} (\cN - \bar \cN)\bar f
\end{equation}
 that is
 \begin{eqnarray}
  f_{AB \Lambda} &\equiv& (f^{-1})_{AB \Lambda} =
{\rm i} (\cN - \bar \cN)_{\Lambda\Sigma}\bar f^\Sigma_{AB}\\
f_{I \Lambda} &\equiv& (f^{-1})_{I \Lambda} = {\rm i} (\cN - \bar
\cN)_{\Lambda\Sigma}\bar f^\Sigma_{I}
\end{eqnarray}
 It can be shown \cite{noi} that the dressed graviphotons and matter self--dual
field--strengths appearing in the transformation
 law of gravitino (\ref{tragra}), dilatino (\ref{tradil})
 and gaugino (\ref{tragau})
  can be constructed as a symplectic invariant  using the $f$ and $h$ matrices, as follows:
\begin{eqnarray}
T^-_{AB}&=& {\rm i} (\bar f^{-1})_{AB \Lambda}F^{- \Lambda} =
f^\Lambda_{AB}(\cN - \bar \cN)_{\Lambda\Sigma}F^{-\Sigma}=
 h_{\Lambda AB} F^{-\Lambda} -f^\Lambda_{AB} \cG^-_\Lambda  \nonumber\\
  T^-_{I}&=&{\rm i} (\bar f^{-1})_{I\Lambda}F^{- \Lambda} =  f^\Lambda_{I}(\cN - \bar \cN)_{\Lambda\Sigma}F^{-\Sigma}=
 h_{\Lambda I} F^{-\Lambda} - f^\Lambda_{I} \cG^-_\Lambda \nonumber\\
\bar T^{+ AB} &=& (T^-_{AB})^* \nonumber\\
\bar T^{+ I} &=& (T^-_{I})^* \label{gravi}
\end{eqnarray}
(For $N>4$, supersymmetry does not allow matter multiplets and $f
_{\Lambda I}= T_I =0$).
 To construct the dressed charges one integrates
 $T_{AB} = T^+_{AB} + T^ -_{AB}  $ and  (for $N=3, 4$)
 $T_I = T^+_I + T^ -_I $ on a large 2-sphere.
 For this purpose we note that
 \begin{eqnarray}
 T^+_{AB} & = &  h_{\Lambda
 AB}F^{+\Lambda} - f^\Lambda_{AB} \cG_\Lambda^+  =0  \label{tiden0}\\
   T^+_I & = &  h_{\Lambda
 I}F^{+\Lambda}-f^\Lambda_{I} \cG_\Lambda^+   =0 \label{tiden}
\end{eqnarray}
as a consequence of eqs. (\ref{nfh-1}), (\ref{trasform}).
Therefore we can introduce the ``dressed'' charges:
\begin{eqnarray}
Z_{AB} (\Phi_0)& = & \int_{S^2} T_{AB} = \int_{S^2} (T^+_{AB} + T^
-_{AB}) = \int_{S^2} T^ -_{AB} =
  h_{\Lambda AB} g^\Lambda- f^\Lambda_{AB} e_\Lambda
\label{zab}\\
Z_I(\Phi_0) & = & \int_{S^2} T_I = \int_{S^2} (T^+_I + T^ -_I) =
\int_{S^2} T^ -_I =
 h_{\Lambda I} g^\Lambda - f^\Lambda_I e_\Lambda   \quad (N\leq 4)
\label{zi}
\end{eqnarray}
where:
\begin{equation}
e_\Lambda = \int_{S^2} \cG_\Lambda , \quad g^\Lambda = \int_{S^2}
F^\Lambda \label{charges}
\end{equation}
and the sections $(f^\Lambda, h_\Lambda)$ on the right hand side
now depend on the {\it v.e.v.}'s $\Phi_0\equiv \Phi(r=\infty )$ of
the scalar fields $\Phi^I$. We see that the  central and matter
charges are given in this case by symplectic invariants and that
the presence of dyons in $D=4$ is related to the
 symplectic embedding.
\par
The scalar field dependent combinations of fields strengths
appearing in the fermion supersymmetry transformation rules have a
profound meaning and play a key role in the physics of BPS
black--holes. The combination $T_{AB\mu\nu}$ is named the
graviphoton field strength and its integral over a $2$--sphere at
infinity gives the value of {\it the central charge $Z_{AB}$} of
the $N=2$ supersymmetry algebra. The combination $T^{I}_{\mu\nu}$
is named the matter field strength. Evaluating its integral on a
$2$--sphere at infinity one obtains the so called {\it matter
charges} $Z^I$.
\par
We are now able to derive some  differential relations among the
central and matter charges using the Maurer--Cartan equations
obeyed by the scalars through the embedded coset representative
$U$. Indeed, let $\Gamma = U^{-1} dU$ be the $Usp(n_V,n_V)$ Lie
algebra left invariant one form satisfying:
\begin{equation}
  d\Gamma +\Gamma \wedge \Gamma = 0
\label{int}
\end{equation}
In terms of $(f,h)$ $\Gamma$ has the following form:
\begin{equation}
  \label{defgamma}
  \Gamma \equiv U^{-1} dU =
\left(\begin{matrix}{\rm i} (f^\dagger dh - h^\dagger df) & {\rm
i} (f^\dagger d\bar h - h^\dagger d\bar f) \cr -{\rm i} (f^t dh -
h^t df) & -{\rm i}(f^t d\bar h - h^t d\bar f)
\cr\end{matrix}\right) \equiv \left(\begin{matrix}\Omega^{(H)}&
\bar \cP \cr \cP & \bar \Omega^{(H)} \cr \end{matrix}\right)
\end{equation}
where the $n_V \times n_V$ subblocks $ \Omega^{(H)}$ and  $\cP$
embed the $H$ connection and the vielbein of ${\rm U}/H$
respectively . This identification follows from the Cartan
decomposition of the $Usp(n_V,n_V)$ Lie algebra. Explicitly, if we
define the $H_{Aut} \times H_{matter}$--covariant derivative of a
vector $V= (V_{AB},V_I)$ as:
\begin{equation}
\nabla V = dV - V\omega , \quad \omega =
\left(\begin{matrix}\omega^{AB}_{\ \ CD} & 0 \cr 0 & \omega^I_{\
J} \cr\end{matrix}\right) \label{nablav}
\end{equation}
we have:
\begin{equation}
\Omega^{(H)} = {\rm i} [ f^\dagger (\nabla h + h \omega) -
h^\dagger (\nabla f + f \omega)] = \omega \bfone \label{defomega}
\end{equation}
where we have used:
\begin{equation}
\nabla h = \bar \cN \nabla f ; \quad h = \cN f
\end{equation}
and  the fundamental identity (\ref{specdef}). Furthermore, using
the same relations, the embedded vielbein  $\cP$ can be written as
follows:
\begin{equation}
\cP = - {\rm i} ( f^t \nabla h - h^t \nabla f) = {\rm i} f^t (\cN
- \bar \cN) \nabla f
\end{equation}
From  (\ref{defu}) and (\ref{defgamma}),  we obtain the $(n_V
\times n_V)$ matrix equation:
\begin{eqnarray}
  \nabla(\omega) (f+{\rm i} h) &=& (\bar f + {\rm i} \bar h) \cP \nonumber \\
\nabla(\omega) (f-{\rm i} h) &=& (\bar f - {\rm i} \bar h) \cP
\end{eqnarray}
together with their complex conjugates. Using further the
definition (\ref{deffh}) we have:
\begin{eqnarray}
  \nabla(\omega) f^\Lambda_{AB} &=&  \bar f^\Lambda_{I}  P^I_{AB} + \frac{1}{2}\bar f^{\Lambda CD} P_{ABCD} \nonumber \\
\nabla(\omega) f^\Lambda_{I} &=&  \frac{1}{2} \bar f^{\Lambda AB}
P_{AB I} +\bar f^{\Lambda J}  P_{JI}
 \label{df}
\end{eqnarray}
where we have decomposed the embedded vielbein $\cP$ as follows:
\begin{equation}
  \label{defp}
  \cP = \left(\begin{matrix}P_{ABCD} & P_{AB J} \cr P_{I CD} & P_{IJ}\cr \end{matrix}\right)
\end{equation}
 the subblocks being related to the vielbein of ${\rm U}/H$, $P = L^{-1} \nabla^{(H)} L$,
 written in terms of the indices of $H_{Aut} \times H_{matter}$.
 In particular, the component $P_{ABCD}$ is completely
 antisymmetric in its indices.
Note that, since $f$ belongs to the unitary matrix $U$, we have:
$(  f^\Lambda _{AB},  f^\Lambda_I)^\star = (\bar f^{\Lambda
AB},\bar f^{\Lambda I})$. Obviously, the same differential
relations that we wrote for $f$ hold true for the dual matrix $h$
as well.
\par
Using the definition of the charges (\ref{zab}), (\ref{zi}) we
then get the following differential relations among charges:
\begin{eqnarray}
  \nabla(\omega) Z_{AB} &=&  \bar Z_{I}  P^I_{AB} +  \frac{1}{2} \bar Z^{CD} P_{ABCD} \nonumber \\
\nabla(\omega) Z_{I} &=& \frac{1}{2}  \bar Z^{AB}  P_{AB I} + \bar
Z_{J}  P^J_{I}
 \label{dz1}
\end{eqnarray}
Depending on the coset manifold, some of the subblocks of
(\ref{defp}) can be actually zero. For example in $N=3$ the
vielbein of ${\rm U}/H = \frac{SU(3,n)}{SU(3)
 \times SU(n)\times U(1)}$ \cite{ccdffm} is $P_{I AB}$ ($AB $ antisymmetric),
 $I=1,\cdots,n;
A,B=1,2,3$ and it turns out that $P_{ABCD}= P_{IJ} = 0$.
\par
In $N=4$, ${\rm U}/H= \frac{SU(1,1)}{U(1)} \times \frac{ O(6,n)}{
O(6) \times O(n)}$ \cite{bekose}, and we have $P_{ABCD}=
\epsilon_{ABCD} P$, $P_{IJ} = \bar P \delta_{IJ}$, where $P$ is
the K\"ahlerian vielbein of $\frac{SU(1,1)}{U(1)}$,  ($A,\cdots ,
D $ $SU(4)$ indices and $I,J$ $O(n)$ indices) and $P_{I AB}$ is
the vielbein of $\frac{ O(6,n)}{O(6) \times O(n)}$.
\par
For $N>4$ (no matter indices) we have that $\cP $ coincides with
the vielbein $P_{ABCD}$ of the relevant ${\rm U}/H$.
\par
For the purpose of comparison of the previous formalism with the
$N=2$ supergravity case, where the $\sigma$-model is in general
not a coset, it is interesting to note that, if the connection
$\Omega^{(H)}$ and the vielbein $\cP$ are regarded as data of
${\rm U}/H$, then the Maurer--Cartan equations (\ref{df}) can be
interpreted as an integrable system of differential equations for
a section $V=(V_{AB}, V_I, \bar V^{AB} , \bar V^I)$ of the
symplectic fiber bundle constructed over ${\rm U}/H$. Namely the
integrable system:
\begin{equation}
 \nabla \left(\begin{matrix}V_{AB} \cr V_I \cr \bar V^{AB} \cr \bar V^I \end{matrix}\right) =
  \left(\begin{matrix}0 & 0 & \frac{1}{2} P_{ABCD} & P_{AB J} \cr
0 & 0 & \frac{1}{2} P_{I CD} & P_{IJ} \cr \frac{1}{2}P^{ABCD} &
P^{ AB J} & 0 & 0 \cr \frac{1}{2}P^{I CD} & P^{IJ} & 0 & 0
\cr\end{matrix}\right)
 \left(\begin{matrix}V_{CD} \cr V_J \cr \bar V^{CD} \cr \bar V^J \end{matrix}\right)
 \label{intsys}
\end{equation}
 has $2n$ solutions given by $V = (f^\Lambda_{\ AB},f^\Lambda_{\ I}),( h_{\Lambda AB}, h_{\Lambda_I})$, $\Lambda = 1,\cdots,n$.
The integrability condition (\ref{int}) means that $\Gamma$ is a
flat connection of the symplectic bundle. In terms of the geometry
of ${\rm U}/H$ this in turn implies that the $\IH$--curvature (and
hence, since the manifold is a symmetric space, also the
Riemannian curvature) is constant, being proportional to the
wedge product of two vielbein.
\par
Besides the differential relations (\ref{dz1}), the charges also
satisfy sum rules.
\par
The sum rule has the following form:
\begin{equation}
 \frac{1}{2} Z_{AB} \bar Z^{AB} + Z_I \bar Z^I = -\frac{1}{2} P^t \cM (\cN) P
\label{sumrule}
\end{equation}
where $\cM(\cN)$ and $P$ are:
\begin{equation}
\cM = \left( \begin{matrix} \bfone & - Re \cN \cr 0
&\bfone\cr\end{matrix}\right) \left( \begin{matrix} Im \cN & 0 \cr
0 &Im \cN^{-1}\cr\end{matrix}\right) \left( \begin{matrix} \bfone
& 0 \cr - Re \cN & \bfone \cr\end{matrix}\right) \label{m+}
\end{equation}
\begin{equation}
P=\left(\begin{matrix}g^\Lambda \cr e_ \Lambda \cr\end{matrix}
\right) \label{eg}
\end{equation}
In order to obtain this result we just need to observe that from
the
 fundamental identities (\ref{specdef}) and  from the definition of the
 kinetic matrix given in (\ref{nfh-1}) it  follows:
 \begin{eqnarray}
ff^\dagger &= &-{\rm i} \left( \cN - \bar \cN \right)^{-1} \\
hh^\dagger &= &-{\rm i} \left(\bar \cN^{-1} - \cN^{-1}
\right)^{-1}\equiv
-{\rm i} \cN \left( \cN - \bar \cN \right) ^{-1}\bar \cN \\
hf^\dagger &= & \cN ff^\dagger  \\
fh^\dagger & = & ff^\dagger \bar \cN .
\end{eqnarray}

\subsubsection{The $N=2$ theory} The formalism we have developed so
far for the $D=4$, $N>2$ theories is completely determined by the
embedding of the coset representative of ${\rm U}/H$ in
$Sp(2n,\IR)$ and by the embedded Maurer--Cartan equations
(\ref{df}). We want now to show that this formalism, and in
particular the identities (\ref{specdef}), the differential
relations among charges (\ref{dz1}) and the sum rules
(\ref{sumrule}) of $N=2$ matter-coupled supergravity
\cite{spegeo},\cite{str} can be obtained in a way completely
analogous to the coset space $\sigma$-model cases discussed in
the previous subsection. This follows essentially from the fact
that, though the scalar manifold $\cM_{N=2}$ of the $N=2$ theory
is not in general a coset manifold, nevertheless it has a
symplectic structure identical to the $N>2$ theories, as a
consequence of the Gaillard--Zumino duality.
\par
In the case of $N=2$ supergravity the requirements imposed by
supersymmetry on the scalar manifold ${\cal M}_{scalar}$ of the
theory is that it should be the following direct product: $ {\cal
M}_{scalar}={\cal M}^{SK}\, \otimes \, {\cal M}^Q$ where ${\cal
M}^{SK}$ is a {\it special K\"ahler} manifold of complex dimension
$n$ and ${\cal M}^Q$ a {\it qua\-ter\-nio\-nic} manifold of real
dimension $4n_H$. Note that $n$ and $n_H$ are respectively the
number of {\it vector multiplets} and {\it hypermultiplets}
contained in the theory. The direct product structure imposed by
supersymmetry precisely reflects the fact that the quaternionic
and special K\"ahler scalars belong to different supermultiplets.
In the construction of BPS black--holes it turns out that the
hyperscalars are spectators playing no dynamical role. Hence we
do not discuss here the hypermultiplets any further and we
confine our attention to an $N=2$ supergravity where the graviton
multiplet, containing, besides the graviton $g_{\mu \nu}$, also a
graviphoton $A^0_{\mu}$, is coupled to $n$  {\it vector
multiplets}. Such a theory has an action of type \eqn{genact}
where the number of gauge fields is ${n_V}=1+n$ and the number of
scalar fields is $m=2 \, {n}$. Correspondingly the indices have
the following ranges
\begin{eqnarray}
 \Lambda,\Sigma,\Gamma,\dots & = & 0,1,\dots, n \nonumber\\
 I,J,K,\dots &=&1,\dots,2 \, n \nonumber\\
 \label{indin}
\end{eqnarray}
To make the action \eqn{genact} fully explicit, we need to discuss
the geometry of the vector multiplets scalars, namely special
K\"ahler geometry. We refer to \cite{n=2} for a detailed analysis.
A special K\"ahler manifold is a K\"ahler-Hodge manifold endowed
with an extra symplectic structure. A K\"ahler manifold ${\cal
M}$ is a Hodge manifold if and only if there exists a $U(1)$
bundle ${\cal L} \, \longrightarrow \, {\cal M}$ such that its
first Chern class equals the cohomology class of the K\"ahler
2-form K:
\begin{equation}
c_1({\cal L} )~=~\left [ \, K \, \right ] \label{chernclass25}
\end{equation}
In local terms this means that there is a holomorphic section
$W(z)$ such that we can write
\begin{equation}
K\, =\, {\rm i} \, g_{i\bar\jmath} \, dz^{i} \, \wedge \, d{\bar
z}^{j^{\star}} \label{chernclass26}
\end{equation}
where $z^i$ are $n$ holomorphic coordinates on ${\cal M}^{SK}$ and
$g_{i\bar\jmath}$ its metric.\\ In this case the $U(1)$ K\"ahler
connection is given by
\begin{equation}
\cQ = -\frac{\rm i}2 \left(\partial_i \cK dz^i -
\partial_{\bar\imath}\cK d\bar z^{\bar\imath}\right)
\end{equation}
where  $\cK$ is the K\"ahler potential, so that $K = d\cQ$.
\par
 Let now
 $\Phi (z, \bar z)$ be a section of the $U(1)$ bundle.  By definition its
covariant derivative is
\begin{equation}
\nabla \Phi = (d + i p {\cal Q}) \Phi
\end{equation}
or, in components,
\begin{equation}
\begin{array}{ccccccc}
\nabla_i \Phi &=&
 (\partial_i + \frac{1}{2} p \partial_i {\cal K}) \Phi &; &
\nabla_{\bar\imath}\Phi &=&(\partial_{\bar\imath}-\frac{1}{2} p
\partial_{\bar\imath} {\cal K}) \Phi \cr
\end{array}
\label{scrivo2}
\end{equation}
A covariantly holomorphic section is defined by the equation: $
\nabla_{\bar\imath} \Phi = 0  $. Setting:
\begin{equation}
\tilde{\Phi} = e^{-p {\cal K}/2} \Phi  \,   . \label{mappuccia}
\end{equation}
we get:
\begin{equation}
\begin{array}{ccccccc}
\nabla_i    \tilde{\Phi}&    =& (\partial_i   +   p   \partial_i
{\cal K}) \tilde{\Phi}& ; & \nabla_{i^*}\tilde{\Phi}&=&
\partial_{i^*} \tilde{\Phi}\cr
\end{array}
\end{equation}
so that under this map covariantly holomorphic sections $\Phi$
become truly holomorphic sections.
\par
There are several equivalent ways of defining what a special
K\"ahler manifold is. An intrinsic definition is the following. A
special K\"ahler manifold can be given by constructing a
$2n+2$-dimensional symplectic bundle over the K\"ahler--Hodge
manifold whose generic sections (with weight $p=1$)
\begin{equation}V = (f^\Lambda,h_\Lambda)\ \ \ \ \
\Lambda=0,\ldots,n\ , \label{sezio} \end{equation}are covariantly
holomorphic
\begin{equation}\nabla_{\bar\imath} V =
(\del_{\bar\imath}-\half\del_{\bar\imath} {\cal K}) V=0
\label{ddue}
\end{equation}and satisfy the further condition \begin{equation}i
<V,\bar V
>=i(\bar f^\Lambda h_\Lambda-\bar h_\Lambda f^\Lambda)=1\ ,
\label{dtre} \end{equation}where $<\ ,\ >$ denotes a symplectic
inner product with metric chosen to be $\left(\begin{matrix}0
&-\bfone\cr\bfone&0\end{matrix}\right)$.
\par
Defining $U_i =D_i V=(f_i^\Lambda,h_{\Lambda i})$, and introducing
a symmetric three-tensor $ C_{ijk}$ by
\begin{equation}D_i U_j=i
C_{ijk} g^{k\bar k}\bar U_{\bar k}\ , \label{dqua}
\end{equation}
the set of differential equations \ba
D_i V &=& U_i\nonumber\\
D_i U_j &=& i C_{ijk} g^{k\bar k}\bar U_{\bar k}\nonumber\\
D_i U_{\bar\jmath} &=& g_{i \bar\jmath } \bar V\nonumber\\
D_{i} \bar V &=& 0 \label{geospec} \ea defines a symplectic
connection. Requiring that the differential system \eqn{geospec}
is integrable is equivalent to require that the symplectic
connection
 is flat.
Since the integrability condition of \eqn{geospec} gives
constraints on the base K\"ahler--Hodge manifold, we define
special-K\"ahler a manifold whose associated symplectic
connection is flat. At the end of this section we will give the
restrictions on the manifold imposed by the flatness of the
connection.
\par
It must be noted that, for special K\"ahler manifolds, the
K\"ahler potential can be computed as a symplectic invariant from
eq. (\ref{dtre}). Indeed, introducing also the holomorphic
sections \ba
\Omega &=& e^{-{\cal K}/2} V=e^{-{\cal K}/2} (f^\Lambda, h_\Lambda)=(X^\Lambda, F_\Lambda)\nonumber\\
\del_{\bar\imath}\Omega &=& 0 \ea eq. (\ref{dtre})  gives
\begin{equation}
{\cal K}=-\ln {\rm i}<\Omega,\bar \Omega >=-\ln i(\bar X^\Lambda
F_\Lambda- X^\Lambda \bar F_\Lambda)\ . \label{dtrd}
\end{equation}
If we introduce the complex symmetric $(n+1)\times (n+1)$ matrix
$\cN_{\Lambda\Sigma}$  defined through the relations
\begin{equation}h_\Lambda=\cN_{\Lambda\Sigma} f^\Sigma\ \ , \ \
h_{i^\star\Lambda}=\cN_{\Lambda\Sigma} f^\Sigma_{i^\star}\ ,
\label{defi} \end{equation} then we have: \ba <V,\bar V >&=&
\left(\cN- \bar \cN \right)_{\Lambda\Sigma} f^\Lambda \bar
f^\Sigma = -{\rm i}
 \ \ \to\ \ {\cal K}=-\ln[{\rm i}(\bar X^\Lambda
\left(\cN- \bar \cN \right)_{\Lambda\Sigma} X^\Sigma)] \\
g_{i \bar\jmath} &=& -i <U_i, U_{ \bar\jmath} >=-2 f^\Lambda_i
\mbox{Im} \cN_{\Lambda\Sigma}
f^\Sigma_{\bar\jmath}\ ,\label{metri}\\
C_{ijk} &=& <D_i U_j, U_k> = 2{\rm i} {\rm Im}\cN_{\Lambda\Sigma}
f^\Lambda_i \nabla_j f^\Sigma_k \ . \label{dott} \ea The matrix
$\cN_{\Lambda\Sigma}$ turns out to be the matrix appearing in the
kinetic lagrangian of the vectors in $N=2$ supergravity.   Under
coordinate transformations, the sections $\Omega$ transform as
\begin{equation}\tilde\Omega =e^{-f_{\cal S} (z)}{\cal S}\Omega\ ,
\label{ddoc} \end{equation}where ${\cal S}
=\left(\begin{matrix}A&B\cr C & D\cr\end{matrix}\right)$ is an
element of $Sp(2n_V,\IR)$ and the factor $e^{-f_{\cal S} (z)}$ is
a $U(1)$ K\"ahler transformation. We also note that, from the
definition of $\cN$, eq. (\ref{defi}):
\begin{equation}\tilde\cN (\tilde X,\tilde F)=(C+D\cN
(X,F))(A+B\cN (X,F))^{-1}\ , \label{dvdu} \end{equation}
\par
Let us now  set $n_V=n+1$ and define  the $n_V \times n_V$
matrices:
\begin{equation}
f^\Lambda_\Sigma \equiv \left( f^\Lambda,  \bar f^{\Lambda i}
\right); \quad h_{\Lambda \Sigma} \equiv \left( h_ \Lambda,
 \bar h_{\Lambda}^{\ i} \right)
\label{matrici}
\end{equation}
where $\bar f^{\Lambda i} \equiv \bar f^\Lambda_{\
\bar\jmath}g^{\bar\jmath i}$, $\bar h_{\Lambda}^{\ i}\equiv \bar
h_{\Lambda \bar\jmath}g^{\bar\jmath i}$ , the set of algebraic
relations of special geometry can  be written  in matrix form as:
\begin{equation}
\left\lbrace\begin{matrix}{\rm i}(f^\dagger h - h^\dagger f) &=&
\bfone \cr (f^t h - h^t f) &=& 0\cr\end{matrix} \right.
\label{specdef1}
\end{equation}
Recalling equations (\ref{specdef}) we see that the previous
relations imply that the matrix
 $U$:
\begin{equation}
U={\frac{1}{\sqrt{2}}}\left ( \begin{matrix}f+{\rm i} h & \bar f +
{\rm i} \bar h \cr
 f-{\rm i} h & \bar f - {\rm i} \bar h \cr\end{matrix} \right)
\end{equation}
is a pseudo-unitary symplectic matrix. In fact if we set
$f^\Lambda \to f^\Lambda \epsilon_{AB}\equiv f^\Lambda_{AB}$ and
flatten the world-indices of $( f^\Lambda_i,\bar
f^\Lambda_{\bar\imath})$ (or $(\bar h_i, h_{\bar\imath})$) with
the K\"ahlerian vielbein $P^I_i, \bar P^{\bar I}_{\bar \imath}$:
\begin{equation}
  ( f^\Lambda_I, \bar f^\Lambda_{\bar I}) = ( f^\Lambda_i P^i_I,
  \bar f^\Lambda_{\bar \imath}\bar P^{\bar \imath}_{\bar I} ),
\quad  \quad P^I_{i} \bar P^{\bar J}_{\bar\jmath}\eta_{I\bar J} =
g_{i\bar\jmath}
\end{equation}
where $\eta_{I\bar J}$ is the flat K\"ahlerian metric and $P^i_{\
I} = (P^{-1})^I_{\ i}$, the relations (\ref{specdef1}) are just a
particular case of equations (\ref{specdef}) since, for $N=2$,
$H_{Aut}= SU(2) \times U(1)$, so that $f^\Lambda_{AB}$ is actually
an $SU(2)$ singlet.
\par
Let us now consider the analogous of the embedded Maurer--Cartan
equations of ${\rm U}/H$. We introduce, as before, the matrix
one--form $\Gamma= U^{-1} dU$ satisfying the relation $d\Gamma+
\Gamma\wedge\Gamma =0$. We further introduce the covariant
derivative of the symplectic section $(f^\Lambda,\bar
f^\Lambda_{\bar I}, \bar f^\Lambda , f^\Lambda_I)$ with respect
to the $U(1)$--K\"ahler connection $\cQ$ and the spin connection
$\omega^{IJ}$ of $\cM_{N=2}$:
 \begin{eqnarray}
& \nabla (f^\Lambda,\bar  f^\Lambda_{\bar I}, \bar f^\Lambda,  f^\Lambda_I) = & \nonumber\\
& d (f^\Lambda,\bar f^\Lambda_{\bar I},  f^\Lambda,\bar
f^\Lambda_I) + (f^\Lambda,\bar f^\Lambda_{\bar J}, \bar f^\Lambda,
f^\Lambda_J)\left(\begin{matrix}-{\rm i} \cQ   & 0 & 0 & 0 \cr 0 &
{\rm i} \cQ \delta^{\bar I} _{\bar J} + \omega^{\bar I}_{\ \bar
J} & 0 & 0 \cr  0 & 0 &  {\rm i} \cQ & 0 \cr 0 & 0 & 0 &-{\rm i}
\cQ \delta^{ I} _{ J} + \omega^{I}_{\ J} \cr\end{matrix}\right) &
\end{eqnarray}
 the K\"ahler weight of $(f^\Lambda, \bar f^\Lambda_I)$ and
$(\bar f^\Lambda,\bar f^\Lambda_{\bar I})$ being $p=1$ and $p=-1$
respectively. Using the same decomposition as in equation
(\ref{defgamma}) and eq.s (\ref{nablav}), (\ref{defomega}) we
have in the $N=2$ case:
\begin{eqnarray}
  \label{gammaspec}
  \Gamma &=& \left(\begin{matrix}\Omega & \bar \cP \cr \cP & \bar \Omega \cr\end{matrix}\right)
  ,\nonumber\\
  \Omega &=& \omega = \left(\begin{matrix}-{\rm i}  \cQ  & 0 \cr 0 &
  {\rm i}   \cQ \delta^I_J+ \bar \omega^I_{\ J} \cr\end{matrix}\right)
   \end{eqnarray}
For the subblock $\cP$ we obtain:
\begin{equation}
\cP = - {\rm i} (f^t \nabla h - h^t \nabla f)  = {\rm i} f^t (\cN
- \bar \cN) \nabla f = \left(\begin{matrix}0 &  P_{\bar I} \cr
P^{ J} & P^{ J}_{\ \bar I} \cr\end{matrix}\right)  \label{viel2}
\end{equation}
where $\bar P^J \equiv \eta^{J\bar I} P_{\bar I} $ is the
$(1,0)$--form K\"ahlerian vielbein while $ P^{ J}_{\ \bar I}
\equiv {\rm i} \left( f^t (\cN - \bar \cN) \nabla f \right)^{
J}_{\ \bar I}$ is a one--form which in general cannot be expressed
in terms of the vielbein $P^I$, since the manifold is in general
not a coset, and therefore represents a new geometrical quantity
on $\cM_{N=2}$. Note that we get zero in the first entry of
equation (\ref{viel2}) by virtue of the fact that the identity
(\ref{specdef1}) implies $f^\Lambda(\cN - \bar
\cN)_{\Lambda\Sigma}f^\Sigma_i =0 $ and that $f^\Lambda$ is
covariantly holomorphic. If $\Omega $ and $\cP $ are considered
as data on $\cM_{N=2}$ then we may interpret $\Gamma =U^{-1} dU$
as an integrable system of differential equations, namely:
\begin{equation}
  \label{picfuc}
  \nabla (V ,\bar U_{\bar I} , \bar V ,   U_I ) =  (V ,\bar U_{\bar J} , \bar V ,  U_J )
  \left(\begin{matrix}0& 0 & 0 & \bar P_{I} \cr
0 & 0 & \bar P^{\bar J} & \bar P^{\bar J}_{\ I} \cr 0 &  P_{\bar
I} & 0 & 0 \cr  P^J & P^J_{\ \bar I} & 0 & 0
\cr\end{matrix}\right)
\end{equation}
where the flat K\"ahler indices $I,\bar I, \cdots $ are raised and
lowered with the flat K\"ahler metric $\eta_{I\bar J}$. Note that
the equation \eqn{picfuc} coincides with the set of relations
\eqn{geospec} if we trade world indices $i, \bar\imath$ with flat
indices $I,\bar I$, provided we also identify:
\begin{equation}
\bar P^{\bar J}_{\ I}=\bar P^{\bar J}_{\ I k}dz^k= P^{\bar J, i}
P_I^{\ j} C_{ijk} dz^k .
\end{equation}
Then, the integrability condition $d\Gamma + \Gamma\wedge
\Gamma=0$ is equivalent to the flatness of the special K\"ahler
symplectic connection and it gives the following three constraints
on the K\"ahler base manifold:
\begin{eqnarray}
  d( {\rm i} \cQ) + \bar P_I \wedge  P^{I} &=& 0
  \to \partial_{\bar\jmath}\partial_i \cK = P^I_{\ ,i} \bar P_{  I,\bar \jmath}=g_{i\bar\jmath}
  \label{spec1} \\
  (d\omega +\omega\wedge\omega)_{\ \bar I}^{\bar J} &=& P_{\bar I}
  \wedge\bar P^{\bar J} -{\rm i} d\cQ
  \delta_{\bar I}^{\bar J}
-\bar P^{\bar J}_{\ L} \wedge  P^L_{\ \bar I}
\label{spec2}\\
\nabla P_{\ \bar I}^{J} &=& 0
\label{spec3}\\
\bar P_J \wedge P^J_{\ \bar I}=0 \label{spec4}
\end{eqnarray}
Equation (\ref{spec1}) implies that $\cM_{N=2}$ is a
K\"ahler--Hodge manifold. Equation (\ref{spec2}), written with
holomorphic and antiholomorphic curved indices, gives:
\begin{equation}
  \label{curvspec}
  R_{\bar\imath j \bar k l} = g_{\bar\imath l} g_{j \bar k} + g_{\bar k l}
  g_{\bar\imath j} - \bar C_{\bar\imath\bar k \bar m}
C_{jln} g^{\bar m n}
\end{equation}
which is the usual constraint on the Riemann tensor of the special
geometry. The further special geometry constraints on the three
tensor $C_{ijk}$  are then consequences of
 equations (\ref{spec3}), (\ref{spec4}), which imply:
\begin{eqnarray}
  \nabla_{[l} C_{i]jk} &=& 0 \nonumber\\
\nabla_{\bar l} C_{ijk}&=&0 \label{cijk}
\end{eqnarray}
In particular, the first of eq. (\ref{cijk}) also implies that $
C_{ijk} $ is a completely symmetric tensor.
\par
In summary, we have seen that the $N=2$ theory and the higher $N$
theories have essentially the same symplectic structure, the only
difference being that since the scalar manifold of $N=2$ is not in
general a coset manifold the symplectic structure allows the
presence of a new geometrical quantity which physically
corresponds to the anomalous magnetic moments of the $N=2$ theory.
It goes without saying that, when $\cM_{N=2}$ is itself a coset
manifold \cite{cp}, then the anomalous magnetic moments $C_{ijk}$
must be expressible in terms of the vielbein of ${\rm U}/H$. \par
To complete the analogy between the $N=2$ theory and the higher
$N$ theories in $D=4$,
 we also give for completeness the supersymmetry transformation laws, the central and matter
 charges, the differential
 relations among them and the sum rules.
\par
The transformation laws for the chiral gravitino $\psi_A$ and
gaugino $\lambda^{iA}$ fields are:
\begin{equation}
 \delta \psi_{A \mu}={ D}_{\mu}\,\epsilon _A\,+ \epsilon _{AB}
T_{\mu\nu} \gamma^\nu \epsilon^B + \cdots \label{trasfgrav}
 \end{equation}
 \begin{equation}
 \delta\lambda^{iA} = {\rm i} \partial_\mu z^i \gamma^\mu\epsilon^A +
 {\frac{\rm i} 2}
T_{\bar \jmath\mu\nu} \gamma^{\mu \nu}
g^{i\bar\jmath}\epsilon^{AB}\epsilon_B + \cdots
\label{gaugintrasfm}
\end{equation}
where:
\begin{equation}
T \equiv  h_\Lambda F ^{\Lambda}  - f^\Lambda \cG_\Lambda
\label{T-def}
\end{equation}
\begin{equation}
 T_{\bar \imath} \equiv
 \bar h_{\Lambda_{\bar \imath}} F^{\Lambda}  - \bar f^\Lambda_{\bar \imath}
  \cG_\Lambda
  \label{G-def}
\end{equation}
are respectively the graviphoton and the matter-vectors,
 $z^i$ ($i=1,\cdots,n$) are the complex scalar fields and the position of
 the $SU(2)$ automorphism index A (A,B=1,2) is related to chirality
 (namely $(\psi_A, \lambda^{iA})$ are chiral, $(\psi^A,
 \lambda^{\bar\imath}_A)$ antichiral).
 In principle only the (anti) self dual part of $ F$ and $\cG$ should
 appear in the transformation laws of the (anti)chiral fermi fields; however,
 exactly as in eqs. (\ref{tiden0}),(\ref{tiden}) for $N>2$ theories,
 from equations (\ref{geospec}) it follows that :
 \begin{equation}
T^+ = h_\Lambda  F^{+\Lambda} - f^\Lambda \cG_\Lambda ^+   =0
\end{equation}
so that $T=T^-$ (and $\bar T = \bar T^+$).
 Note that both the graviphoton and the matter vectors are symplectic
  invariant according to the fact that the fermions do not
 transform under the duality group (except for a possible R-symmetry
 phase).
To define the physical charges let us note that in presence of
electric and magnetic sources we can write:
\begin{equation}
\int_{S^2}  F^\Lambda = g^\Lambda , \quad \int_{S^2} \cG _ \Lambda
= e_ \Lambda .
\end{equation}
The central charges and the matter charges are now defined as the
integrals over a $S^2$ of the physical graviphoton and matter
vectors:
\begin{equation}
Z= \int_{S^2} T= \int_{S^2} ( h_\Lambda  F ^{\Lambda} - f^\Lambda
\cG_\Lambda ) = ( h_\Lambda (z,\bar z) g^{\Lambda} -
f^\Lambda(z,\bar z) e_\Lambda )
\end{equation}
where $z^i, \bar z^{\bar \imath}$ denote the v.e.v. of the moduli
fields in a given background. Owing to eq. (\ref{geospec}) we get
immediately:
\begin{equation}
Z_i= \nabla_i Z \label{fundeq}
\end{equation}
 As a consequence of the symplectic structure, one can derive
  two sum
 rules for  $Z$ and $Z_i$:
  \begin{equation}
  \vert Z \vert ^2 +  \vert Z_i \vert ^2 \equiv
  \vert Z \vert ^2 +   Z_i g^{i\bar \jmath} \bar Z_{\bar \jmath} =
  -\frac{1}{2} P^t \cM P
  \label{sumrules}
\end{equation}
  where:
  \begin{equation}
\cM = \left( \begin{matrix} \bfone & - Re \cN \cr 0
&\bfone\cr\end{matrix}\right) \left( \begin{matrix}Im \cN & 0 \cr
0 &Im \cN^{-1}\cr\end{matrix}\right) \left( \begin{matrix} \bfone
& 0 \cr - Re \cN & \bfone \cr\end{matrix}\right) \label{m+2}
\end{equation}
and:
\begin{equation}
P=\left(g^\Lambda, e_ \Lambda \right) \label{eg2}
\end{equation}
Equation (\ref{m+2}) is obtained by using exactly the same
procedure as in (\ref{m+}).
\section{$N=2$ BPS black holes: general discussion}
 \label{n2bpsh} Recently
\cite{feka1},\cite{feka2}, S. Ferrara and R. Kallosh gave a
general rule for finding the values of fixed scalars, and then the
Bekenstein--Hawking entropy, in $N=2$ theories through an {\it
extremum principle}. It states that the fixed scalars $\Phi_{fix}$
are the ones (among all the possible values taken by scalar
fields) that extremize the ADM mass of the black hole in moduli
space:
\begin{equation}
 \Phi_{fix} \, : \quad {\frac{\partial
M_{ADM}(\Phi)}{\partial \Phi_i}}\left\vert_{\Phi_{fix}}=0 \right.
\label{fixed}
\end{equation}
 Correspondingly, the
Bekenstein--Hawking entropy is given in terms of that extremum
among the possible ADM masses (given by all possible boundary
conditions that one can impose on scalars at spatial infinity),
identified with the Bertotti--Robinson mass $M_{BR}$:
 \begin{equation}
M_{BR}\equiv M_{ADM}(\Phi_{fix}) =\mbox{extr}
\{M_{ADM}(\Phi(\infty))\} \end{equation}
 The {\it extremum
principle} \eq{fixed} can be explained for the $N=2$ theory
\cite{feka1},\cite{feka2} by means of the special geometry
relations on the Killing spinor equations near the horizon.
\par
Indeed, the Killing spinor equations expressing the existence of
unbroken supersymmetries is obtained, for the gauginos in the
$N=2$ case \cite{n=2}, setting equal to zero the r.h.s. of
equation  \eqn{gaugintrasfm} that is, using flat indices:
\begin{equation}
\delta \lambda_{IA}=P_{I,i}\partial_\mu z^i \gamma^\mu
\varepsilon_{AB} \epsilon^B +T_{I\vert \mu\nu}\gamma^{\mu\nu}
\epsilon_A +\cdots =0. \label{kill}
\end{equation}
Approaching the black-hole horizon, the scalars $z^i$ reach their
fixed values so that \begin{equation}\partial_\mu z^i =0
\end{equation}and equation \eq{kill} is satisfied for
\begin{equation}T_I =0 \end{equation}that is, using integrated quantities:
 \begin{equation}Z_I = Z_i P_I^i =\int_{S^2_\infty} T_I = h_{\Lambda
I}(z(\infty))g^\Lambda -f^\Lambda_I(z(\infty)) e_\Lambda =
0.\label{mat}
\end{equation}That is, the Killing spinor equation imposes the
vanishing of the matter charges near the horizon. Now, eq.
\eq{mat} shows that the matter charges $Z_I$ are linear in the
scalar functions $f^\Lambda_I(z(\infty))$, $h_{\Lambda
I}(z(\infty))$ so that, remembering eq. \eq{fundeq}, we then have,
near the horizon:
\begin{equation}
Z_I= \nabla_I Z =0 \label{charges1}
\end{equation}
where $Z$ is the central charge appearing in the $N=2$
supersymmetry algebra, so that:
\begin{equation}\partial_i \vert Z \vert =0 \end{equation}which, for an
extremal black hole ($\vert Z \vert=M_{ADM}$), coincides with  eq.
\eq{fixed} giving the fixed scalars $\Phi_{fix}\equiv z_{fix}$.
We see that the entropy of the black--hole is related to the
central charge, namely to the integral of the graviphoton field
strength evaluated for very special values of the scalar fields
$z^i$. These special values, the {\it fixed scalars} $z^i_{fix}$,
are functions solely of the electric and magnetic charges $\{ q_
\Sigma,p^\Lambda\}$ of the black hole  and are attained by the
scalars $z^i(r)$ at the black-hole horizon $r=0$.
\par
Let us discuss the explicit solution of the Killing spinor
equation and the general properties of BPS saturated black--holes
in the context of $N=2$ supergravity. As our analysis will reveal,
these properties are completely encoded in the special K\"ahler
geometric structure of the mother theory. To illustrate in more
detail what happens, let us consider a black-hole
ansatz\footnote{This ansatz is dictated by the general p-brane
solution of supergravity bosonic equations in any dimensions
\cite{mbrastelle}.} for the metric:
\begin{equation}
ds^2 = e^{2U(r)} \, dt^2 - e^{-2U(r)} \, d{\vec x}^2 \quad ; \quad
\left( r^2 = {\vec x}^2 \right) \label{ds2U}
\end{equation}
and  for the vector field strengths:
\begin{equation}
F^{\Lambda}\,=\,\frac{p^{\Lambda}}{2r^3}\epsilon_{abc} x^a
dx^b\wedge dx^c-\frac{\ell^{\Lambda}(r)}{r^3}e^{2\cal {U}}dt
\wedge \vec{x}\cdot d\vec{x}. \label{flambda}
\end{equation}
 It is convenient to rephrase the same ansatz in the complex
formalism well-adapted to the $N=2$ theory. To this effect we
begin by constructing a 2--form which is {\it anti--self--dual} in
the background of the metric \eqn{ds2U} and whose integral on the
$2$--sphere at infinity $ S^2_\infty$ is normalized to $ 2 \pi $.
A short calculation yields:
\begin{eqnarray}
E^-   &=&   \mbox{i} \frac{e^{2U(r)}}{r^3} \, dt \wedge {\vec
x}\cdot d{\vec x}  +  \frac{1}{2}   \frac{x^a}{r^3} \, dx^b
\wedge
dx^c   \epsilon_{abc} \nonumber\\
2 \, \pi  &=&    \int_{S^2_\infty} \, E^- \label{eaself}
\end{eqnarray}
from which one obtains:
\begin{equation}
E^-_{\mu\nu} \, \gamma^{\mu\nu}  = 2 \,\mbox{i}
\frac{e^{2U(r)}}{r^3}  \, \gamma_a x^a \, \gamma_0 \,
\frac{1}{2}\left[ {\bf 1}+\gamma_5 \right] \label{econtr}
\end{equation}
which will prove of great help in the unfolding of the
supersymmetry transformation rules.
\par
Next, introducing the following complex combination of the
magnetic charge $ p^\Lambda $ and of the radial function
$\ell^\Sigma(r)$ defined by eq. \eqn{flambda}:
\begin{equation}
\label{tlamb} t^\Lambda(r)\,=\, 2\pi(p^\Lambda+{\rm i}\ell
^\Lambda (r))
\end{equation}
we can rewrite the ansatz \eqn{flambda} as:
\begin{eqnarray}
\label{strenghtsans}
F^{-\vert \Lambda}\,&=&\, \frac{t^\Lambda}{4\pi}E^-\nonumber\\
\end{eqnarray}
and we retrieve the original formulae from:
\begin{equation}
\begin{array}{rcccl}
F^{\Lambda}\,&=&\,2{\rm Re}F^{-\vert \Lambda}&= &
\frac{p^{\Lambda}}{2r^3}\epsilon_{abc} x^a dx^b\wedge
dx^c-\frac{\ell^{\Lambda}(r)}{r^3}e^{2\cal {U}}dt
\wedge \vec{x}\cdot d\vec{x} \\
\tilde{F}^{\Lambda}&=&-2{\rm Im}F^{-\vert \Lambda}& = &
-\frac{\ell^{\Lambda}(r)}{2r^3}\epsilon_{abc} x^a dx^b\wedge dx^c
-\frac{p^{\Lambda}}{r^3}e^{2\cal {U}}dt\wedge \vec{x}\cdot
d\vec{x}.
\end{array}
\label{fedfd}
\end{equation}
Before proceeding further it is convenient to define the electric
and magnetic charges of the black hole as it is appropriate in any
electromagnetic theory. Recalling the general form of the field
equations and of the Bianchi identities as given in \eqn{biafieq}
we see that the field strengths ${F}_{\mu \nu}$ and ${ G}_{\mu
\nu}$ are both closed 2-forms, since their duals are
divergenceless. Hence, for Gauss theorem, their integral on a
closed space--like $2$--sphere does not depend on the radius of
the sphere. These integrals are the electric and magnetic charges
of the hole that, in a quantum theory, we expect to be quantized.
We set:
\begin{eqnarray}
q_\Lambda & \equiv & \frac{1}{4 \pi} \, \int_{S^2_\infty}
\,G_{\Lambda\vert \mu\nu}\, dx^\mu \wedge dx^\nu
\label{holele}\\
p^\Sigma & \equiv & \frac{1}{4 \pi} \, \int_{S^2_\infty}
\,F^{\Sigma}_{ \mu\nu}\,\, dx^\mu \wedge dx^\nu \label{holmag}
\end{eqnarray}
If rather than the integral of $G_{\Lambda }$ we were to
calculate the integral of ${\tilde F}^\Lambda$, which is not a
closed form, we would obtain a function of the radius:
\begin{eqnarray}
4\pi \ell^\Lambda (r)\,&=&\,-\int_{S^2_r}\tilde{F}^{\Lambda}\,=\,2
{\rm Im}\, t^\Lambda .\label{cudierre}
\end{eqnarray}
Consider now the Killing spinor equations obtained from the
supersymmetry transformations rules \eqn{trasfgrav},
\eqn{gaugintrasfm}:
\begin{eqnarray}
0 &=& {\nabla}_{\mu}\,\xi _A\, + \epsilon_{AB} \, T^-_{\mu \nu}\,
\gamma^{\nu}\xi^B
 \label{kigrav} \\
0 &=& {\rm i}\,    \nabla _ {\mu}\, z^i  \, \gamma^{\mu} \xi^A
+G^{-i}_{\mu \nu} \gamma^{\mu \nu} \xi _B \epsilon^{AB}
\label{kigaug}
\end{eqnarray}
where the killing spinor $\xi _A(r)$  is of the form of a single
radial function times a constant spinor satisfying:
\begin{eqnarray}
\xi_A (r)\,&=&\,e^{f(r)} \chi _A~~~~~~~~~\chi_A=\mbox{constant}\nonumber\\
\gamma_0 \chi_A\,&=&\,\pm {\rm i}\epsilon_{AB}\chi^B
\label{quispi}
\end{eqnarray}
We observe that the condition \eq{quispi} halves the number of
supercharges preserved by the solution. Inserting eq.s
\eqn{T-def},\eqn{G-def},\eqn{quispi} into eq.s\eqn{kigrav},
\eqn{kigaug} and using  the result \eqn{econtr}, with a little
work we obtain the first order differential equations:
\begin{eqnarray}
\frac{dz^i}{dr}\, &=&\, \mp \left(\frac{e^{U(r)}}{4\pi r^2}\right)
g^{ij^\star}\bar{f}_{j^\star}^\Lambda
({\cal N}-\bar{{\cal N}})_{\Lambda\Sigma}t^\Sigma\,=\,\nonumber\\
&&\mp \left(\frac{e^{U(r)}}{4\pi r^2}\right) g^{ij^\star}
\nabla_{j^\star}\bar{Z}(z,\bar{z},{p},{q}) \label{zequa}\\
\frac{dU}{dr}\, &=&\,\mp \left(\frac{e^{U(r)}}{r^2}\right)
(M_\Sigma {p}^\Sigma- L^\Lambda {q}_\Lambda)\,=\, \mp
\left(\frac{e^{U(r)}}{r^2}\right)Z(z,\bar{z},{p},{q})
\label{Uequa}
\end{eqnarray}
where ${\cal N}_{\Lambda\Sigma}(z,\bar{z})$ is the kinetic matrix
of special geometry defined by eq.\eqn{nfh-1},
 the vector $V= \left(L^\Lambda(z,\bar{z}),
M_\Sigma(z,\bar{z})\right)$ is the covariantly holomorphic section
of the symplectic bundle entering the definition of a Special
K\"ahler manifold,
\begin{equation}
 Z(z,\bar{z},{p},{q}) \equiv \left( M_\Sigma {p}^\Sigma-
L^\Lambda {q}_\Lambda\right) \label{zentrum}
\end{equation}
is the local realization on the scalar manifold ${\cal SM}$ of the
central charge of the $N=2 $ superalgebra,
\begin{equation}
 Z^i(z,\bar{z},{p},{q}) \equiv g^{ij^\star}\nabla_
{j^\star}\bar{Z}(z,\bar{z},{p},{q}) \label{zmatta}
\end{equation}
are the central charges associated with the matter vectors, the
so--called matter central charges. \par
 To obtain eqs.
(\ref{zequa}),(\ref{Uequa}) we made use of  the properties :
\begin{eqnarray}
0\,&=&\, \bar{h}_{j^\star\vert\Lambda} t^{\star \Sigma}-
\bar{f}_{j^\star}^\Lambda{\cal N}_{\Lambda\Sigma}t^{\star \Sigma}\nonumber\\
0\,&=&\, M_\Sigma t^{\star \Sigma}-L^\Lambda {\cal
N}_{\Lambda\Sigma} t^{\star \Sigma}
\end{eqnarray}
which are a direct consequence of the definition \eqn{nfh-1} of
the kinetic matrix. The electric charges $\ell^\Lambda (r)$
defined in (\ref{cudierre}) are {\it moduli dependent} charges
which are functions of the radial direction through the moduli
$z^i$. On the other hand, the {\it moduli independent} electric
charges $q_\Lambda$ in eqs. (\ref{Uequa}),(\ref{zequa}) are those
defined by eq.\eqn{holele} which, together with $p^\Lambda$ fulfil
a Dirac quantization condition. Their definition allows them to be
expressed in terms of of $t^\Lambda (r)$ as follows:
\begin{equation}
q_\Lambda\,=\, \frac{1}{2\pi}{\rm Re}({\cal N}(z(r),\bar{z}(r))t
(r))_\Lambda \label{ncudierre}
\end{equation}
Equation (\ref{ncudierre}) may be inverted in order to find the
moduli dependence of $\ell_\Lambda (r)$. The independence of
$q_\Lambda$ on $r$ is a consequence of one of the Maxwell's
equations:
\begin{equation}
\partial_a \left(\sqrt{-g}\tilde{G}^{a0\vert \Lambda}(r)\right)\,=\,0\Rightarrow
\partial_r {\rm Re}({\cal N}(z(r),\bar{z}(r))t(r))^\Lambda\,=\,0
\label{prione}
\end{equation}
In this way we have reduced the  condition that the black-hole
should be a BPS saturated state to a pair of first order
differential equations for the metric scale factor $U(r)$ and for
the scalar fields $z^i(r)$. To obtain explicit solutions one
should specify the special K\"ahler manifold one is working with,
namely the specific Lagrangian model. There are, however, some
very general and interesting conclusions that can be drawn in a
model--independent way. They are just consequences of the fact
that these BPS conditions are {\it first order differential
equations}. Because of that there are fixed points (see the
papers \cite{ferkal2,feka1,strom3}) namely values either of the
metric or of the scalar fields which, once attained in the
evolution parameter $r$ (= the radial distance ) persist
indefinitely. The fixed point values are just the zeros of the
right hand side in either of the coupled eq.s \eqn{Uequa} and
\eqn{zequa}. The fixed point for the metric equation is $r=\infty
$, which corresponds to its asymptotic flatness. The fixed point
for the moduli is $r=0$.  So, independently from the initial data
at  $r=\infty$ that determine the details of the evolution,  the
scalar fields flow into their fixed point values at $r=0$, which,
as we will show, turns out to be a horizon. Indeed in the
vicinity of $r=0$ also the metric takes the universal form of an
$AdS_2 \, \times \, S^2$, Bertotti Robinson metric.
\par
Let us see this more closely.
\par
To begin with we consider the equations determining the fixed
point values for the moduli and the universal form attained by
the metric at the moduli fixed point:
\begin{eqnarray}
0 &=& -g^{ij^\star} \, {\bar f}^\Gamma_{j^\star} \left(
\mbox{Im}{\cal N}\right)_{\Gamma\Lambda} \, t^{\Lambda }(0)
\label{zequato} \\
\frac{dU}{dr}\, & \cong & \mp \left(\frac{e^{U(r)}}{r^2}\right)
Z\left (z_{fix},{\bar{z}}_{fix} ,{p},{q} \right) \label{Uequato}
\end{eqnarray}
Multiplying eq.\eqn{zequato} by $f^\Sigma_i$ using the identity
\eqn{metri} and the definition \eqn{zentrum} of the central charge
 we conclude that
at the fixed point the following condition is true:
\begin{equation}
0=-\frac{1}{2} \, \frac{t^\Lambda}{4\pi} \, -\, \frac{Z_{fix} \,
{\bar L}_{fix}^\Lambda}{8\pi} \label{passetto}
\end{equation}
In terms of the previously defined electric and magnetic charges
(see eq.s \eqn{holele},\eqn{holmag}, \eqn{ncudierre})
eq.\eqn{passetto} can be rewritten as:
\begin{eqnarray}
p^\Lambda & = & \mbox{i}\left( Z_{fix}\,{\bar L}^\Lambda_{fix}
- {\bar Z}_{fix}\,L^\Lambda_{fix} \right)\\
q_\Sigma & = & \mbox{i}\left( Z_{fix}\,{\bar M}_\Sigma^{fix}
- {\bar Z}_{fix}\,M_\Sigma^{fix} \right)\\
Z_{fix} &=& M_\Sigma^{fix} \, p^\Sigma \, - L^\Lambda_{fix} \,
q_\Lambda \label{minima}
\end{eqnarray}
which can be regarded as algebraic equations determining the value
of the scalar fields at the fixed point as functions of the
electric and magnetic charges $p^\Lambda, q_\Sigma$:
\begin{equation}
L_{fix}^\Lambda = L^\Lambda(p,q) \, \longrightarrow \,
Z_{fix}=Z(p,q)=\mbox{const} \label{ariacalda}
\end{equation}
\par
In the vicinity of the fixed point the differential equation for
the metric becomes:
\begin{equation}
\pm \, \frac{dU}{dr}=\frac{Z(p,q)}{  \, r^2} \, e^{U(r)}
\end{equation}
which has the approximate solution:
\begin{equation}
\exp[U(r)]\, {\stackrel{r \to 0}{\longrightarrow}}\, \mbox{const}
+ \frac{Z(p,q)}{  \, r} \label{approxima}
\end{equation}
Hence, near $r=0$   the metric \eqn{ds2U} becomes of the Bertotti
Robinson type (see eq.\eqn{br} ) with Bertotti Robinson mass given
by:
\begin{equation}
m_{BR}^2 = \vert  {Z(p,q)}  \vert^2 \label{brmass}
\end{equation}
In the metric \eqn{br} the surface $r=0$ is light--like and
corresponds to a horizon since it is the locus where the Killing
vector generating time translations $\frac{\partial}{\partial t}
$, which is time--like at spatial infinity $r=\infty$, becomes
light--like. The horizon $r=0$ has a finite area given by:
\begin{equation}
\mbox{Area}_H = \int_{r=0} \,
\sqrt{g_{\theta\theta}\,g_{\phi\phi}} \,d\theta \,d\phi \, = \,
4\pi \, m_{BR}^2 \label{horiz}
\end{equation}
Hence, independently from the details of the considered model,
the BPS saturated black--holes in an N=2 theory have a
Bekenstein--Hawking entropy given by the following horizon area:
\begin{equation}
 \frac{\mbox{Area}_H}{4\pi} = \,     \vert Z(p,q) \vert^2
 \label{ariafresca}
\end{equation}
the value of the central charge being determined by eq.s
\eqn{minima}. Such equations can also be seen as the variational
equations for the minimization of the horizon area as given by
\eqn{ariafresca}, if the central charge is regarded as a function
of both the scalar fields and the charges:
\begin{eqnarray}
 \mbox{Area}_H (z,{\bar z})&=& \,  4\pi \, \vert Z(z,{\bar z},p,q) \vert^2
 \nonumber\\
 \frac{\delta \mbox{Area}_H }{\delta z}&=&0 \, \longrightarrow \,  z = z_{fix}
\end{eqnarray}
\subsection{Extension to the $N>2$ case}
 Let us now
observe that the extremum principle just described, although
shown to be true for $N=2$ four dimensional black holes, has
however a more general validity, being true for all $N$-extended
supergravities in four dimensions (where the Bekenstein--Hawking
entropy for black holes is in general
different from zero) \cite{noi}.\\
Indeed, the general discussion of section \ref{sugras} has shown
that the coset structure of extended supergravities in four
dimensions (with $N>2$) induces the existence, in every theory, of
differential relations among central and matter charges that
generalize eq. \eq{charges1} given in equation \eq{dz1}.
Furthermore, the Killing spinor equations for gauginos and
dilatinos, analogous to eq. \eq{dz1}, are obtained by setting
equal to zero the r.h.s of equations \eq{tradil}, \eq{tragau}.
Correspondingly, at the fixed point $\partial_\mu \Phi^i=0$ one
gets again some conditions that allow to find the value of fixed
scalars and hence of the Bekenstein--Hawking entropy.
\\
The first condition, from the gaugino transformation law, is as
before:
\begin{equation}T_I=0 \to Z_I=0 \end{equation}
Moreover there is a further condition, from the dilatino equation:
\begin{equation}T_{[AB}\epsilon_{C]}=0 \to Z_{[AB}\epsilon_{C]}=0
\label{dil} \end{equation} Inserting these relations in eq.
\eq{dz1} one has that fixed scalars are found by solving:
\begin{eqnarray}
 Z_{[AB}\epsilon_{C]}&=&0 \label{zabe}\\
  Z_I &=& 0 \Rightarrow
  \nabla Z_{AB} = \frac 12 P_{ABCD}\bar Z^{CD} \label{zabc}
\end{eqnarray}
From a case by case analysis of equations \eq{zabe}, \eq{zabc}
the explicit form of fixed scalars and then of the entropy is
easily obtained for each theory.
\par
Let us now look first at equation \eq{zabe}. We work in the normal
frame, where the central charge matrix $Z_{AB}$ is written in
terms of its skew diagonal eigenvalues (we analyze the $N=8$
case, that includes all lower $N$ theories):
\begin{equation}Z_{AB}^{(N)}  = \left(\begin{matrix}Z_1 \epsilon &0&0&0 \cr
                         0&Z_2 \epsilon &0&0 \cr
                         0&0&Z_3 \epsilon &0 \cr
                         0&0&0&Z_4 \epsilon  \cr\end{matrix}\right)
\, ; \quad   Z_i \epsilon  = \left(\begin{matrix}0 & Z_i \cr -
Z_i & 0\end{matrix}\right)
\end{equation}If only two Killing spinors, say $\epsilon_1,
\epsilon_2$ are different from zero, then eq. \eq{zabe} implies
that the three central charge eigenvalues $Z_2, Z_3, Z_4$ must be
zero, the only non vanishing eigenvalue being $Z_1$, and we are
left with an $N=2$ unbroken theory. On the other hand, if one more
Killing spinor, say $\epsilon_3$, is different from zero, then
from eq. \eq{dil} we get that all the central charge eigenvalues
are zero, so that this becomes the same Minkowski vacuum
background that describes spatial infinity. Let us then consider
the case $Z_{12} \neq 0$, $Z_{AB}=0$ for $A,B \neq 1,2$. Equation
\eq{zabc} now reduces to $\nabla Z_{12}=0$, which gives the fixed
scalars as an extremum of  $(Z_{AB}\bar Z^{AB})^{\frac 12} \equiv
|Z_{12}|$.
\subsection{The geodesic potential} The results of the previous
section can be retrieved in an alternative way, that has the
advantage of being covariant, not referring explicitly
to the horizon properties for finding the entropy \cite{gkk},\cite{fegika}.\\
Let us consider the field equations for the metric components
$e^U$ (see eq. \eq{ds2U}) and for the scalar fields $\Phi^i$,
written in terms of the evolution parameter $\tau = \frac{1}{
\rho}= \frac{1}{ r-r_H}$ \footnote{ The equations in \eq{geoeq}
are valid for extremal black holes. For non extremal ones similar
relations hold, where however in the third eq. in \eq{geoeq}
there is one further contribution proportional to the surface
gravity $\kappa$ (that is to the black-hole temperature, which is
zero only for extremal configurations).}:
\begin{equation}\left\{\begin{matrix}\frac{d^2 U}{d\tau^2}  =  2
V(\Phi,e,g)e^{2U} \cr \frac{D^2 \Phi_i}{D\tau^2}  = \half
\frac{\partial V(\Phi,e,g)}{\partial \Phi^i} e^{2U}  \cr
\left(\frac{d U}{d\tau}\right)^2 + G_{ij}\frac{d
\Phi^i}{d\tau}\frac{d \Phi^j}{d\tau} - V(\Phi,e,g)e^{2U}=0
\cr\end{matrix} \right. \label{geoeq}
\end{equation}Here $G_{ij}$ is the metric of the sigma-model
described by scalars while $V(\Phi,e,g)$ is a function of scalars
and of the electric and magnetic charges of the theory defined by:
\begin{equation}
V=-\frac{1}{2}P^t\cM(\cN)P \label{geopot}
\end{equation}
where $P$ is the symplectic vector $P=(g^\Lambda, e_\Lambda) $ of
quantized electric and magnetic charges and $\cM(\cN)$ is a
symplectic $2n_V \times 2n_V$ matrix,
 whose $n_V\times n_V$ blocks are given in terms of an $n_V\times n_V$
 matrix $\cN_{\Lambda\Sigma}(\Phi)$
\begin{equation}
\cM(\cN) = \left( \begin{matrix} \bfone & - Re \cN \cr 0
&\bfone\cr\end{matrix}\right) \left( \begin{matrix} Im \cN & 0 \cr
0 &Im \cN^{-1}\cr\end{matrix}\right) \left( \begin{matrix} \bfone
& 0 \cr - Re \cN & \bfone \cr\end{matrix}\right) \label{mn}
\end{equation}
 The real and imaginary components of the
matrix $\cN$ appear in the vector kinetic terms of the
supergravity lagrangian describing the black hole:
\begin{equation}\cL=- \frac{1}{2} R + \frac{1}{
2}G_{ij}\partial_\mu \Phi^i\partial^mu \Phi^j - \frac{1}{4} Re
\cN_{\Lambda\Sigma}F_{\mu\nu}^\Lambda F^{\mu\nu |\Sigma} +
\frac{1 }{4} Im \cN_{\Lambda\Sigma}F_{\mu\nu}^\Lambda F^{*\mu\nu
|\Sigma} + \mbox{fermions} \label{lagrbh} \end{equation} Let us
note that the field equations \eq{geoeq} can be extracted from the
effective 1--dimensional lagrangian:
\begin{equation}\cL_{eff}= \left(\frac{d U}{d\tau}\right)^2 +
G_{ij}\frac{d \Phi^i}{d\tau}\frac{d \Phi^j}{d\tau} +
V(\Phi,e,g)e^{2U}. \label{effect} \end{equation}From equation
\eq{effect} we see that the properties of extreme black holes are
completely encoded in the metric of the scalar manifold $G_{ij}$
and on the scalar effective potential $V$, known as {\it geodesic
potential} \cite{gkk},\cite{fegika}. In particular it was shown in
\cite{gkk},\cite{fegika} that the area of the event horizon is
proportional to the value of $V$ at the horizon:
\begin{equation}\frac{A}{ 4\pi}=V(\Phi_h,e,g) \label{mingeo} \end{equation}
where $\Phi_h$ denotes the value taken by scalar fields at the
horizon. To see this, let us consider the set of equations
\eq{geoeq}: it is possible to show that the field equations for
the scalars give, near the horizon, the solution:
\begin{equation}\Phi^i = \left(\frac{2\pi}{ A}\right)\frac{\partial V
}{ \partial \Phi^i}
        \mbox{log}\tau + \Phi^i_h .
        \label{phih}
\end{equation}From eq. \eq{phih} we see that the request that the horizon is
a fixed point ($\frac{d\Phi^i}{d \tau}=0$) implies that the
geodesic potential is extremized in moduli space:
\begin{equation}\Phi_h \, : \quad \frac{d\Phi^i}{ d \tau}=0 \quad
\leftrightarrow \quad \frac{\partial V}{\partial \Phi^i}|_{\Phi_h}
=0    . \label{min} \end{equation}Furthermore, let us consider the
third of \eq{geoeq}. Near the horizon, introducing the results
\eq{min}, it becomes:
\begin{equation}\left(\frac{dU}{d\tau}\right)^2 \sim
V(\Phi_h(e,g),e,g)e^{2U} \end{equation}from which it follows, for
the metric components near the horizon: \begin{equation}e^{2U}
\sim \frac{1}{\tau^2 V(\Phi_h)} = \frac{\rho^2}{V(\Phi_h)}  ,
\end{equation}that is: \begin{equation}ds^2_{hor} =
\frac{\rho^2}{V(\Phi_h)} dt^2 - \frac{V(\Phi_h)}{\rho^2}\left(
d\rho^2+\rho^2 d\Omega \right). \label{hor} \end{equation}By
comparing eq. \eq{hor} with eq. \eq{br} we see that
$V(\Phi_h)=M_{BR}^2$ and therefore, remembering  \eq{bert}, we get
the result \eq{mingeo}.

To summarize, we have just found that the area of the event
horizon (and hence the Bekenstein--Hawking entropy of the black
hole) is given by the geodesic potential evaluated at the horizon,
and we also gave a tool for finding this value: the geodesic
potential gets an extremum at the horizon.
\par
However, the geodesic potential $V(\Phi, e,g)$ defined in eq.s
\eq{geopot} and \eq{mn} has a particular meaning in supergravity
theories, that allows to find its extremum in an easy way. Indeed,
an expression exactly coinciding with \eq{geopot} has been found
in section 3 in an apparently different context, as the result of
a sum rule among central and matter charges in supergravity
theories \eq{m+}. So, in every supergravity theory, the geodesic
potential has the general form: \begin{equation}V\equiv -\frac{1}{
2}P^t\cM(\cN)P =\frac{1}{2} Z_{AB}\bar Z^{AB}+Z_I \bar Z^I
\end{equation}Then, to find the extremum of $V$ we can apply the
differential relations among central and matter charges found in
Section 3, that in general read:
\begin{eqnarray}
  \nabla Z_{AB} &=&  \bar Z_{I}  P^I_{AB} +  \frac{1}{2} \bar Z^{CD} P_{ABCD} \nonumber \\
\nabla Z_{I} &=& \frac{1}{2}  \bar Z^{AB}  P_{AB I} + \bar Z^{J}
P_{JI}
 \label{diffcharge1}
\end{eqnarray}
where the matrices $P_{ABCD}$, $P_{AB I}$, $P_{IJ}$ are the
subblocks of the
 vielbein of  the scalar manifolds $U/H$ \cite{noi}, already defined
 in Section 3:
\begin{equation}
  \label{vielbein}
  \cP \equiv L^{-1} \nabla L = \left(\begin{matrix}P_{ABCD}& P_{AB I} \cr
P_{I AB} & P_{IJ} \cr \end{matrix}\right)
\end{equation}
written in terms of the indices of $H=H_{Aut} \times H_{matter}$.

Applying eq.s \eq{diffcharge1} to the geodesic potential, we find
that the extremum is given by: \ba
dV &=& \frac{1}{2}\nabla Z_{AB}\bar Z^{AB}+\nabla Z_I \bar Z^I  +c.c. =\nn\\
   &=& \frac{1}{2}\left(\frac{1}{2} \bar Z^{CD} P_{ABCD}+
        \bar Z_{I}  P^I_{AB} \right)\bar Z^{AB}  \nn\\
   &+& \left(\frac{1}{2}  \bar Z^{AB}  P_{AB I} + \bar Z^{J}  P_{JI}\right)
        \bar Z^I + c.c. =0
\ea that is $dV=0$ for: \begin{equation}Z_I=0 \, ; \, \bar Z^{AB}
\bar Z^{CD} P_{ABCD} =0 \label{extgeo} \end{equation}Let us note
that the conditions \eq{extgeo}, defining the extremum of the
geodesic potential and so the fixed scalars, have the same
content, and are therefore completely equivalent, to the former
relations found in the previous subsection from the Killing spinor
conditions. However, with this latter procedure it is not
necessary to specify explicitly horizon parameters (like the
metric and the fixed values of scalars at that point), $V$ being a
well defined quantity over all the space--time.
\par
As an exemplification of the method, let us analyze in detail the
$D=4$, $N=4$ pure supergravity. The field content is given by the
gravitational multiplet, that is by the graviton $g_{\mu\nu}$,
four gravitini $\psi_{\mu A}$, $A=1,\cdots ,4 \in SU(4)$, six
vectors $A_\mu^{[AB]}$, four dilatini $\chi^{[ABC]}$ and a
complex scalar $\phi =a+{\rm i}e^\varphi$ parametrizing the coset
manifold $U/H=SU(1,1)/U(1)$. The symplectic $Sp(12)$--sections
$(f^\Lambda_{AB}, h_{\Lambda AB})$ ($\Lambda \equiv[AB]=1,\cdots
,6$) over the scalar manifold are given by: \ba
f^\Lambda_{AB}&=& e^{-\varphi /2} \delta^\Lambda_{AB} \nn\\
h_{\Lambda AB}&=& \phi e^{-\varphi /2} \delta_{\Lambda AB} \ea so
that: \begin{equation}\cN_{\Lambda\Sigma}=(h\cdot
f^{-1})_{\Lambda\Sigma}=\phi \delta_{\Lambda\Sigma}
\end{equation}The central charge matrix is then given by:
\begin{equation}Z_{AB}=h_{\Lambda AB}g^\Lambda -
f^\Lambda_{AB}e_\Lambda =
        e^{-\varphi /2} (\phi g_{AB}-e_{AB})
\end{equation}where: \begin{equation}g^\Lambda =\int F^\Lambda \equiv \int
dA^\Lambda \,\quad e_\Lambda =\int \cN_{\Lambda\Sigma} F^\Sigma
\end{equation}The geodesic potential is therefore: \bea V(\phi ,
e,g,)&=&\half e^{-\varphi} (\phi g_{AB}-e_{AB})
                (\bar\phi g^{AB}-e^{AB}) \nn\\
              &=&\half (a^2 e^{-\varphi}+e^{\varphi} ) g_{AB}g^{AB}
                  + e^{-\varphi} e_{AB}e^{AB}-2ae^{-\varphi}e_{AB}g^{AB}\nn\\
              &\equiv &\half  (g,e)\left(\begin{matrix}1&0 \cr -a &
              1\cr\end{matrix}\right)
                     \left(\begin{matrix}e^\varphi & 0 \cr 0 & e^{-\varphi} \cr\end{matrix}\right)
                     \left(\begin{matrix}1& -a \cr 0 & 1\cr\end{matrix}\right)
                     \left(\begin{matrix}g \cr e\end{matrix}\right)
\eea By extremizing the potential in the moduli space we get: \bea
\frac{\del V}{\del a}=0 & \to & a_{h}=\frac{e_{AB}g^{AB}}{g_{AB}g^{AB}}\nn\\
\frac{\del V}{\del \varphi}=0 & \to & e^{\varphi_{h}} =
\frac{\sqrt{e_{AB}e^{AB}g_{CD}g^{CD}-
(e_{AB}g^{AB})^2}}{g_{AB}g^{AB}} \eea from which it follows that
the entropy is: \begin{equation}S_{B-H}=4\pi V(\phi_h, e,g)=4\pi
\sqrt{e_{AB}e^{AB}g_{CD}g^{CD}- (e_{AB}g^{AB})^2}
\end{equation}\vskip 5mm As a final observation, let us note,
following \cite{fegika}, that the extremum reached by the geodesic
potential at the horizon is in particular a minimum, unless the
metric of the scalar fields change sign, corresponding to some
sort of phase transitions, where the effective lagrangian
description \eq{lagrbh} of the theory breaks down. This can be
seen from the properties of the Hessian of the geodesic potential.
It was shown in \cite{fegika} for the $N=2$, $D=4$ case that at
the critical point $\Phi=\Phi_{fix}\equiv \Phi_h$, from the
special geometry properties it follows: \begin{equation}\left(\bar
\partial_{\bar\imath}
\partial_j |Z| \right)_{fix}= \frac{1}{2}G_{\bar \imath j}|Z|_{fix}
\end{equation}and then, remembering, from the above discussion, that
$V_{fix}=|Z_{fix}|^2$: \begin{equation}\left(\bar
\partial_{\bar\imath}
\partial_j V \right)_{fix}= \frac{1}{2}G_{\bar \imath j}V_{fix}
\label{wein} \end{equation}From eq. \eq{wein} it follows, for the
$N=2$ theory, that the minimum is unique. In the next section we
will see that a result similar to \eq{wein} still hold for higher
$N$ theories, but that in general the Hessian of $V$ has some
degenerate directions.
 \\
 Moreover, in the next subsection we will show one more technique for finding
 the entropy, exploiting the fact that it is a `topological quantity' not
 depending on scalars. This last procedure is particularly interesting
 because it refers only to group theoretical properties of the coset manifolds
 spanned by scalars, and do not need the knowledge of any details of the
 black-hole horizon.

\subsection{Central charges, U-invariants and entropy}
Extremal   black-holes   preserving   one supersymmetry
correspond  to $N$-extended multiplets with
\begin{equation}
M_{ADM} = \vert Z_1 \vert >  \vert Z_2 \vert  \cdots > \vert
Z_{[N/2]} \vert
         \end{equation}
         where $Z_\alpha$, $\alpha =1,\cdots, [N/2]$, are the proper values of
 the
         central charge antisymmetric matrix written in normal form
 \cite{fesaz}.
The central charges $Z_{AB}= -Z_{BA}$, $A,B=1,\cdots,N$, and
matter charges
 $Z_I$, $I= 1,\cdots , n$ are
those (moduli-dependent) symplectic invariant combinations of
field strenghts and their duals (integrated over a large
two-sphere)
 which appear
in the gravitino and gaugino supersymmetry variations respectively
\cite{CDFp},\cite{noi},\cite{noi1}. Note that the total number of
vector fields is $n_V=N(N-1)/2+n$ (with the
 exception of $N=6$
in which case there is an extra singlet graviphoton)\cite{cj}.
         \\
 As we discussed in the previous section, at the attractor point, where
  $M_{ADM}$ is extremized, supersymmetry requires that $Z_\alpha$, $\alpha >1$,
  vanish toghether with the matter charges $Z_I$, $I= 1, \cdots , n$
($n$ is the number of matter multiplets, which can exist only for
$N=3,4$)
\par
This result can be used to show that for ``fixed scalars'',
corresponding to the attractor point, the scalar ``potential'' of
the geodesic
 action \cite{bmgk},\cite{gkk},\cite{fegika}
\begin{equation}
V=-\frac{1}{2}P^t\cM(\cN)P \label{sumrule1}
\end{equation}
is extremized in moduli space. The main purpose of this subsection
is to provide  particular expressions which
 give the
entropy formula as a moduli--independent quantity in the entire
moduli space and not just at the critical points. Namely, we are
looking for quantities $S\left(Z_{AB}(\phi), \bar Z^{AB}
 (\phi),Z_{I}(\phi), \bar Z^{I} (\phi)\right)$
such that $\frac{\partial}{\partial \phi ^i} S =0$, $\phi ^i$
being the moduli
 coordinates
 \footnote{The Bekenstein-Hawking entropy $S_{BH} =\frac{A}{4}$ is actually
$\pi S$ in our notation.}. To this aim, let us first consider
invariants $I_\alpha$ of the isotropy group $H$ of the scalar
manifold $U/H$, built with the central and matter charges. We will
take all possible $H$-invariants up to quartic ones for four
dimensional theories (except for the $N=3$ case, where the
invariants of order higher than quadratic are not irreducible).
Then, let us consider a linear combination $S^2=\sum_\alpha
C_\alpha I_\alpha$ of the $H$-invariants, with arbitrary
coefficients $C_\alpha$. Now, let us extremize $S$ in the moduli
space $\frac{\del S}{\del \Phi^i}=0$, for some set of
$\{C_\alpha\}$. Since $\Phi^i \in U/H$, the quantity found in this
way (which in all cases turns out to be unique) is a
$U$-invariant, and is indeed proportional to the
Bekenstein--Hawking entropy.

 These formulae generalize the quartic $E_{7(-7)}$ invariant of $N=8$
supergravity \cite{kako} to all other cases. We will show in the
appendix how these invariants can be computed in an almost trivial
fashion by using the (non compact) Cartan elements of $U/H$.
\footnote{Our analysis is based on general properties of scalar
coset manifolds. As a consequence, it can be applied
straightforwardly also to the $N=2$ cases, whenever one considers
special coset manifolds.}
\par
Let us first consider the theories $N=3,4$, where  matter can be
present \cite{ccdffm},\cite{bks}.
\par
The U--duality groups \footnote{Here we denote by U-duality group
the isometry
 group $U$
acting on the scalars, although only a restriction of it to
integers is the
 proper U-duality group \cite{huto}.}
 are, in these cases, $SU(3,n)$ and $SU(1,1)
\times SO(6,n)$ respectively. The central and matter charges
$Z_{AB}, Z_I$ transform in an obvious way under the isotropy
groups
\begin{eqnarray}
H&=& SU(3) \times SU(n) \times U(1) \qquad (N=3) \\
 H&=& SU(4) \times O(n) \times U(1) \qquad (N=4)
\end{eqnarray}
Under the action of the elements of $U/H$ the charges get mixed
with their complex conjugate. The infinitesimal transformation
 can be read from the differential relations satisfied by the charges
 \eq{diffcharge1} \cite{noi} .
\par
For $N=3$:
\begin{equation}
P^{ABCD}=P_{IJ}=0 , \quad P_{ AB I} \equiv \epsilon_{ABC}P^C_I
\quad Z_{AB}
 \equiv \epsilon_{ABC}Z^C
 \label{viel3}
\end{equation}
Then  the variations are:
\begin{eqnarray}
\delta Z^A  &=& \xi^A _I \bar Z^I  \\
\delta Z_{I}  &=&\xi^A _{ I} \bar Z_A \label{deltaz3}
\end{eqnarray}
where $\xi^A_I$ are infinitesimal parameters of $K=U/H$.
\par
The possible quadratic $H$-invariants are: \ba
I_1 &=& Z^A\bar Z_A \nn\\
I_2 &=& Z_I \bar Z^I \ea So, the U-invariant expression is:
\begin{equation}
S =   Z^A \bar Z_A - Z_I \bar Z^I \label{invar3}
\end{equation}
In other words, $\nabla_i S = \partial_i S =0 $, where the
covariant derivative is defined in ref. \cite{noi}.
\par
Note that at the attractor point ($Z_I =0$) it coincides with the
moduli-dependent potential (\ref{sumrule1}) computed at its
extremum.
\par
For $N=4$
\begin{equation}
P_{ABCD} = \epsilon_{ABCD}P ,\quad P_{IJ} = \eta_{IJ}P,\quad P_{AB
I}=\frac{1}{2} \eta_{IJ} \epsilon_{ABCD}\bar P^{CD J}
\label{viel4}
\end{equation}
and the transformations of $K= \frac{SU(1,1)}{U(1)} \times
\frac{O(6,n)}{O(6) \times O(n)}$ are:
 \begin{eqnarray}
\delta Z_{AB}  &=& \frac{1}{2} \xi  \epsilon_{ABCD} \bar Z^{CD}  +
 \xi _{AB I} \bar Z^I \\
\delta Z_{I}  &=& \xi \eta_{IJ} \bar Z^J + \frac{1}{2}\xi _{AB I}
\bar Z^{AB} \label{deltaz4}
\end{eqnarray}
with $\bar \xi^{AB I} =  \frac{1}{2} \eta^{IJ}   \epsilon^{ABCD}
\xi_{CD J}$.
\par
The possible $H$-invariants are: \ba
I_1 &=& Z_{AB} \bar Z^{AB}\nn\\
I_2 &=& Z_{AB} \bar Z^{BC} Z_{CD} \bar Z^{DA} \nn\\
I_3 &=& \epsilon^{ABCD}  Z_{AB}   Z_{CD} \nn\\
I_4 &=& Z_I Z^I \ea There are three $O(6,n)$ invariants given by
$S_1$, $S_2$, $\bar S_2$ where:
\begin{eqnarray}
  S_1 &=& \frac{1}{2}  Z_{AB} \bar Z^{AB} - Z_I  Z_I
\label{invar41} \\
S_2 &=& \frac{1}{4} \epsilon^{ABCD}  Z_{AB}   Z_{CD} - \bar Z_I
Z_I \label{invar42}
\end{eqnarray}
and the unique   $SU(1,1)  \times O(6,n) $  invariant $S$, $\nabla
S =0$, is given by:
\begin{equation}
S= \sqrt{(S_1)^2 - \vert S_2 \vert ^2 } \label{invar4}
\end{equation}
At the attractor point $Z_I =0 $ and $\epsilon^{ABCD} Z_{AB}
Z_{CD}
=0$ so that $S$ reduces to the square of the BPS mass.\\
Note that, in absence of matter multiplets, one recovers the
expression found in the previous subsecion by extremizing the
geodesic potential.
\par
For $N=5,6,8$ the U-duality invariant expression $S$ is the square
root of a unique invariant under the corresponding U-duality
groups $SU(5,1)$, $O^*(12)$ and $E_{7(-7)}$. The strategy is to
find a quartic expression $S^2$    in terms of $Z_{AB}$ such that
$\nabla S=0$, i.e. $S$ is moduli-independent.
\par
As before, this quantity is a particular combination of the $H$
quartic invariants.
\par
For $SU(5,1)$ there are only two  $U(5)$ quartic invariants. In
terms of the matrix $A_A^{\ B} = Z_{AC} \bar Z^{CB}$ they are:
$(Tr A)^2$, $Tr(A^2)$, where
\begin{eqnarray}
 Tr A & = & Z_{AB} \bar Z^{BA} \\
 Tr (A^2) & = & Z_{AB} \bar Z^{BC} Z_{CD} \bar Z^{DA}
\end{eqnarray}
As before, the relative coefficient is fixed by the transformation
properties of $Z_{AB}$ under $\frac{SU(5,1)}{U(5) } $ elements of
infinitesimal parameter $\xi^C$:
\begin{eqnarray}
  \delta Z_{AB} = \frac{1}{2} \xi^C \epsilon_{CABPQ} \bar Z^{PQ}
\end{eqnarray}
It then follows that the required invariant is:
\begin{equation}
S= \frac{1}{2} \sqrt{  4 Tr(A^2) - (Tr A)^2 } \label{invar5}
\end{equation}

For $N=8$ the $SU(8)$ invariants are \footnote{The Pfaffian of an
 $(n\times n)$ ($n$ even) antisymmetric
matrix is defined as $Pf Z=\frac{1}{ 2^n n!} \epsilon^{A_1 \cdots
A_n}
 Z_{A_1A_2}\cdots Z_{A_{N-1}A_N}$, with the property:
$ \vert Pf Z \vert = \vert det Z \vert ^{1/2}$.}:
\begin{eqnarray}
I_1 &=& (Tr A) ^2 \\
I_2 &=& Tr (A^2) \\
I_3 &=& Pf \, Z
 =\frac{1}{2^4 4!} \epsilon^{ABCDEFGH} Z_{AB} Z_{CD} Z_{EF} Z_{GH}
\end{eqnarray}
The $\frac{E_{7(-7)}}{ SU(8)}$ transformations are:
\begin{equation}
\delta Z_{AB} =\frac{1}{2} \xi_{ABCD} \bar Z^{CD}
\end{equation}
where $\xi_{ABCD}$ satisfies the reality constraint:
\begin{equation}
\xi_{ABCD} = \frac{1}{24} \epsilon_{ABCDEFGH} \bar \xi^{EFGH}
\end{equation}
One finds the following $E_{7(-7)}$ invariant \cite{kako}:
\begin{equation}
S= \frac{1}{2} \sqrt{4 Tr (A^2) - ( Tr A)^2 + 32 Re (Pf \, Z) }
\end{equation}

The $N=6$ case is the more complicated because under $U(6)$ the
left-handed spinor of $O^*(12)$ splits into:
\begin{equation}
32_L \to (15,1) + ( \bar {15}, -1) + (1, -3) + (1,3)
\end{equation}
The transformations of $\frac{O^*(12)}{U(6)}$ are:
\begin{eqnarray}
\delta Z_{AB} &=& \frac{1}{4} \epsilon_{ABCDEF} \xi^{CD} \bar
Z^{EF} +
\xi_{AB} \bar X \\
\delta X &=& \frac{1}{2} \xi _{AB} \bar Z^{AB}
\end{eqnarray}
 where we denote by $X$ the $SU(6)$ singlet.

The quartic $U(6)$ invariants are:
\begin{eqnarray}
I_1&=& (Tr A)^2 \label{invar61}\\
I_2&=& Tr(A^2)\label{invar62} \\
I_3 &=& Re (Pf \, ZX)= \frac{1}{2^3 3!}
Re( \epsilon^{ABCDEF}Z_{AB}Z_{CD}Z_{EF}X)\label{invar63}\\
I_4 &=& (Tr A) X \bar X\label{invar64}\\
I_5&=& X^2 \bar X^2\label{invar65}
\end{eqnarray}

The unique $O^*(12)$ invariant is:
\begin{eqnarray}
S&=&\frac{1}{2} \sqrt{4 I_2 - I_1 + 32 I_3 +4I_4 + 4 I_5 }
\label{invar6}   \\
\nabla S &=& 0
\end{eqnarray}
Note that at the attractor point $Pf\,Z =0$, $X=0$ and $S$ reduces
to the square of the BPS mass.
\subsubsection{ A simple determination of the U-invariants }
 In order to determine the quartic U-invariant expressions $S^2$ , $\nabla S
 =0$, of the $N>4$ theories, it is useful to use, as a calculational tool,
transformations
 of the coset which preserve the normal form of the $Z_{AB}$ matrix.
 It turns out that these transformations are certain Cartan elements
 in $K=U/H$ \cite{solv1}, that is they belong to $O(1,1)^p\in K$, with $p=1$
for $N=5$,
 $p=3$ for $N=6,8$.
 \par
 These elements act only on the $Z_{AB}$ (in normal form), but they
 uniquely determine the U-invariants since they mix the eigenvalues
 $e_i$ ($i=1,\cdots,[N/2]$).
 \par
 For $N=5$, $SU(5,1)/U(5)$ has rank one (see ref. \cite{hel}) and
 the element is:
 \begin{equation}
\delta e_1 = \xi e_2 ; \quad \delta e_2 = \xi e_1
\end{equation}
which is indeed a $O(1,1)$ transformation with unique invariant
\begin{equation}
\vert (e_1)^2 - (e_2) ^2\vert = \frac{1}{2} \sqrt{ 8 \left((e_1)^4
+ (e_2)^4 \right) - 4 \left( (e_1)^2 + (e_2)^2 \right)^2 }
\end{equation}

For $N=6$, we have $\xi_1 \equiv \xi_{12}; \xi_2 \equiv \xi_{34};
\xi_3 \equiv \xi_{56}$ and we obtain the 3 Cartan elements of
$O^*(12) /U(6)$, which has rank 3, that is it is a $O(1,1)^3$ in
$O^* (12)/U(6)$. Denoting by $e$ the singlet charge, we have the
following $O(1,1)^3$ transformations:
\begin{eqnarray}
\delta e_1 &=& \xi_2 e_3  + \xi _3 e_2   + \xi _1 e \label{trans61}\\
 \delta e_2 &=& \xi_1 e_3  + \xi _3 e_1   + \xi _2 e\label{trans62} \\
 \delta e_3 &=& \xi_1 e_2  + \xi _2 e_1   + \xi _3 e\label{trans63} \\
 \delta e &=& \xi_1 e_1  + \xi _2 e_2   + \xi _3 e_3\label{trans64}
\end{eqnarray}
these transformations fix uniquely the $O^*(12)$ invariant
constructed out of the five $U(6)$ invariants displayed in
(\ref{invar61}-\ref{invar65}).

For $N=8$ the infinitesimal parameter is $\xi_{ABCD}$ and, using
the reality condition, we get again a $O(1,1)^3$ in $E_{7(-7)}
/SU(8)$. Setting $\xi_{1234}= \xi_{5678} \equiv \xi_{12}$,
$\xi_{1256}= \xi_{3478} \equiv \xi_{13}$, $\xi_{1278}= \xi_{3456}
\equiv \xi_{14}$, we have the following set of transformations:
\begin{eqnarray}
 \delta e_1 &=& \xi_{12} e_2  + \xi _{13} e_3   + \xi _{14} e_4\label{trans81}
 \\
 \delta e_2 &=& \xi_{12} e_1  + \xi _{13} e_4   + \xi _{14} e_3\label{trans82}
 \\
 \delta e_3 &=& \xi_{12} e_4  + \xi _{13} e_1   + \xi _{14} e_2\label{trans83}
 \\
 \delta e_4 &=& \xi_{12} e_3  + \xi _{13} e_2   + \xi _{14} e_1\label{trans84}
\end{eqnarray}

These transformations fix uniquely the relative coefficients of
the three
 $SU(8)$
invariants:
\begin{eqnarray}
I_1&=& e_1^4 + e_2^4 + e_3^4 + e _4^4 \\
I_2&=& (e_1^2 + e_2^2 + e_3^2 + e _4^2)^2 \\
I_3&=& e_1e_2e_3e _4 \\
\end{eqnarray}

It is easy to see that the transformations
(\ref{trans61}-\ref{trans64}) and
 (\ref{trans81}-\ref{trans84}) correspond to three commuting matrices
(with square equal to $\bfone$):
\begin{equation}
\left(\begin{matrix}0&0&0&1 \cr 0&0&1&0 \cr 0&1&0&0 \cr 1&0&0&0
\cr\end{matrix}\right) ; \left(\begin{matrix}0&1&0&0 \cr 1&0&0&0
\cr 0&0&0&1 \cr 0&0&1&0 \cr\end{matrix}\right) ;
\left(\begin{matrix}0&0&1&0 \cr 0&0&0&1 \cr 1&0&0&0 \cr 0&1&0&0
\cr\end{matrix}\right)
\end{equation}
which are proper non compact Cartan elements of $K$. The reason we
get the same transformations for $N=6$ and $N=8$ is because the
extra singlet $e$ of $N=6$ can be identified with the fourth
eigenvalue of the central charge of $N=8$.

\subsubsection{Extrema of the BPS mass and fixed scalars}
In this subsection we would like to extend the analysis of the
extrema of the black-hole induced potential
\begin{equation}
V= \frac{1}{2} Z_{AB}\bar Z^{AB} + Z_I \bar Z^ I
\end{equation}
which was performed in ref \cite{fegika} for the $N=2$ case to all
$N>2$
 theories.

We recall that, in the case of $N=2$ special geometry with metric
 $g_{i\bar\jmath}$,
 at the fixed scalar critical point
$\partial _i V=0$ the Hessian matrix reduces to:
\begin{eqnarray}
( \nabla_i \nabla _{\bar\jmath} V)_{fixed}&=&
( \partial_i \partial _{\bar\jmath} V)_{fixed} = 2g_{i\bar\jmath}V_{fixed}\\
( \nabla_i \nabla _j V)_{fixed}&=&0
\end{eqnarray}
The Hessian matrix is strictly  positive-definite if the critical
point is not at the singular point of the vector multiplet
moduli-space. This matrix was related to the Weinhold metric
earlier  introduced in the geometric approach to thermodynamics
and used for the study of critical phenomena \cite{fegika}.

For $N$-extended supersymmetry, a form of this matrix was also
given and
 shown to be
equal to \footnote{Generically the indices $i,j$ refer to real
coordinates, unless the manifold is K\"ahlerian, in which case we
use holomorphic coordinates and
 formula (\ref{hessian})
reduces to the hermitean $i\bar\jmath$ entries of the Hessian
matrix.}:
\begin{equation}
V_{ij}= ( \partial_i \partial _j V)_{fixed} = Z_{CD} Z^{AB}
(\frac{1}{2}
 P^{CDPQ}_{\ \ \ \ ,j}P_{ABPQ,i} +
P^{CD}_{I,j} P^I_{AB,j}). \label{hessian}
\end{equation}

It is our purpose to further investigate properties of the
Weinhold metric
 for fixed scalars.

Let us first observe that the extremum conditions $\nabla_i V=0$,
using the
 relation
between the covariant derivatives of the central charges, reduce
to the
 conditions:
\begin{equation}
\epsilon^{ABCDL_1 \cdots L_{N-4}}Z_{AB} Z_{CD} = 0,\quad Z_I=0
\label{fixedsc}
\end{equation}
These equations give the fixed scalars in terms of electric and
magnetic charges and also show that the topological invariants of
the previous section reduce
 to the extremum of the square of the ADM mass since, when the above
 conditions are fulfilled,
$(Tr A)^2 = 2Tr(A^2)$, where $A_A^{\ B} = Z_{AB} \bar Z^{BC}$.

On the other hand, when these conditions are fulfilled, it is easy
to see that the Hessian matrix is degenerate. To see this, it is
sufficient to go, making an $H$ transformation, to the normal
frame in which these conditions imply $Z_{12} \neq 0 $ with the
other
 charges vanishing.
Then we have:
\begin{equation}
\partial_i \partial_j V \vert_{fixed} = 4 \vert Z_{12}\vert ^2
(\frac{1}{2} P^{12ab} _j P_{12ab,i} +  P^{ 12 I }_{,j} P_{12 I,i})
,\quad a,b \neq 1,2 \label{normetr}
\end{equation}

To understand the pattern of degeneracy for all $N$, we observe
that when only
 one central
charge in not vanishing the theory effectively reduces to an $N=2$
theory. Then the actual degeneracy respects $N=2$ multiplicity of
the scalars degrees of freedom in the sense that the degenerate
directions will correspond to the hypermultiplet
 content
of $N>2$ theories when decomposed with respect to $N=2$
supersymmetry.

Note that for $N=3$, $N=4$, where $P_{AB I} $ is present, the
Hessian is block diagonal.

For $N=3$, referring to eq. (\ref{viel3}), since the  scalar
manifold is K\"ahler, $P_{AB I}$ is a (1,0)-form while $P^{ AB I}=
\bar P_{AB I}$ is a (0,1)-form.
\par
The scalars appearing in the $N=2$ vector multiplet and
hypermultiplet content of the vielbein are $P_{3I}$ for the vector
multiplets and $P_{aI}$  ($a=1,2$) for the hypermultiplets.
 From equation (\ref{normetr}), which for the $N=3$ case reads
 \begin{equation}
  \partial_{\bar \jmath} \partial_i V \vert_{fixed} = 2 \vert Z_{12}\vert ^2
 P_{3I,\bar \jmath} P^{3I}_{,i}
\end{equation}
we see that the metric has $4n$ real directions corresponding to
$n$ hypermultiplets  which are degenerate.
\par
For $N=4$, referring to (\ref{viel4}),
 $P$ is the $SU(1,1)/U(1)$
vielbein which gives one matter vector multiplet scalar while $P_{
12 I}$ gives $n$ matter vector multiplets. The directions which
are hypermultiplets correspond to $P_{ 1a I}, P_{ 2a I}$
 ($
a=3,4$). Therefore the ``metric'' $V_{i j}$ is of rank $2n+2$.
\vskip 5mm
\par
For $N>4$, all the scalars are in the gravity multiplet and
correspond to
 $P_{ABCD}$.
\par
The splitting in vector and hypermultiplet scalars proceeds as
before. Namely, in the $N=5$ case we set
$P_{ABCD}=\epsilon_{ABCDL}P^L $ ($A,B,C,D,L =1,\cdots 5$). In this
case the vector multiplet scalars are $P^a$ ($a=3,4,5$) while the
hypermultiplet scalars are $P^1, P^2$ ($n_V = 3$, $n_h =1$).
\par
For $N=6$, we set $P_{ABCD} =\frac{1}{2} \epsilon_{ABCDEF}
P^{EF}$. The vector multiplet scalars are now described by
$P^{12}, P^{ab} $ ($A,B,...=1,...,6$; $a,b = 3,\cdots 6$), while
the hypermultiplet scalars are given in terms of $P^{1a}, P^{2a}
$. Therefore we get $n_V = 6+1 =7$, $n_h =4$.
\par
This case is different from the others because, besides the
hypermultiplets $P^{1a}, P^{2a}$, also the vector multiplet
direction $P^{12}$ is degenerate.

Finally, for $N=8$ we have $P_{1abc}, P_{2abc}  $ as
hypermultiplet scalars and $P_{abcd}$ as vector multiplet scalars,
which give $n_V=15$, $n_h = 10$ (note that in this case the
vielbein satisfies a reality condition: $P_{ABCD}= \frac{1}{ 4! }
\epsilon_{ABCDPQRS} \bar P ^{PQRS}$). We have in this case 40
degenerate directions.
\par
In conclusion we see that the rank of the matrix $V_{ij} $ is
$(N-2) (N-3) + 2 n$ for all the four dimensional theories.

\section{BPS black holes in $N=8$ supergravity}
 In this section we
consider BPS extremal black--holes in the context of $N=8$
supergravity.
\par
$N=8$ supergravity is the $4$--dimensional effective lagrangian
of both type IIA and type IIB superstrings compactified on a
torus $T^6$. Alternatively it can be viewed as the $4D$ effective
lagrangian of $11$--dimensional M--theory compactified on a torus
$T^7$. For this reason its $U$--duality group $E_{7(7)}(\ZZ)$,
which is defined as the discrete part of the isometry group of
its scalar manifold:
\begin{equation}
{\cal M}^{(N=8)}_{scalar} \, = \, \frac{E_{7(7)}}{SU(8)} \, ,
\label{e7su8n}
\end{equation}
unifies all superstring dualities relating the various consistent
superstring models. The {\it non perturbative BPS states} one
needs to adjoin  to the string states in order to complete linear
representations of the $U$--duality group are, generically, {\it
BPS black--holes}.
\par
These latter can be viewed as intersections of several $p$--brane
solutions of the higher dimensional theory {\it wrapped} on the
homology cycles of the $T^6$ (or $T^7$) torus. Depending on how
many $p$--branes intersect, the residual supersymmetry can be:
\begin{enumerate}
\item $ \frac{1}{2}$ of the original supersymmetry
 \item $\frac{1}{4}$ of the original supersymmetry
 \item $\frac{1}{8}$ of the original supersymmetry
\end{enumerate}
 The distinction between these three kinds of BPS solutions can be
considered directly in a $4$--dimensional setup and it is related
to the structure of the central charge eigenvalues and to the
behaviour of the scalar fields at the horizon. BPS black--holes
with a finite horizon area are those for which the scalar fields
are regular at the horizon and reach a fixed value there. These
can only be the $1/8$--type black holes, whose structure is that
of $N=2$ black--holes embedded into the $N=8$ theory. For $1/2$
and $1/4$ black--holes the scalar fields always diverge at the
horizon and the entropy is zero.
\par
The nice point, in this respect, is that we can make a complete
classification of all BPS black holes belonging to the three
possible types. Indeed the distinction of the solutions into these
three classes can be addressed a priori and, as we are going to
see, corresponds to a classification into different orbits of  the
possible ${\bf 56}$--dimensional vectors $Q=\{ p^\Lambda ,
q_\Sigma \}$ of magnetic-electric charges of the hole. Indeed
$N=8$ supergravity contains ${\bf 28}$ gauge fields
$A^\Lambda_\mu $ and correspondingly the hole can carry ${\bf
28}$ magnetic $p^\Lambda$ and ${\bf 28}$ electric $q_\Sigma$
charges. Through the symplectic embedding of the scalar manifold
\eqn{e7su8n} it follows that the field strengths $F^\Lambda_{\mu
\nu}$ plus their duals $G_{\Sigma \vert \mu \nu}$ transform in
the fundamental ${\bf 56}$ representation of $E_{7(7)}$ and the
same is true of their integrals, namely the charges.
\par
The Killing spinor equation that imposes preservation of either
$1/2$, or $1/4$, or $1/8$ of the original supersymmetries enforces
two consequences different in the three cases:
\begin{enumerate}
\item a different decomposition of the scalar field manifold into
two sectors: {\begin{itemize}
\item a sector of {\it dynamical scalar fields} that evolve in the radial
parameter $r$
\item a sector of {\it spectator scalar fields} that do not evolve in
$r$ and are constant in the BPS solution.
\end{itemize}
}
\item a different orbit structure for the charge vector $Q$
\end{enumerate}
Then, up to $U$--duality transformations, for each case one can
write
 a fully general {\it generating solution} that contains the
minimal necessary number of excited dynamical fields and the
minimal necessary number of non vanishing charges. All other
solutions of the same supersymmetry type can be obtained from the
generating one by the action of   $E_{7(7)}$--rotations.
\par
Such an analysis is clearly group--theoretical and requires the
use of appropriate techniques.
\subsection{Summary of $N=8$ supergravity.} \label{41}
 Let us now summarize the structure of
$N=8$ supergravity. The action is of the general form \eqn{genact}
with $g_{IJ}$ being the invariant metric of $E_{7(7)}/SU(8)$ and
the kinetic matrix ${\cal N}$ being determined via the appropriate
symplectic embedding of $E_{7(7)}$ into $Sp(56,\IR)$ (see table
\ref{topotable}). Hence, according to the general formalism
discussed in section 3 and to eq.\eqn{defu} we introduce the coset
representative $\IL$ of $\frac{E_{7(7)}}{ SU\left(8\right)}$ in
the ${\bf 56}$ representation of $E_{7(7)}$:
\begin{equation}
\label{Lcoset} \IL=\frac{1}{\sqrt{2}}\left(\begin{array}{c|c}
                             f+\mbox{\rm i} h & \bar{f}+\mbox{\rm i}\bar{h}\\
                             \hline\\
                             f-\mbox{\rm i} h & \bar{f}-\mbox{\rm i}\bar{h}
                             \end{array}
                     \right)
\end{equation}
where the submatrices $\left(h,f\right)$ are $28\times28$ matrices
labeled by antisymmetric pairs $\Lambda,\Sigma, A,B$ (
$\Lambda,\Sigma=1,\dots,8$, $A,B=1,\dots,8 $) the first pair
transforming under $E_{7(7)}$ and the second one under
$SU\left(8\right)$:
\begin{equation}
\left(h,f\right)=\left(h_{\Lambda\Sigma|AB},
f^{\Lambda\Sigma}_{~~AB}\right)
\end{equation}
As expected from the general formalism we have $\IL\in
Usp\left(28,28\right)$. The vielbein $P_{ABCD}$ and the
$SU\left(8\right)$ connection $\Omega_{A}^{~B}$ of
$\frac{E_{7(7)}}{SU\left(8\right)}$ are computed from the left
invariant $1$-form $\IL^{-1}d\IL$:
\begin{equation}
\label{L-1dL} \IL^{-1}d\IL=\left(\begin{array}{c|c}
                             \delta^{[A}_{~~[C}\Omega^{B]}_{~~D]} & \bar{P}^{ABCD}\\
                             \hline\\
                             P_{ABCD} & \delta_{[A}^{~~[C}\bar{\Omega}_{B]}^{~~D]}\end{array}
                     \right)
\end{equation}
where $P_{ABCD}\equiv P_{ABCD,
i}d\Phi^{i}~~~\left(i=1,\dots,70\right)$ is completely
antisymmetric and satisfies the reality condition
\begin{equation}
\label{realcondP} P_{ABCD}=\frac{1}{ 24}\epsilon_{ABCDEFGH}\bar
P^{EFGH}
\end{equation}
The bosonic lagrangian of $N=8$ supergravity is \cite{cre}
\begin{eqnarray}
\label{lN=8} {\cal L}&=&\int\sqrt{-g}\, d^4x\left(2R+\Im{\cal
N}_{\Lambda\Sigma|\Gamma\Delta}F_{\mu\nu}^{~~\Lambda\Sigma}F^{\Gamma\Delta|\mu\nu}+
\frac{1}{6}P_{ABCD,i}\bar{P}^{ABCD}_{j}\partial_{\mu}
\Phi^{i}\partial^{\mu}\Phi^{j}+\right.\nonumber\\
&+&\left.\frac{1}{2}\Re{\cal N}_{\Lambda\Sigma|\Gamma\Delta}
\frac{\epsilon^{\mu\nu\rho\sigma}}{\sqrt{-g}}
F_{\mu\nu}^{~~\Lambda\Sigma}F^{\Gamma\Delta}_{~~\rho\sigma}\right)
\end{eqnarray}
where the curvature two-form is defined as
\begin{equation}
R^{ab}=d\omega^{ab}-\omega^a_{~c}\wedge\omega^{cb}.
\end{equation}
and the kinetic matrix ${\cal N}_{\Lambda\Sigma|\Gamma\Delta}$ is
given by the usual general formula:
\begin{equation}
\label{defN} {\cal N}=hf^{-1}~~\rightarrow~~ {\cal
N}_{\Lambda\Sigma|\Gamma\Delta}=
h_{\Lambda\Sigma|AB}f^{-1~AB}_{~~~~~\Gamma\Delta}.
\end{equation}
Note that the $56$ dimensional (anti)self-dual vector
$\left(F^{\pm~\Lambda\Sigma},G^{\pm}_{\Lambda\Sigma}\right)$
transforms covariantly under $U\in Sp\left(56,\IR\right)$
\begin{eqnarray}
&&U\left(\begin{array}{c} F \\
G\end{array}\right)=\left(\begin{array}{c} F' \\
G'\end{array}\right)~~;~~
U=\left(\begin{matrix}A&B\cr C&D\cr\end{matrix}\right)\nonumber\\
&&A^tC-C^tA=0\nonumber\\
&&B^tD-D^tB=0\nonumber\\
&&A^tD-C^tB={\bf 1}
\end{eqnarray}
The matrix transforming the coset representative $\IL$ from the
$Usp\left(28,28\right)$ basis, eq.\eqn{Lcoset}, to the real
$Sp\left(56,\IR\right)$ basis is the Cayley matrix:
\begin{equation}
\IL_{Usp}={\cal C}\IL_{Sp}{\cal C}^{-1}~~~~~{\cal
C}=\left(\begin{matrix} \bfone & \mbox{\rm i} \bfone \cr \bfone
&-\mbox{\rm i}\bfone \cr\end{matrix}\right). \label{caylone}
\end{equation}
 Having established our definitions and notations, let us now
write down the Killing spinor equations obtained by equating to
zero the SUSY transformation laws of the gravitino $\psi_{A\mu}$
and dilatino $\chi_{ABC}$ fields of $N=8$ supergravity in a purely
bosonic background:
\begin{eqnarray}
\label{2} \delta\chi_{ABC}&=&~4 \mbox{\rm i}
~P_{ABCD|i}\partial_{\mu}\Phi^i\gamma^{\mu}
\epsilon^D-3T^{(-)}_{[AB|\rho\sigma}\gamma^{\rho\sigma}
\epsilon_{C]}=0\\
\label{psisusy2} \delta\psi_{A\mu}&=&\nabla_{\mu}\epsilon_A~
~-\frac{\rm 1}{4}\,
T^{(-)}_{AB|\rho\sigma}\gamma^{\rho\sigma}\gamma_{\mu}\epsilon^{B}=0
\end{eqnarray}
 where
$\nabla_{\mu}$ denotes the derivative covariant both with respect
to Lorentz and $SU\left(8\right)$ local transformations
\begin{equation}
\nabla_{\mu}\epsilon_A=\partial_{\mu}\epsilon_A-\frac{1}{
4}\gamma_{ab} \, \omega^{ab}\epsilon_A-\Omega_A^{~B}\epsilon_B
\label{delepsi}
\end{equation}
and where $T^{\left(-\right)}_{AB}$ is the ''dressed
graviphoton''$2-$form,
  defined according to the general formulae \eqn{gravi}
\begin{equation}
T^{\left(-\right)}_{AB}=\left(h_{\Lambda\Sigma AB}
\left(\Phi\right)F^{-\Lambda\Sigma}-f^{\Lambda\Sigma}_{~~AB}
\left(\Phi\right)G^-_{\Lambda\Sigma}\right) \label{tab}
\end{equation}
From equations (\ref{defN}), (\ref{specdef}) we have the following
identities that are the particular $N=8$ instance of
eq.\eqn{tiden0}:
\begin{eqnarray}
&&T^+_{AB}=0\rightarrow T^-_{AB}=T_{AB}\nonumber ~~~~~~
\bar{T}^-_{AB}=0\rightarrow \bar{T}^+_{AB}=\bar{T}_{AB}
\end{eqnarray}
Following the general procedure indicated by eq.\eqn{zab} we can
define the central charge:
\begin{equation}
Z_{AB}=\int_{S^2}{T_{AB}}=h_{\Lambda\Sigma|AB}
p^{\Lambda\Sigma}-f^{\Lambda\Sigma}_{~~AB}q_{\Lambda\Sigma}
\label{zab2}
\end{equation}
which in our case is an antisymmetric tensor transforming in the
${\bf {28}}$ irreducible representation of $SU(8)$. In
eq.\eqn{zab2}
 the integral of the two-form $T_{AB}$ is evaluated on a large
two-sphere at infinity and the quantized charges
($p_{\Lambda\Sigma},~q^{\Lambda\Sigma}$) are defined, following
the general eq.s \eqn{charges} by
\begin{eqnarray}
p^{\Lambda\Sigma}&=&\int_{S^2}{F^{\Lambda\Sigma}}\nonumber\\
q_{\Lambda\Sigma}&=&\int_{S^2}{\cal
N}_{\Lambda\Sigma|\Gamma\Delta}\star F^{\Gamma\Delta}.
\end{eqnarray}
\subsection{The Killing spinor equation and its covariance
group} \label{ksecov} In order to translate eq.\eqn{2} and
\eqn{psisusy2} into first order differential equations
 on the bosonic fields of supergravity we consider
a configuration where all the fermionic fields are zero and a
SUSY parameter that satisfies the following conditions:
\begin{equation}
\begin{array}{rclcl}
\chi^\mu \, \gamma_\mu \,\epsilon_{A} &=& \mbox{i}\, \IC_{AB}
\,  \epsilon^{B}   & ; &   A,B=1,\dots,n_{max}\\
\epsilon_{A} &=& 0  &;&   A> n_{max} \\
\end{array}
\label{kilspieq}
\end{equation}
Here $\chi^\mu$ is a time--like Killing vector for the
space--time metric ( in the following we just write
 $\chi^\mu \gamma_{\mu} = \gamma^0$) and
$ \epsilon _{A}, \epsilon^{A}$ denote the two chiral projections
of a single Majorana spinor: $ \gamma _5 \, \epsilon _{A} \, = \,
\epsilon _{A} $ , $ \gamma _5 \, \epsilon ^{A} \, = - \epsilon
^{A} $. We name such an equation the {\sl Killing spinor equation}
and the investigation of its group--theoretical structure is the
main task we face in order to derive the three possible types of
BPS black--holes, those preserving $1/2$ or $1/4$ or $1/8$ of the
original supersymmetry.
To appreciate the distinction among the three types of $N=8$
black--hole solutions we need to recall the results of
\cite{FGun} where a classification was given of the ${\bf
56}$--vectors of quantized electric and magnetic charges ${\vec
Q}$ characterizing such solutions. The basic argument is provided
by the reduction of the central charge skew--symmetric tensor
$\ZZ_{AB}$ to normal form. The reduction can always be obtained
by means of local $SU(8)$ transformations, but the structure of
the skew eigenvalues depends on the orbit--type of the $\bf
56$--dimensional charge vector which can be described by means of
its stabilizer subgroup $G_{stab}({\vec Q}) \subset E_{7(7)}$:
\begin{equation}
g \, \in \, G_{stab}({\vec Q}) \subset E_{7(7)} \quad
\Longleftrightarrow \quad g \,  {\vec Q}  = {\vec Q}
\end{equation}
There are three possibilities:
\begin{equation}
\begin{array}{cccc}
\null & \null & \null & \null \\
 \mbox{SUSY} & \mbox{Central Charge} & \mbox{Stabilizer}\equiv G_{stab} &
 \mbox{Normalizer}\equiv G_{norm}\\
  \null & \null & \null & \null \\
   1/2 & Z_1 = Z_2 =Z_3 =Z_4  & E_{6(6)} & O(1,1) \\
 \null & \null & \null & \null \\
 1/4 & Z_1 = Z_2 \neq Z_3=Z_4 & SO(5,5)&
 SL(2,\IR) \times O(1,1) \\
 \null & \null & \null & \null \\
 1/8 & Z_1 \ne Z_2 \ne Z_3 \ne Z_4 & SO(4,4) & SL(2,\IR)^3 \\
 \null & \null & \null & \null \\
\label{classif}
\end{array}
\end{equation}
where the normalizer $G_{norm}({\vec Q})$ is defined as the
subgroup of $E_{7(7)}$ that commutes with the stabilizer:
\begin{equation}
   \left[ G_{norm} \, , \,  G_{stab} \right] = 0
\end{equation}
The main result of \cite{mp1} is that the most general $1/8$
black--hole solution of $N=8$ supergravity is related to the
normalizer group $SL(2,\IR)^3$. In the subsequent paper \cite{mp2}
the $1/2$ and $1/4$ cases were completely worked out. Finally in
\cite{mp3} the explicit form of the generating solutions was
discussed for the $1/8$ case. In this paper we review the simlest
case, corresponding to 1/2 preserved supersymmetry. For the other
cases, we refer the reader to references \cite{mp3},\cite{pieric}.
\par
In all three cases the Killing spinor equation  breaks the
original $SU(8)$ automorphism group of the supersymmetry algebra
to the subgroup $Usp(2 \,n_{max})\times SU(8-2\,n_{max})\times
U(1)$
\par
We then have to decompose $N=8$ supergravity into multiplets of
the lower supersymmetry $N^\prime = 2 \, n_{max}$. This is easily
understood by recalling that close to the horizon of the black
hole one doubles the supersymmetries holding in the bulk of the
solution. Hence the near horizon supersymmetry is precisely
$N^\prime = 2 \, n_{max}$ and the black solution can be
interpreted as a soliton that interpolates between {\it ungauged}
$N=8$ supergravity at infinity and some form of $N^\prime$
supergravity at the horizon. \subsection{The $1/2$ SUSY case}
Here we have $n_{max} = 8$ and correspondingly the covariance
subgroup of the Killing spinor equation is $Usp(8)\, \subset \,
SU(8)$. Indeed
 condition \eqn{kilspieq} can be rewritten as follows:
 \begin{equation}
\label{urcunmez} \gamma^0 \,\epsilon_{A}  =  \mbox{i}\, \IC_{AB}
\, \epsilon^{B} \quad ; \quad A,B=1,\dots ,8
\end{equation}
where $\IC_{AB}= - \IC_{BA}$ denotes an $ 8 \times 8$
antisymmetric matrix satisfying $\IC^2 = -\bfone$. The group
$Usp(8)$ is the subgroup of unimodular, unitary $ 8 \times 8$
matrices that are also symplectic, namely that preserve the matrix
$\IC$. Relying on eq. (\ref{classif}) we see that in the present
case $G_ {\it stab}=E_{6(6)}$ and $G_{\it norm}=O(1,1)$.
Furthermore we have the following decomposition of the ${\bf 70}$
irreducible representation of $SU(8)$ into irreducible
representations of $Usp(8)$:
\begin{equation}
{\bf 70} \, \stackrel{Usp(8)}{\longrightarrow} \, {\bf 42} \,
\oplus \, {\bf 1} \, \oplus \, {\bf 27} \label{uspdecompo1}
\end{equation}
\par
Furthermore, we also decompose the $\bf 56$ charge representation
of $E_{7(7)}$ with respect to $O(1,1)\times E_{6(6)}$ obtaining
\begin{equation}
{\bf 56}\stackrel{Usp(8)}{\longrightarrow} ({\bf
1,27})\oplus({\bf 1,27})\oplus({\bf 2,1}) \label{visto}
\end{equation}
 \par
In order to  single out the content of the first order Killing
spinor equations we need to decompose them into irreducible
$Usp(8)$ representations. The gravitino equation \eqn{psisusy2}
is an ${\bf 8}$ of $SU(8)$ that remains irreducible under
$Usp(8)$ reduction. On the other hand the dilatino equation
\eqn{2} is a ${\bf 56}$ of $SU(8)$ that reduces as follows:
\begin{equation}
{\bf 56} \, \stackrel{Usp(8)}{\longrightarrow} \, {\bf 48} \,
\oplus \, {\bf 8} \label{uspdecompo2}
\end{equation}
Hence altogether we have that $3$ Killing spinor equations in the
representations ${\bf 8}$, ${\bf 8}^\prime$ , ${\bf 48}$
constraining the scalar fields
 parametrizing  the three subalgebras
${\bf 42}$, ${\bf 1}$ and ${\bf 27}$. In the sequel we will work
out the consequences of these constraints and find which scalars
are set to constants, which are instead evolving and how many
charges are different from zero. As we will explicitly see, the
content of the Killing spinor equations after $Usp(8)$
decomposition, is such as to set to a constant $69$ scalar fields:
indeed in this case $G_{norm} = O(1,1)$ and $H_{norm} =\bf 1$, so
that there is just one
 surviving field parametrizing $G_{norm}= O(1,1)$.
Moreover, the same Killing spinor equations tell us that the $54$
belonging to the two $({\bf 1},{\bf 27})$ representation of eq.
(\ref{visto}) are actually zero, leaving only two non--vanishing
charges transforming as a doublet of $O(1,1)$.
Let us now discuss the explicit solution.
 The $N=1/2$ SUSY preserving black hole
solution of $N=8$ supergravity has $4$ equal skew eigenvalues in
the normal frame for the central charges. The stabilizer of the
normal form is $E_{6(6)}$ and the normalizer of this latter in
$E_{7(7)}$ is $O\left(1,1\right)$:
\begin{equation}
E_{7(7)}\supset E_{6(6)}\times O\left(1,1\right) \label{orfan1}
\end{equation}
According to our previous discussion, the relevant subgroup of the
$SU\left(8\right)$ holonomy group is $Usp\left(8\right)$, since
the BPS Killing spinor conditions involve supersymmetry parameters
$\epsilon_A,~\epsilon^A$ satisfying eq.\eqn{urcunmez}. As
discussed in the introduction, it is natural to guess that modulo
$U-$duality transformations the complete solution is given in
terms of a single scalar field parametrizing $O\left(1,1\right)$.
Indeed, we can  now demonstrate that according to the previous
discussion there is just one scalar field, parametrizing the
normalizer $O\left(1,1\right)$, which appears in the final
lagrangian, since the Killing spinor equations imply that $69$ out
of the $70$ scalar fields are actually  constants. In order to
achieve this result, we have to decompose the $SU\left(8\right)$
tensors appearing in the equations \eqn{2},\eqn{psisusy2} with
respect to $Usp\left( 8\right)$ irreducible representations.
According to the decompositions
\begin{eqnarray}
\label{70-28decomp1/2}
{\bf 70}& \stackrel{Usp(8)}{=} &{\bf 42}\oplus{\bf 27}\oplus{\bf 1}\nonumber\\
{\bf 28}& \stackrel{Usp(8)}{=} &{\bf 27}\oplus{\bf 1}
\end{eqnarray}
we have
\begin{eqnarray}
\label{usp8dec} P_{ABCD}&=&\stackrel{\circ}{P}_{ABCD}+\frac{3}{2}
C_{[AB}\stackrel{\circ}{P}_{CD]}+\frac{1}{16}C_{[AB}C_{CD]}P\nonumber\\
T_{AB}&=&\stackrel{\circ}{T}_{AB}+\frac{1}{ 8}C_{AB}T
\end{eqnarray}
where the notation $ \stackrel{\circ}{t}_{A_1 \dots , A_n}$ means
that the antisymmetric tensor is  $Usp\left(8\right)$
irreducible, namely has vanishing $C$-traces: $C^{A_1 A_2}
\,\stackrel{\circ}{t}_{A_1 A_2 \dots , A_n}=0$.
\par
Starting from  equation \eqn{2} and using equation
(\ref{urcunmez}) we easily find:
\begin{equation}
4  P_{,a}\gamma^a\gamma^0-6T_{ab} \gamma^{ab}=0\, ,
\end{equation}
where we have twice contracted the free $Usp(8)$ indices with the
$Usp(8)$ metric $C_{AB}$. Next, using the decomposition
(\ref{usp8dec}), eq. \eqn{2} reduces to
\begin{equation}
-4  \left(  \stackrel{\circ}{P}_{ABCD,a}+ \frac{3}{2}
\stackrel{\circ}{P}_{[CD,a}C_{AB]}\right) C^{DL}\gamma^a\gamma^0-3
\stackrel{\circ}{T}_{[AB}\delta^L_{C]}\gamma^{ab}=0\, .
\label{basta}
\end{equation}
Now we may alternatively contract equation (\ref{basta}) with
$C^{AB}$ or $\delta^L_C$ obtaining two relations on $
\stackrel{\circ}{P}_{AB} $ and   $\stackrel{\circ}{T}_{AB}$ which
imply that they are separately zero:
\begin{equation}
\stackrel{\circ}{P}_{AB}=\stackrel{\circ}{T}_{AB}=0\, ,
\end{equation}
which also imply, taking into account (\ref{basta})
\begin{equation}
\stackrel{\circ}{P}_{ABCD}=0\, .
\end{equation}
Thus we have  reached the conclusion
\begin{eqnarray}
\label{42-27=0}
\stackrel{\circ}{P}_{ABCD|i}\partial_{\mu}\Phi^i\gamma^{\mu}\epsilon^D&=&0\nonumber\\
\stackrel{\circ}{P}_{AB|i}\partial_{\mu}\Phi^i\gamma^{\mu}\epsilon^B&=&0\\
\stackrel{\circ}{T}_{AB}&=&0 \label{tnot}
\end{eqnarray}
implying that $69$ out the $70$ scalar fields are actually
constant, while the only surviving central charge is that
associated with the singlet two-form $T$. Since $T_{AB}$ is a
complex combination of the electric and magnetic field strengths
\eqn{tab}, it is clear that eq. \eqn{tnot} implies the vanishing
of $54$ of the quantized charges $p^{\Lambda \Sigma},q_{\Lambda
\Sigma}$, the surviving two charges transforming as a doublet of
$O(1,1)$ according to eq. (\ref{visto}). The only non-trivial
evolution equation relates $P$ and $T$ as follows:
\begin{equation}
\label{singleteqn1/2} \left( \hat
P\partial_{\mu}\Phi\gamma^{\mu}-\frac{3}{ 2} \mbox{\rm i}\,
T^{(-)}_{\rho\sigma}\gamma^{\rho\sigma}\gamma^0
\right)\epsilon_A=0
\end{equation}
where we have set $P = \hat P d\Phi $  and $\Phi$ is the unique
non trivial scalar field parametrizing $O(1,1)$.
\par
In order to make this equation explicit we perform the usual
static ans\"atze. For the metric we set the ansatz \eqn{ds2U}. The
scalar fields are assumed to be radial dependent and for the
vector field strengths we assume the ansatz of
eq.\eqn{strenghtsans} which adapted to the $E_{7(7)}$ notation
reads as follows:
\begin{eqnarray}
F^{-\Lambda\Sigma}&=&\frac{1 }{ 4\pi}t^{\Lambda\Sigma}
\left(r\right)E^{\left(-\right)}\label{F}\\
t^{\Lambda\Sigma}\left(r\right)&=& 2\pi\left(g+\mbox{\rm
i}\ell\left(r\right)\right)^{\Lambda\Sigma} \label{smallt}
\end{eqnarray}
The anti self dual form $E^{\left(-\right)}$ was defined in eq.
\eqn{eaself}. Using (\ref{tab}), (\ref{F}) we have
\begin{equation}
\label{Tab} T_{ab}^- = {\rm i}\ t^{\Lambda\Sigma}(r) E_{ab}^-
C^{AB} {\rm Im}{\cal N }_{\Lambda
\Sigma,\Gamma\Delta}f^{\Gamma\Delta}_{~~AB} \, .
\end{equation}
A simple gamma matrix manipulation gives further
\begin{equation}
\label{gammaalgebra} \gamma_{ab}\, E^{\mp}_{ab}= 2 \mbox{\rm i}
\frac{e^{2U}}{ r^3}x^i\gamma^0\gamma^i\left(\frac{\pm 1+\gamma_5}{
2}\right)
\end{equation}
and we arrive at the final equation
\begin{equation}
\label{finale} \frac{d \Phi}{dr}= -\frac{\sqrt{3}}{4 }
\ell(r)^{\Lambda\Sigma}\, {\rm Im}{\cal N
}_{\Lambda\Sigma|\Gamma\Delta}\, f^{\Gamma\Delta}_{~~AB}\,
\frac{e^U}{r^2}\, .
\end{equation}
In eq. (\ref{finale}), we have set $p^{\Lambda\Sigma}=0$ since
reality of  the l.h.s. and of $f_{AB}^{\Gamma\Delta} $(see eq.
(\ref{f})) imply the vanishing of the magnetic charge.
Furthermore, we have normalized the vielbein component of the
$Usp(8)$ singlet as follows
\begin{equation}
\label{P} \hat P=4 \sqrt{3}
\end{equation}
which corresponds to normalizing the $Usp(8)$ vielbein as
\begin{equation}
\label{Pabcdnor} P_{ABCD}^{\left({\it singlet}\right)} = \frac{1}{
16}P C_{[AB} C_{CD]}= \frac{\sqrt{3}}{ 4}C_{[AB} C_{CD]}\, d
\Phi\, .
\end{equation}
This choice agrees with the normalization of the scalar fields
existing in the current literature. Let us now consider the
gravitino equation \eqn{psisusy2}. Computing the spin connention
$\omega^a_{~b}$ from equation (\ref{ds2U}), we find
\begin{eqnarray}
\omega^{0i}&=&\frac{dU}{ dr}\frac{x^i}{ r}
e^{U\left(r\right)}V^0\nonumber\\
\omega^{ij}&=&2\frac{dU}{ dr}\, \frac{x_k}{ r}\,
 \eta^{k[i}\, V^{j]} \, e^U
\end{eqnarray}
where $V^0=e^U \, dt$, $V^i = e^{-U} \, dx^i$. Setting
$\epsilon_A=e^{f\left(r\right)}\zeta_A$, where $\zeta_A$ is a
constant chiral spinor, we get
\begin{eqnarray}
\label{deltapsi} &&\left\{\frac{df}{dr}\frac{x^i}{
r}e^{f+U}\delta_A^BV^i+
\Omega_{A,\alpha}^{~B}\partial_i\Phi^{\alpha}e^fV^i\right.
\nonumber\\
&&\left.-\frac{1}{ 4}\left(2\frac{dU}{ dr}\frac{x^i}{ r}e^Ue^f
\left(\gamma^0\gamma^iV^0+\gamma^{ij}V_j\right) \right)\delta_A^B
\, + \, \delta_A^B \,T^-_{ab}\gamma^{ab}\gamma^c\gamma^0
V_c\right\}\zeta_B=0
\end{eqnarray}
where we have used eqs.\eqn{psisusy2},(\ref{delepsi}),
(\ref{usp8dec}). This equation has two sectors; setting  to zero
the coefficient of $V^0$ or of $V^i\gamma^{ij}$ and tracing over
the $A,B$ indices we find two identical equations, namely:
\begin{equation}
\frac{dU}{ dr}= -\frac{1}{ 8} \ell(r)^{\Lambda\Sigma}\frac{e^U}{
r^2} C^{AB}{\rm Im}{\cal N }_{\Lambda \Sigma,\Gamma\Delta}
f^{\Gamma\Delta}_{~~AB}. \label{dudr}
\end{equation}
Instead, if we set  to zero the coefficient of $V^i$, we find a
differential equation for the function $f\left(r\right)$, which
is uninteresting for our purposes. Comparing now equations
(\ref{finale}) and (\ref{dudr}) we immediately find
\begin{equation}
\label{phiu} \Phi= 2 \sqrt{3} \, U
\end{equation}
\par
In order to compute the l.h.s. of eq.s (\ref{finale}),
(\ref{dudr}) and the lagrangian of the 1/2 model, we need the
explicit form of the coset representative $\IL$ given in equation
(\ref{Lcoset}). This will also enable us to compute explicitly the
r.h.s. of equations (\ref{finale}), (\ref{dudr}). In the present
case the explicit form of $\IL$ can be retrieved by
exponentiating the $Usp\left(8\right)$ singlet generator. As
stated in equation (\ref{L-1dL}), the scalar vielbein in the
$Usp\left(28,28\right)$ basis is given by the off diagonal block
elements of $\IL^{-1}d\IL$, namely
\begin{equation}
\label{Pvielbein} \IP = \left(
\begin{array}{cc}
0&{\bar P}_{ABCD}\\
P_{ABCD}&0
\end{array}
\right).
\end{equation}
From equation (\ref{Pabcdnor}), we see that the
$Usp\left(8\right)$ singlet corresponds to the generator
\begin{equation}
\label{kappa} \IK =\frac{\sqrt{3}}{ 4}\left( \begin{array}{c|c}
                               0 & C^{[AB} C^{HL]}\\
                               \hline\\
                               C_{[CD}C_{RS]} & 0
                           \end{array}
                \right)
\end{equation}
and therefore, in order to construct the coset representative of
the $O\left(1,1\right)$ subgroup of $E_{7(7)}$, we need only to
exponentiate $\Phi\IK$. Note that $\IK$ is a $Usp\left(8\right)$
singlet in the $\bf 70$ representation of $SU\left(8\right)$, but
it acts non-trivially in the $\bf 28$ representation of the
quantized charges $\left(q_{AB}, p^{AB}\right)$. It follows that
the various powers of $\IK$ are proportional to the projection
operators onto the irreducible $Usp(8)$ representations $\bf 1$
and $\bf 27$ of the charges:
\begin{equation}
\label{P1} \IP_1=\frac{1}{8} C^{AB}C_{RS}
\end{equation}
\begin{equation}
\label{P27} \IP_{27}=(\delta_{RS}^{AB}- \frac{1}{8} C^{AB}C_{RS}).
\end{equation}
Straightforward exponentiation gives
\begin{eqnarray}
\exp(\Phi\IK)&=&\cosh\left(\frac{1}{
2\sqrt{3}}\Phi\right)\IP_{27}+\frac{3}{ 2}
\sinh\left(\frac{1}{ 2\sqrt{3}}\Phi\right)\IP_{27}\IK\IP_{27}+\\
&&+\cosh\left(\frac{\sqrt{3}}{ 2}\Phi\right)\IP_{1}+\frac{1}{2}
\sinh\left(\frac{\sqrt{3}}{2}\Phi\right)\IP_{1}\IK\IP_{1}
\end{eqnarray}
Since we are interested only in the singlet subspace
\begin{equation}
\label{P1proj} \IP_1\exp[\Phi\IK]\IP_1 = \cosh(\frac{\sqrt{3}}{
2}\Phi)\IP_1 + \frac{1}{2} \sinh(\frac{\sqrt{3}}{ 2}\Phi)
\IP_1\IK\IP_1
\end{equation}
\begin{equation}
\label{Lsinglet} \IL_{singlet}= \frac{1}{8}\left(
\begin{array}{c|c}
                \cosh (\frac{\sqrt{3}}{ 2}\Phi) C^{AB}C_{CD} &
                \sinh (\frac{\sqrt{3}}{ 2}\Phi) C^{AB} C^{FG}\\
                \hline\\
                 \sinh(\frac{\sqrt{3}}{ 2}\Phi) C_{CD}C_{LM} &
                 \cosh(\frac{\sqrt{3}}{ 2}\Phi) C_{CM}C^{FG}
                \end{array}
                \right).
\end{equation}
Comparing (\ref{Lsinglet}) with  the equation (\ref{Lcoset}), we
find \footnote{Note that we are we are writing the coset matrix
with the same pairs of indices $AB, CD, \dots$ without
distinction between the pairs $\Lambda\Sigma$ and $AB$ as was
done in sect. (\ref{41})}:
\begin{equation}
\label{f} f=\frac{1}{8 \sqrt{2}} e^{\frac{\sqrt{3}}{ 2}\Phi}
C^{AB}C_{CD}
\end{equation}
\begin{equation}
\label{h} h=- \mbox{\rm i} \, \frac{1}{8 \sqrt{2}}\, e^{-
\frac{\sqrt{3}}{ 2}\Phi} C_{AB}C_{CD}
\end{equation}
and hence, using ${\cal N}=hf^{-1}$, we find
\begin{equation}
\label{Nmatrix} {\cal N}_{AB \, CD} = - \, \mbox{\rm i} \,
\frac{1}{ 8}\, e^{-\sqrt{3}\Phi} C_{AB}C_{CD}
\end{equation}
so that we can compute the r.h.s. of (\ref{finale}), (\ref{dudr}).
Using the relation (\ref{phiu})  we find a single equation for
the unknown functions $U\left(r\right)$,
$\ell\left(r\right)=C_{\Lambda\Sigma}
\ell^{\Lambda\Sigma}\left(r\right)$
\begin{equation}
\label{dudr1} \frac{dU}{dr}=\frac{1}{
8\sqrt{2}}\frac{\ell\left(r\right)}{ r^2} \exp\left(-2U\right)
\end{equation}
At this point to solve the problem completely  we have to consider
also the second order field equation obtained from the lagrangian.
The bosonic supersymmetric lagrangian of the $1/2$ preserving
supersymmetry case is obtained from equation (\ref{lN=8}) by
substituting the values of $P_{ABCD}$ and ${\cal
N}_{\Lambda\Sigma|\Gamma\Delta}$ given in equations
(\ref{Pabcdnor}) and (\ref{defN}) into equation (\ref{lN=8}). We
find
\begin{equation}
\label{lbose1/2} {\cal L} = 2R- e^{- \sqrt{3} \Phi} F_{\mu \nu}
F^{\mu \nu} + \frac{1}{2} \partial_\mu \Phi \partial^\mu \Phi
\end{equation}
Note that this action has the general form of $0$--brane action in
$D=4$. According to this we expect a solution where:
\begin{eqnarray}
U & = &-\frac{1}{4} \, \log \, H(r) \nonumber\\
\Phi &=& - \frac{\sqrt{3}}{2} \, \log \, H(r)\nonumber\\
\ell & = & 2 r^3\, \frac{d}{dr}  \left( H(r)
\right)^{-\frac{1}{2}} = k \, \times \, \left( H(r)
\right)^{-\frac{3}{2}} \label{indovin}
\end{eqnarray}
where $H(r)= 1+k/r$ denotes a harmonic function. \par
 The
resulting field equations are
\\
{\underline {\sl Einstein equation:}}
\begin{equation}
\label{einstein1/2} U'' + \frac{2}{r} U' - (U)^2 = \frac{1}{ 4}
(\Phi ')^2
\end{equation}
{\underline {\sl Maxwell equation:}}
\begin{equation}
\label{maxwell1/2} \frac{d}{dr}(e^{-\sqrt{3}\Phi} \ell(r)) =0
\end{equation}
{\underline {\sl Dilaton equation:}}
\begin{equation}
\label{dilaton1/2} \Phi'' + \frac{2}{r} \Phi' =-e^{-\sqrt{3}\Phi
+ 2 U}
        \ell(r)^2 \frac{1}{r^4}   \, .
\end{equation}
From Maxwell equations one immediately finds
\begin{equation}
\ell\left(r\right)=e^{\sqrt{3}\Phi\left(r\right)}.
\end{equation}
Taking into account (\ref{phiu}), the second order field equation
and the first order Killing spinor equation have the common
solution
\begin{eqnarray}
\label{solution1/2}
U & = & -\frac{1}{4} \log \, H(x) \nonumber\\
\Phi     & = & - \frac{\sqrt{3}}{2} \log \, H(x)\nonumber\\
\ell       & = &  H(x)^{- \frac{3}{2}}
\end{eqnarray}
where:
\begin{equation}
H(x) \equiv 1 + \sum_{ i} \, \frac{k_i}{  {\vec x} - {\vec
x}^0_i  } \label{armonie}
\end{equation}
is a harmonic function describing $0$--branes located at ${\vec
x}^0_\ell$ for $\ell=1, 2,\dots$, each brane carrying a charge
$k_i$. In particular for a single $0$--brane we have:
\begin{equation}
H(x) = 1+ \frac{k}{r} \label{unasola}
\end{equation}
and the solution reduces to the expected form \eqn{indovin}.
\section{Conclusions}
In this paper we have given an account of the deep connection due
to supersymmetry among the central charges of supergravity
theory, BPS states and the Bekenstein--Hawking entropy of
extremal black holes. One of the most relevant results concerns
the structure of the entropy formula, which turns out to depend
only on the quantized electric and magnetic charges of the theory.
Actually, the entropy is proportional to some group theoretical
invariants which can be constructed out of the duality group of
the corresponding supergravity theory.
\par
An important point, not discussed in this paper, is the fact that
four-dimensional black--hole configurations can be interpreted as
the four-dimensional appearence of more general configurations
named ``black--p-branes'' (namely p dimensional extended objects
in D dimensions interpolating between flat Minkowski space at
spatial infinity and the horizon geometry) in higher D dimensional
supergravities. When the extra D$-$4 spatial dimensions are
compactified, the p-brane configuration can be suitably wrapped
over some p-dimensional homology cycles of the compact space,
giving rise to a point-like configuration, namely the
four-dimensional black hole. This approach has been very fruitful
and has a wide range of applications, among which we mention the
fact that it allows a statistical interpretation of the
black-hole entropy. The reader interested in the subject is
referred to the literature
\cite{blackmicro},\cite{malda},\cite{mbrastelle},\cite{duffrep}.
\section*{Acknowledgements}
Work supported in part by the European Community's Human
Potential Programme under contract HPRN-CT-2000-00131 Quantum
Spacetime, in which L. Andrianopoli and R. D'Auria are associated
to Torino University.


\end{document}